\documentclass[11pt]{article}
\pdfoutput=1

\usepackage{jheppub}
\usepackage{url,comment}
\usepackage{times}
\usepackage{latexsym}
\usepackage{graphicx, graphics, hyperref, amsmath, amssymb, slashed, xcolor, bbm,bm,amsthm, array}
\usepackage{subfigure}
\usepackage{listings} 
\usepackage{trimclip}
\usepackage{amsmath}
\usepackage{afterpage}

\usepackage{lineno}
 
 \newcommand{\nc}{\newcommand}
\nc{\denselist}{\setlength{\itemsep}{0cm} \setlength{\parskip}{0cm}}
\nc{\ev}{\mathrm{eV}}
\nc{\mev}{\mathrm{MeV}}
\nc{\gev}{\mathrm{GeV}}
\nc{\kev}{\mathrm{keV}}
\nc{\tev}{\mathrm{TeV}}
\nc{\pev}{\mathrm{PeV}}
\nc{\eev}{\mathrm{EeV}}
\nc{\zev}{\mathrm{ZeV}}
 
\graphicspath{{plots/}}

\title{Cosmic-ray searches with the \mbox{MATHUSLA} detector}

\author{
\includegraphics[width=4cm]{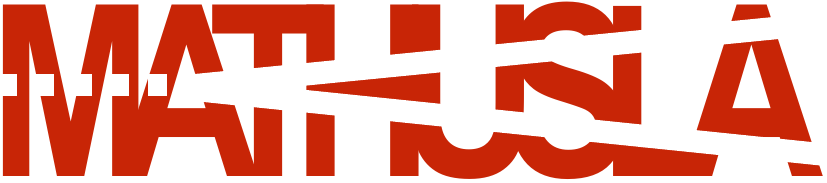}
\\
{\normalfont 
\href{https:mathusla-experiment.web.cern.ch}{\texttt{mathusla-experiment.web.cern.ch}}
\\
\vspace{5mm}}}

\author[a]{\hspace*{-1mm}Cristiano Alpigiani}
\author[b,*]{, J. C. Arteaga-Vel\'azquez}
\author[g]{, Daniela Blanco-Lira}
\author[c]{, Davide Boscherini}
\author[d]{, K. S. Caballero-Mora}
\author[e,f]{, Paolo Camarri}
\author[e]{, Roberto Cardarelli}
\author[i]{, Dennis Cazar Ram\'irez}
\author[e]{, Giuseppe Di Sciascio}
\author[g]{, Arturo Fern\'andez T\'ellez}
\author[a]{, H. J. Lubatti}
\author[d,j]{, O. G. Morales-Olivares}
\author[h]{, Piter Amador Paye Mamani}
\author[b]{, David Rivera Rangel}
\author[g]{, Mario Rodr\'iguez-Cahuantzi}
\author[e,f]{, Rinaldo Santonico}
\author[h]{, Martin Alfonso Subieta V\'azquez}

\affiliation[a]{University of Washington, Seattle, WA, USA}
\affiliation[b]{Instituto de F\'\i sica y Matem\'aticas, Universidad Michoacana, Morelia, Michoacan, Mexico }
\affiliation[c]{Istituto Nazionale di Fisica Nucleare, Sezione di Bologna, Bologna, Italy}
\affiliation[d]{Universidad Aut\'onoma de Chiapas, Tuxtla Guti\'errez, Chiapas, Mexico}
\affiliation[e]{Istituto Nazionale di Fisica Nucleare, Sezione di Roma Tor Vergata, Roma, Italy}
\affiliation[f]{Universit\`{a} degli Studi di Roma Tor Vergata, Roma, Italy}
\affiliation[g]{Benem\'erita Universidad Aut\'onoma de Puebla, Puebla, Mexico}
\affiliation[h]{Universidad Mayor de San Andr\'es, La Paz, Bolivia}
\affiliation[i]{Universidad San Francisco de Quito USFQ, Quito, Ecuador}
\affiliation[j]{Escuela Nacional de Ciencias de la Tierra, UNAM, Ciudad de Mexico, Mexico}

\affiliation[*]{Corresponding author. Email address: juan.arteaga@umich.mx}

\abstract{
The performance of the proposed MATHUSLA detector as an instrument for studying the physics of cosmic rays by measuring extensive air showers is presented. The MATHUSLA detector is designed to observe and study the decay of long-lived particles produced at the pp interaction point of the CMS detector at CERN during the HL-LHC data-taking period. The proposed MATHUSLA detector will be composed of many layers of long scintillating bars that cannot measure more than one hit per bar and correctly report the hit coordinate in case of multiple hits. This study shows that adding a layer of RPC detectors with both analogue and digital readout significantly enhances the capabilities of MATHUSLA to measure the local densities and arrival times of charged particles at the front of air showers.  We discuss open issues in cosmic-ray physics that the proposed MATHUSLA detector with an additional layer of RPC detectors could address and conclude by comparing with other air-shower facilities that measure cosmic rays in the PeV energy range.
}


\begin{document}

\maketitle

\section{Introduction}
This paper describes a study of the potential contributions of the proposed MATHUSLA detector to the understanding of the origin and nature of cosmic rays by means of extended air showers (EAS).  MATHUSLA is designed to search for new physics beyond the standard model, in particular, decays from neutral long-lived particles (LLPs). The detector is to be located on the surface above the CMS (Compact Muon Spectrometer) experiment at the CERN LHC.  The proposed MATHUSLA detector has nearly \mbox 100$\%$ coverage for charged particles over \mbox{100 m$^{2}$} \cite{Chou:2016lxi,Alpigiani:2018fgd,Alpigiani:2022Snowmass}. We show in this paper that by adding a layer of RPC detectors, the MATHUSLA detector can also make important contributions to many unresolved issues in cosmic-ray physics.

 In particular, in the next sections, we present the results of a Monte Carlo (MC) study to analyze the sensitivity of MATHUSLA as a detector of cosmic-ray showers.  We show that the addition of a layer of RPCs combined with the many layers of scintillating detector planes can provide unique spatial and temporal measurements of EAS. We discuss that such a detector provides information for studies of the energy spectrum, arrival-direction distribution and composition of cosmic rays that can be used for the research on the origin and nature of these particles and for tests of hadronic interaction models. We also discuss the potential for studying inclined events. 

 Section~\ref{sec:glance} gives a summary of open questions in the study of cosmic rays and Section~\ref{sec:MATHUSLA} a description of the detector as designed to address the particle physics goals.  In Section~\ref{s.configuration} we discuss the addition of a layer of RPC detectors that would allow for detailed measurements of EAS. Section~\ref{s.simulations} discusses the MC simulations of the production of EAS and the detector response, Section~\ref{EASreco} the event reconstruction methodology, and Section~\ref{selection}, the MC event selection. Results of this study are presented in Section~\ref{results} and discussed in Section~\ref{discussion}.  We present our conclusions in Section~\ref{s.conclusions}.

 \section{A glance at cosmic-ray physics}
\label{sec:glance}
The Earth is being continuously hit by high-energy subatomic particles known as cosmic rays. They are dominated by protons and can have energies as low as some MeV and as high as $10^{20} \mathrm{eV}$, i.e.  orders of magnitude beyond the energies of  the most powerful accelerators on Earth \cite{CosmicRays,CosmicRays1,CosmicRays2}. Their total energy spectrum can be described by a broken power-law, with a softening at around $4 \times 10^{15}$ eV (where the spectral index $\gamma$ changes from $-2.7$ to $-3$) and a hardening close to $5 \times 10^{18}$ eV (at which the spectral index increases to $-2.6$) \cite{CosmicRays2}.  Relative abundances of elemental nuclei in the intensity of cosmic rays are difficult to measure. In spite of this, it has been found that the composition of cosmic rays evolves with the primary energy \cite{CosmicRays2, PDG}. In the region below the knee, protons and He nuclei dominates the composition of cosmic rays, while at higher energies the heavier nuclei are more abundant. This happens up to  $\sim 10^{17}$ eV, where the composition of cosmic rays becomes lighter. This behavior stops at energies close to the ankle, where an additional transition to the heavy mass group is observed. The evolution of one mass group to another one seems to be correlated with the presence of fine structures in the energy spectrum of cosmic rays \cite{HAWCCR23, Antoni2005, Apel2013, PAO_Composition_2023, PDG}, which, in turn, could be related with acceleration \cite{Peters61} and propagation \cite{Giacinti15} issues or the transition between different source populations of cosmic rays \cite{Liu2022, PAO_Composition_2023}. 

The acceleration of cosmic rays may occur in astrophysical plasmas with strong shock waves by means of the so called first-order Fermi acceleration mechanism \cite{Axford77, Krymsky77, Bell78, Blandford78}, which predicts power-law spectra for the intensity of cosmic rays. The propagation in space of these energetic particles is mainly diffusive due to interactions with the random component of magnetic fields in the interstellar and extragalactic media. As a consequence, the angular distribution of the arrival directions of cosmic rays in the sky seems to be mostly isotropic. There are deviations from isotropy in the sky maps of cosmic rays at the level of  $10^{-3}$ and $10^{-4}$, which creates large- and small-scale anisotropic distributions in the sky \cite{Deligny2016, Ahlers2017, Deligny2019, Ahlers2019}. The large-scale anisotropies are thought to be due to propagation effects and the global distribution of the sources of cosmic rays in the galaxy and in our galactic neighborhood. The details behind the origin of the small-scale anisotropic structures are however less clear. At TeV energies, they may be related with the heliosphere or magnetic turbulences. At tens of EeV, they could be due to some extragalactic sources in the direction of Centaurus and the Ursa Major, as well as the Perseus-Pisces region \cite{PAO_TA_Aniso23}. 
A further implication of the diffusive propagation of cosmic rays in the universe is that they loss information about the position of their sources in the sky. Only at extreme energies, the small magnetic deflections of these particles in the space may allow to perform cosmic-ray astronomy \cite{CosmicRays2}. At energies above $10^{18}$ eV, cosmic rays are of extragalactic origin \cite{CosmicRays3}, while below $10^{17}$ eV, of galactic nature \cite{CRModels2019}. Somewhere, between $10^{17}$ and  $10^{18}$, a transition between galactic and extragalactic cosmic rays is expected \cite{Alosio12, PAO_Composition_2023}. In our galaxy, supernova remnants could be one of the astrophysical sources responsible for cosmic rays of high-energy \cite{CosmicRays3}. However, there are some clues that the galactic center \cite{CosmicRays4} and star forming regions \cite{HAWC_Cygnus_2021, LHAASO_ICRC2023_010} could also accelerate cosmic rays of very high energies in the Milky Way \cite{HAWC_Cygnus_2021}. At ultra high-energies, starburst galaxies are promising candidates \cite{PAO_POSICRC2023_252}. 

The existence of cosmic rays faces us with several compelling questions about the universe at very high energies. The quest for satisfactory answers to these phenomena is  pushed to the limit by the combined observations of cosmic rays, gamma rays, cosmic neutrinos and gravitational waves \cite{MultimessAJL2017,MultimessAA2017}. The astroparticle messenger approach calls for scenarios that involve both the domain of elementary particles, the so-called "micro-cosmos", and the realm of the huge celestial objects, what we know as "macro-cosmos". 

There are several open questions as to the astroparticle messengers phenomena such as:
\begin{itemize}
    \item What is the origin of the cosmic-ray particles of extraordinary energy?
    \item What is the mass composition of cosmic rays?
    \item What are the phenomena behind the small- and large-scale cosmic-ray anisotropies?
    \item How are cosmic-rays accelerated to high energies?
\end{itemize}

When cosmic rays of high energies arrive to the Earth, they collide with the molecules in the atmosphere at altitudes that goes from  $\sim 15$ up to $35$ km \cite{EASmodelsReview}  initiating a set of chain reactions that leads to the production of a cascade of particles that travels at relativistic velocities to the ground. This extensive air shower, as it is known, contains a large number of particles, which depends on the primary energy of the cosmic ray and increases according to a power-law function. The air shower has electromagnetic, penetrating and hadronic components. The first one is most abundant and is composed of photons and $e^{\pm}$'s, the second one, by $\mu^{\pm}$'s, and the last one, by nucleons, mesons, and atomic nuclei. During their development in the air, EAS produce radio and Cherenkov radiation \cite{CosmicRays}. Also, the atmosphere emits fluorescence light from the excitation of nitrogen molecules \cite{CosmicRays2}.
 
 At energies above $10^{13}$ eV, cosmic rays are detected indirectly from the measurements of EAS using arrays of particle detectors, radio antennas, Cherenkov and fluorescence telescopes that are displayed over large surfaces \cite{CosmicRays}. At energies smaller than $10^{15}$ eV,  cosmic-ray studies are observed in a direct manner from balloon-borne and satellite instruments  \cite{CosmicRays2, DirectCR23}.  
 
 Data collected by many ground and underground-based observatories as well as by sensitive radiation detectors installed on balloons and also on satellites orbiting Earth \cite{CosmicRays13,CosmicRays14,CosmicRays15,CosmicRays16,CosmicRays17,CosmicRays18,CosmicRays19} have provided much information on the energy spectrum, arrival direction and mass composition of cosmic rays that have revealed very important clues about the nature, propagation, acceleration mechanism and the origin of these energetic particles    \cite{CosmicRays3,CosmicRays5,CosmicRays6,CosmicRays7,CosmicRays8,CosmicRays9,CosmicRays10}. In addition, observations from the northern hemisphere of high-energy gamma rays  reveals many potential sources of cosmic rays within our galaxy \cite{CosmicRays4,CosmicRays11,CosmicRays12,Lhaaso2021}. 

Notwithstanding these excellent measurements, the abovementioned questions still remain without conclusive answers. They will be addressed by a new generation of ground-based observatories, some of which are located at high altitudes ($\sim 4500 \, \mbox{m}$ a.s.l) in the Earth’s southern \cite{CosmicRays20,CosmicRays21} and northern hemispheres \cite{CosmicRays22}, and by new space-based instruments \cite{Olinto21}. In addition, multiwavelength and multimessenger observations will  play an important role in these tasks \cite{MultimessAJL2017,MultimessAA2017}. We demonstrate in this paper that the MATHUSLA detector could also make important contributions to advancing the understanding of the astrophysics and physics of high-energy cosmic rays.

\begin{figure}[hbtp!]
\begin{center}
\includegraphics[width=0.7\textwidth]{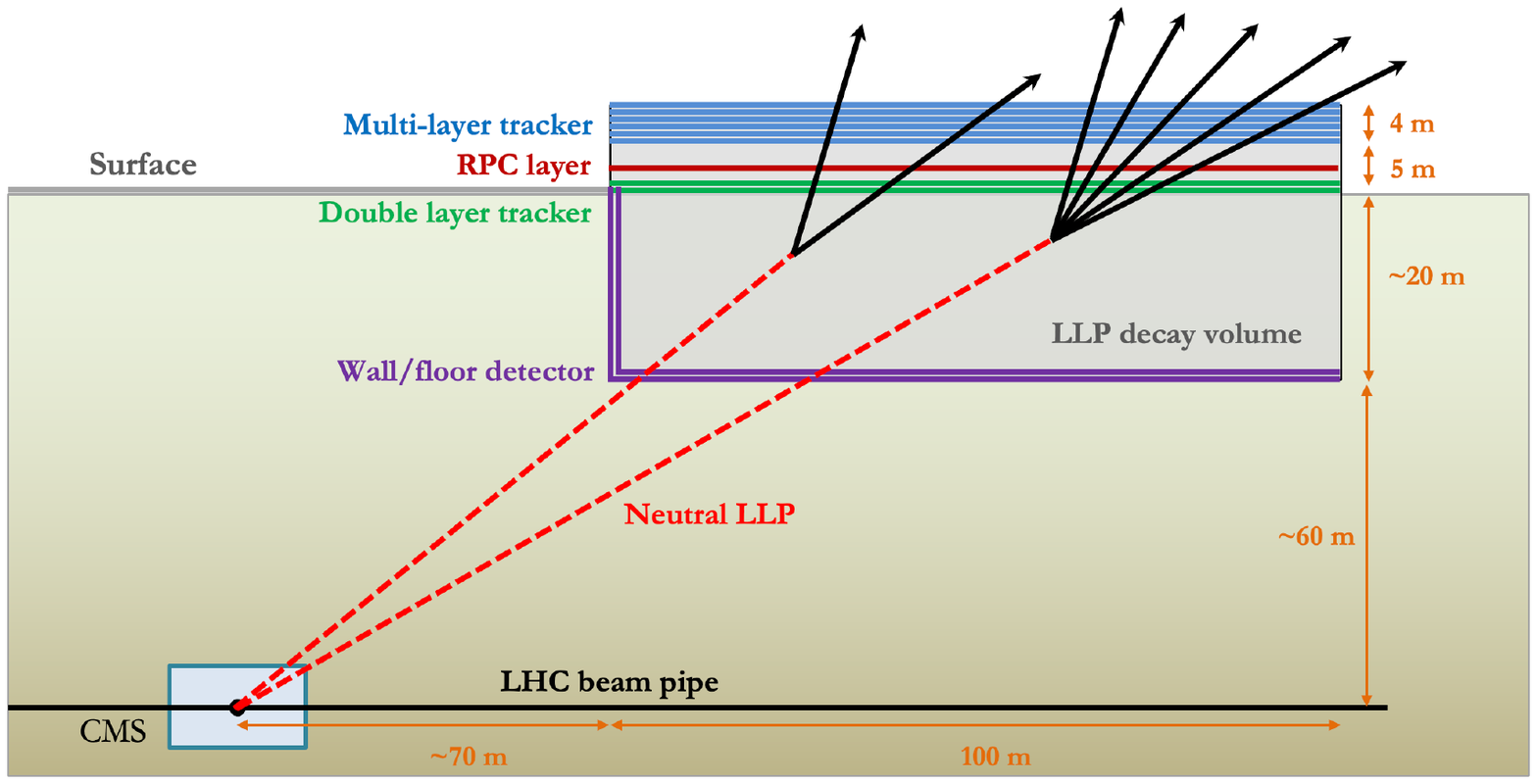}
\end{center}
\caption{MATHUSLA detector at CMS site \cite{Alpigiani:2022Snowmass}.}
\label{fig:layout_v4}
\end{figure}

\section{The MATHUSLA Detector}
\label{sec:MATHUSLA}
The proposed MATHUSLA detector shown in Figure \ref{fig:layout_v4}\ will be located on the surface above the CMS~\cite{CMS} $pp$ interaction point (IP) located on CERN-owned land. The proposed goal is to have MATHUSLA ready to take data when the High-Luminosity LHC (HL-LHC)~\cite{HL-LHC} turns on. 

\begin{figure}[hbtp!]
\begin{center}
\includegraphics[width=0.48\textwidth]{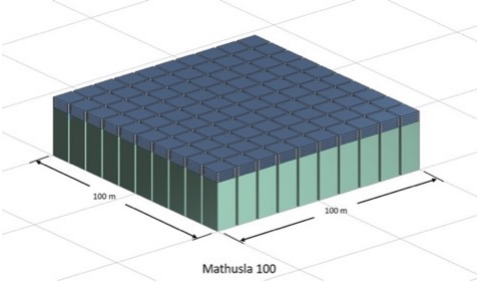}
\includegraphics[width=0.48\textwidth]{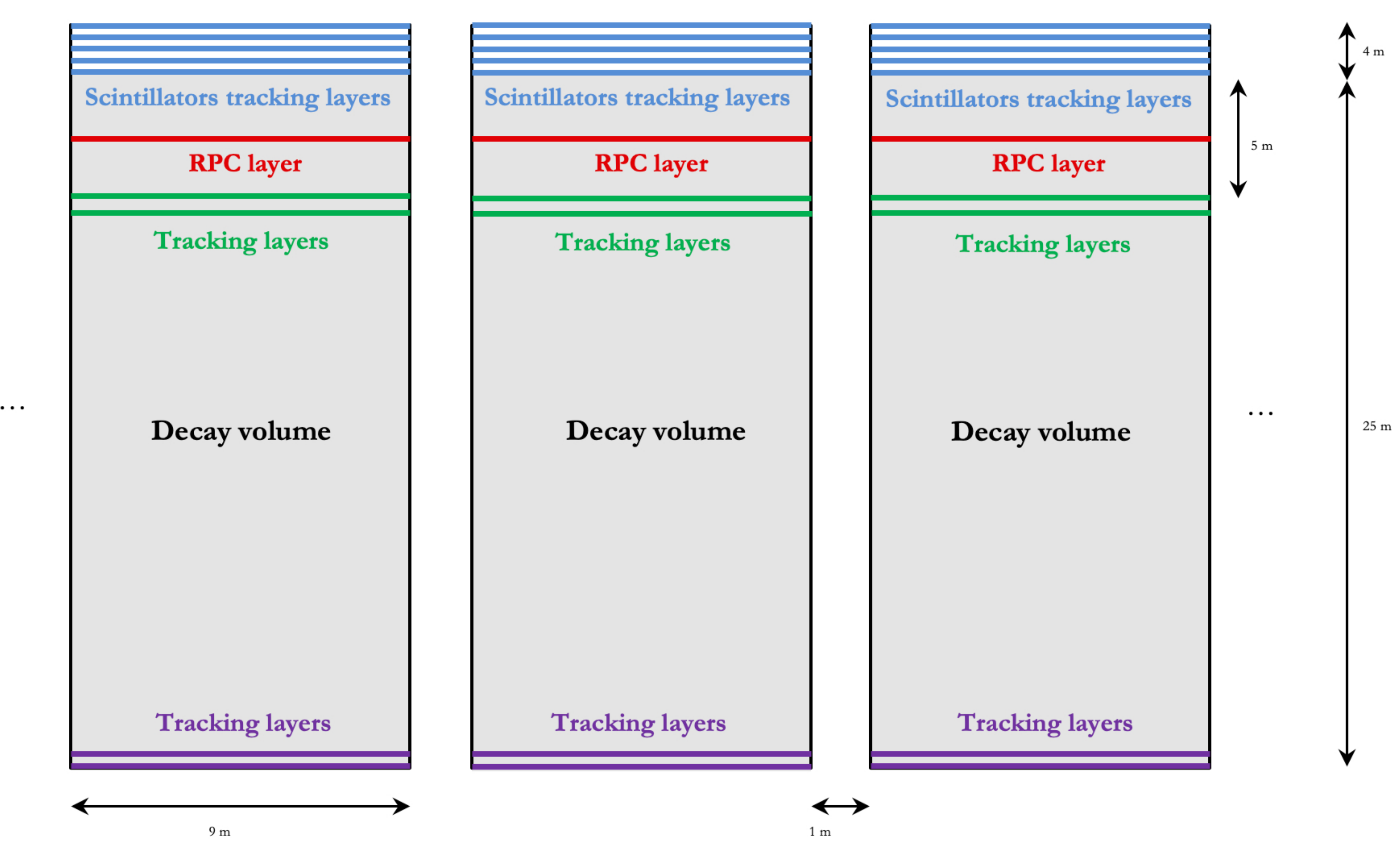}
\end{center}
\caption{Schematic view of the MATHUSLA detector modular concept \cite{Alpigiani:2022Snowmass}.}
\label{fig:geometry}
\end{figure}

To have sufficient solid-angle coverage, the current detector concept, shown in Fig.~\ref{fig:geometry}, is a \mbox{100 $\times$ 100 m$^{2}$} detector consisting of a hundred $9 \,\textrm{m} \times 9 \,\textrm{m}$ units. Each detector unit comprises 9-layers of scintillating-detector planes that provide position and timing coordinates of charged particles resulting from the decay of long-lived particles in the \mbox{MATHUSLA} detector decay volume. There are five detector planes at the top, two additional planes 5 m below (to enhance the particle position measurement precision close to the floor), and two additional planes at the floor (to veto charged particles from the LHC and cosmic-muon backscattering).  The total height of $\sim 40$ m includes a $\sim 25$ m LLP decay volume, 21 m of which is below ground, and 17 m above ground that houses the tracker and the crane system used for assembly and maintenance. 

\begin{figure}[hbtp!]
\begin{center}
\includegraphics[width=0.7\textwidth]{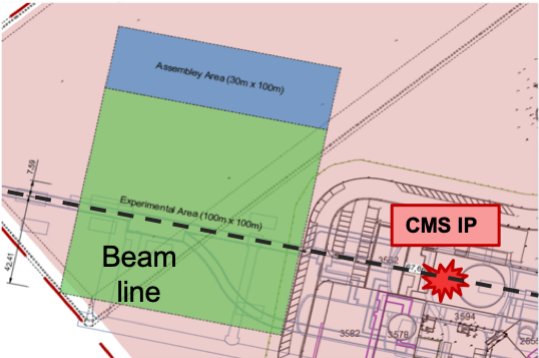}
\caption{MATHUSLA detector at CMS site \cite{Alpigiani:2022Snowmass}.}
\end{center}
\label{fig:layout_P5}
\end{figure}

The layout of the building that houses the 100 m $\times$ 100 m experimental area and the 30 m $\times$ 100 m adjacent area for the detector assembly is shown in Figure \ref{fig:layout_P5}. The structure, which is located on the surface near the CMS IP fits well on CERN-owned land.  Having a large part of the decay volume underground brings it closer to the IP, which increases the solid angle in the acceptance of LLPs generated in the collisions. To adjust to the available land, this proposal has a 7.5 m offset to the centre of the beams. The site allows for the detector to be as close as 68 m from the IP, which is shown in red in Figure \ref{fig:layout_P5}\

\subsection{Scintillating Detector Planes}
\label{Scintillator.plane}

\begin{figure}[hbtp!]
\begin{center}
\includegraphics[width=0.5\textwidth]{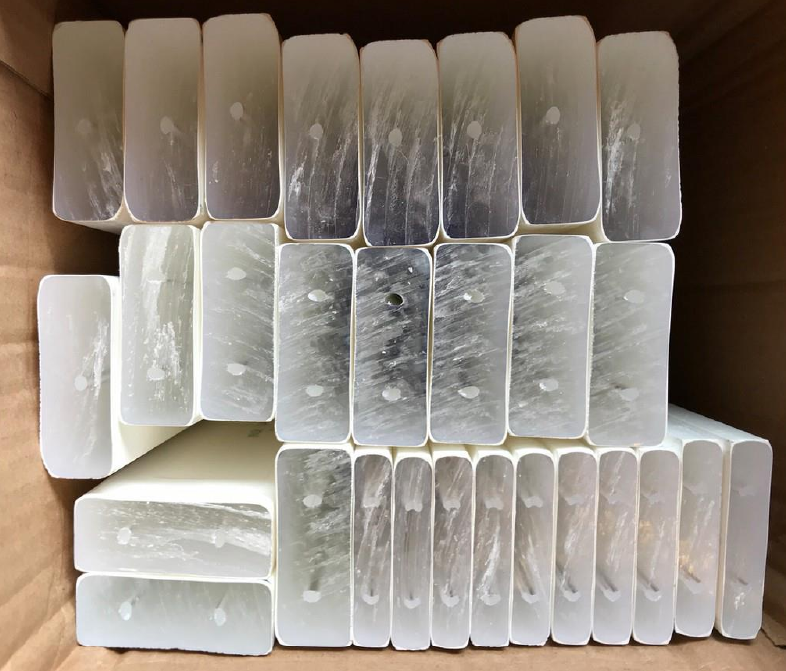}
\end{center}
\caption{Extruded scintillating bars produced at FNAL with one and two holes for WLS fiber insertion.}
\label{fig:FNAL_extrusions}
\end{figure}

Each of the $9 \,\textrm{m} \times 9 \,\textrm{m}$ detector planes consists of an assembly of extruded scintillating bars whose length, width and thickness are typically $4.6$ m, $4.5$ cm, $2$ cm, respectively.  Each bar is extruded with a hole at the centre into which a wave-length shifting (WLS) fibre is inserted and connected at each end to a SiPM. The coordinate along the length of the bar is determined by differential time measurement at the two ends of the bar with resolution $\sigma \sim \pm$ $15$ cm. The width of the bar determines the corresponding transverse coordinate with $\sigma = \pm$ $1.3$ cm. Figure \ref{fig:FNAL_extrusions}\ shows some typical extruded scintillating bars produced at the Fermi National Accelerator Laboratory (FNAL).

To facilitate installation the scintillating bars are assembled into $4$ sub-units that comprise $\sim100$ bars resulting in $\sim4.5 \,\textrm{m} \times 4.5 \,\textrm{m}$ sub-units that allow for overlapping the $4$ sub-units longitudinally and transversely to avoid gaps in coverage. In this arrangement the $4$ sub-units provide hermetic surface-coverage over a $\sim9 \,\textrm{m} \times 9 \textrm{m}$. 

\subsection{Cosmic Rays}

The dominant background to long-lived particle searches in the MATHUSLA detector is cosmic rays. Timing information from the scintillating bars allows for the rejection of down\-ward-going charged cosmic rays. The two-floor detectors provide rejection of inelastic back-scattering of cosmic-ray muons striking the surface and muon decays that send an upward electron through the detector planes. They also serve to reject inelastic interactions of muons near the surface.  Several simulation studies show that these backgrounds can be rejected. A more detailed description of the possible backgrounds can be found in \cite{Alpigiani:2018fgd}.

MATHUSLA, equipped with seven scintillating detector planes at the top of the instrument that has good space and time resolution (see section 3.1) and nearly continuous coverage over an area of $10^{4}$ $\mbox{m}^{2}$, is also sensitive to air showers induced by cosmic rays. However, it is limited as an air shower detector because the long scintillating bars are not able to provide coordinates of several charged tracks hitting the same bar at nearly the same time. Such occurrences lead to saturation and loss of information. 

\section{The RPC detector}
\label{s.configuration}
\subsection{Introduction}

The peculiarity of MATHUSLA, as a cosmic-ray experiment, is that its detector is primarily conceived for searching for very long-lived particles produced by pp interactions in the LHC. Studying its performance as a cosmic-ray detector, and substantially improving it with a special sub-detector dedicated to the study of extensive cosmic-ray showers, is discussed in this section.

The main constraints imposed by the primary purpose of the experiment are the operation at CERN's very modest altitude ($373.6$ m.a.s.l.) and the sensitive area not exceeding $10^4$ m$^2$. On the other hand, MATHUSLA, equipped with $10$ scintillating planes that provide good spatial resolution and one ns time resolution is an unprecedented large area cosmic-ray tracker.

Taking into account the features described above the cosmic-ray search of MATHUSLA may be focused on the following main items:
\begin{itemize}
\item	Cosmic-ray composition, i.e. measurement of the atomic number Z of the primary particle.

\item	Parallel-muon bundles crossing the detector. In this case, a “pure muonic shower” is observed. For inclined or almost horizontal showers that traverse a larger thickness of the atmosphere where the hadronic component is absorbed, purer muonic showers can be observed at high and very high energies.

\end{itemize}

However, a scintillating tracker such as MATHUSLA has the following limits for detecting EAS:
\begin{itemize}
\item Limited resolution for multiple-track detection. The limit is no more than approximately $1$ track in a scintillating bar of area  ($450 \times 4.5$) cm$^2$ = $2025$ cm$^2$. Multiple tracks in the same scintillating bar, in addition to being counted as a single track, also spoil the timing information and do not allow localization of the track position along the scintillating bar. This severely limits the study of parallel-muon bundles, which is an excellent field of investigation where MATHUSLA could be competitive with other cosmic-ray experiments.

The MATHUSLA scintillating bars do not provide energy deposition measurements that are extremely important for the shower-core detection where the hits produced by electro-photonic component can reach a density of $10^6$ m$^{-2}$ for a primary cosmic-ray particle of $1$ PeV.

\item Absence of an obvious muon-identification strategy, which is normally based on a heavy-material layer absorbing secondary particles except muons. This is impossible for MATHUSLA and requires a different strategy. It should be stressed that muon detection is important in any channel of cosmic-ray search, but it is crucial in particular for gamma-hadron discrimination. A possible muon-detection alternative strategy could be based on the identification of the regular linear pattern of a muon track from the random hits produced by the electro-photonic component of the shower. This could be possible thanks to the tracking capability of MATHUSLA. The expected efficiency of this method in identifying a muon would be modest. However, at about $1$ PeV a proton shower produces an order of $10^{4}$ muons, and the identification of just one of them would be sufficient to reject the shower as a gamma ray candidate.

\end{itemize}

\subsection{The RPC sub-detector}
\label{sec:RPC}

The addition of a single RPC layer to the layers of MATHUSLA scintillating detectors can significantly enhance the capability of the instrument to make detailed measurements of EAS. In this section, we describe the RPC chambers, the proposed layout for MATHUSLA, and the front-end electronics.  The proposed RPC configuration profits from the experience obtained during 5 years of operation of the ARGO-YBJ experiment \cite{AIELLI200692}.  
The main difference compared to ARGO-YBJ is that the RPC chambers for MATHUSLA will operate in avalanche mode instead of the streamer mode chosen for ARGO-YBJ. Operating in avalanche mode requires improved front-end electronics to compensate for the smaller charge and lower discrimination threshold, which has the advantage of substantially extending the linear response range of the analogue readout \cite{BARTOLI201547}.

\begin{figure}[hbtp!]
\begin{center}
\includegraphics[width=1.0\textwidth]{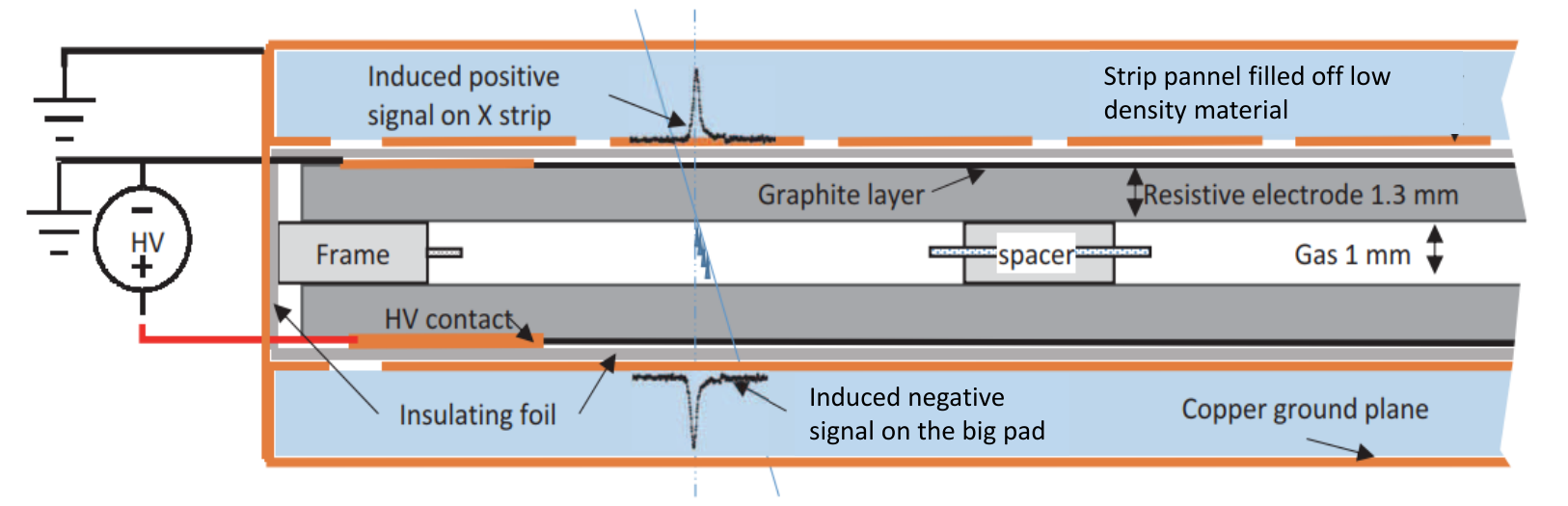}
\end{center}
\caption{RPC gas gap sandwiched between the strip and the ``big pad'' panels. The main components are the high-pressure laminate electrode plates, the graphite electrodes, the insulating PET foil, and the spacers.}
\label{fig:GasGap}
\end{figure}

The gas-gap thickness is another difference with respect to ARGO-YBJ. At the ARGO-YBJ altitude of $4300$ m a.s.l., the atmospheric pressure was about $600$ mbar to be compared with the CERN pressure of $980$ mbar. This makes possible a substantial reduction of the gap with respect to the $2$-mm gap of the ARGO-YBJ RPCs. We are testing $1$-mm gas gaps that will be used for the BI RPCs of the ATLAS muon spectrometer \cite{ATLAS-TDR-026}. The choice of the $1$-mm standard for the gap size of the MATHUSLA RPCs allows us to exploit the research and development of the ATLAS Phase II upgrade. A further advantage of reducing the gap size to $1$ mm is improved time resolution. The MATHUSLA RPC gas gap is schematically shown in Fig.~\ref{fig:GasGap}

\par

The main RPC parameters that can be easily optimized concern the read-out granularity. Following the ARGO-YBJ scheme there are two types of read-out electrodes: the digitally readout strips located on top of the gas gap in Fig.~ \ref{fig:GasGap} and and the “big pads”, located below the gas gap in Fig.~ \ref{fig:GasGap}. The pads are large area capacitors, covering half of the gap area, and integrate the charge of all hits located in the corresponding pad area. The analog readout of the pad signal makes it possible to measure the shower-core multiplicity when it is so high that the digital readout is saturated.
The strips, longitudinally oriented, are $2.2$ m long with an $11.2$-mm pitch with front-end electronics located at both ends in order to determine the hit position along the strip using the signal arrival time at the two ends. The area of the strips, $242$ cm$^2$, is substantially smaller than the area of the scintillating bars, $2025$ cm$^2$, which significantly reduces the probability of multiple hits in the same strip. This is particularly important for the detection of parallel muon bundles where the probability of two or more muons striking the same scintillating bar is large as well as for selecting straight tracks from the background of randomly distributed hits due to the electro-photonic component of the shower.

The  $0.99$ m$^2$“big pads” ($1.1 \times 0.9$) m$^2$), are smaller than those of ARGO-YBJ ($1.75$ m$^2$) and provide better detail of the shower-core profile. This is important for identifying showers generated by different kinds of primary cosmic rays. For example, gamma showers have a very compact single core compared to the less compact multiple-core structure of hadron showers. Moreover, the high-Z primary showers are broader than proton showers and have a substantially larger multiplicity of different cores. 

These qualitative considerations need to be studied in detail with an adequate simulation in order to optimize the granularity of both strips and “big pads”, both of which can be easily tuned for specific physics channels.

The RPC characteristics are the following:
\begin{itemize}
\item Gas gap: a single 1-mm gap of area ($2.2 \times 0.9$) m$^2$ with $2$ gas inlets and $2$ outlets at the corners.
\item Strip read-out panels: $80$ longitudinal strips $2.2$ m long with $11.2$-mm pitch. Front-end boards are located at both ends of the strip for interpolating the position along the strip. The alternative solution with the strips oriented along the short side of the chamber can also be considered but it is less practical for the final assembly of the chamber.
\item ``Big pads’’ for the analogue read-out of area ($1.1 \times 0.9$) m$^2$
\item Front-end electronics embedded in the same Faraday cage of the detector
\item Mechanical support structure made with honeycomb panels about $1$ cm thick.
\end{itemize}

\begin{figure}[hbtp!]
\begin{center}
\includegraphics[width=1.0\textwidth]{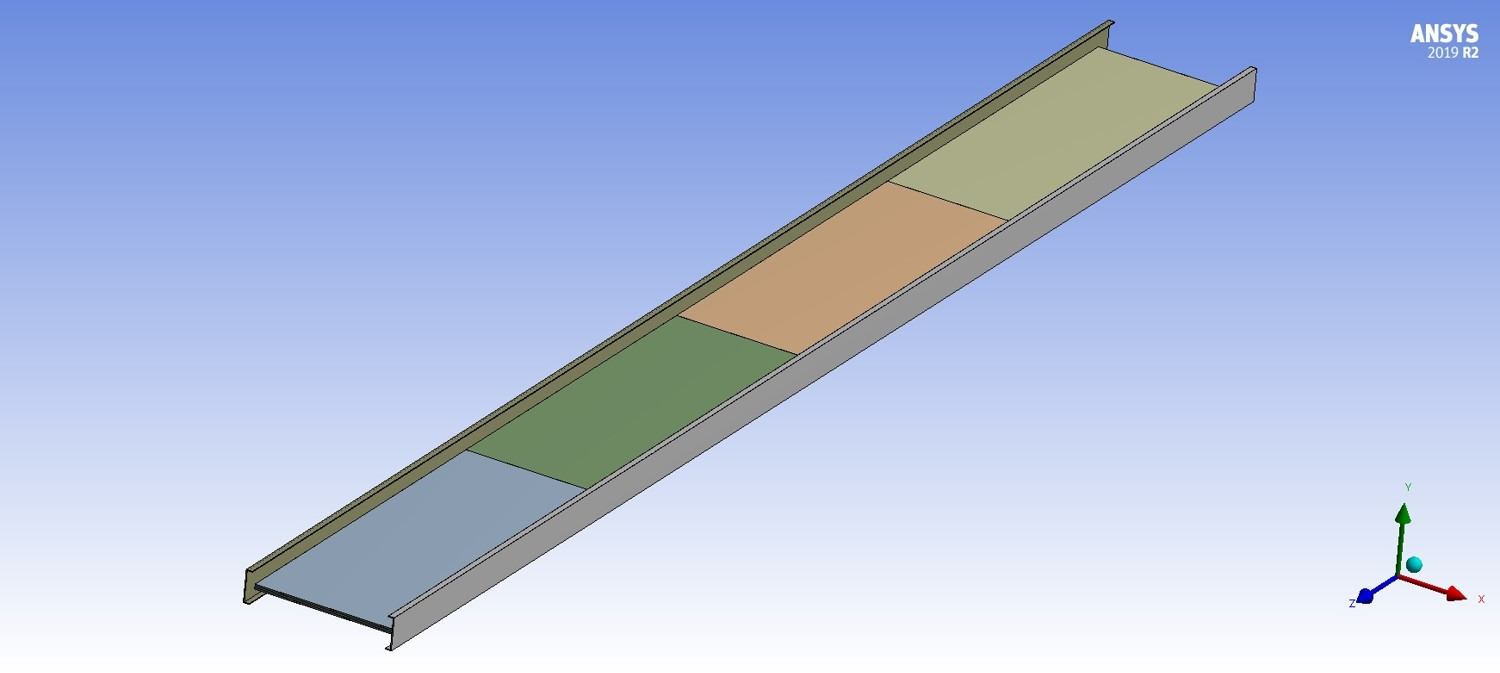}
\end{center}
\caption{Schematic of one RPC unit with four chambers assembled in a line within the same mechanical structure.}
\label{fig:MechStruct1}
\end{figure}

\begin{figure}[hbtp!]
\begin{center}
\includegraphics[width=1.0\textwidth]{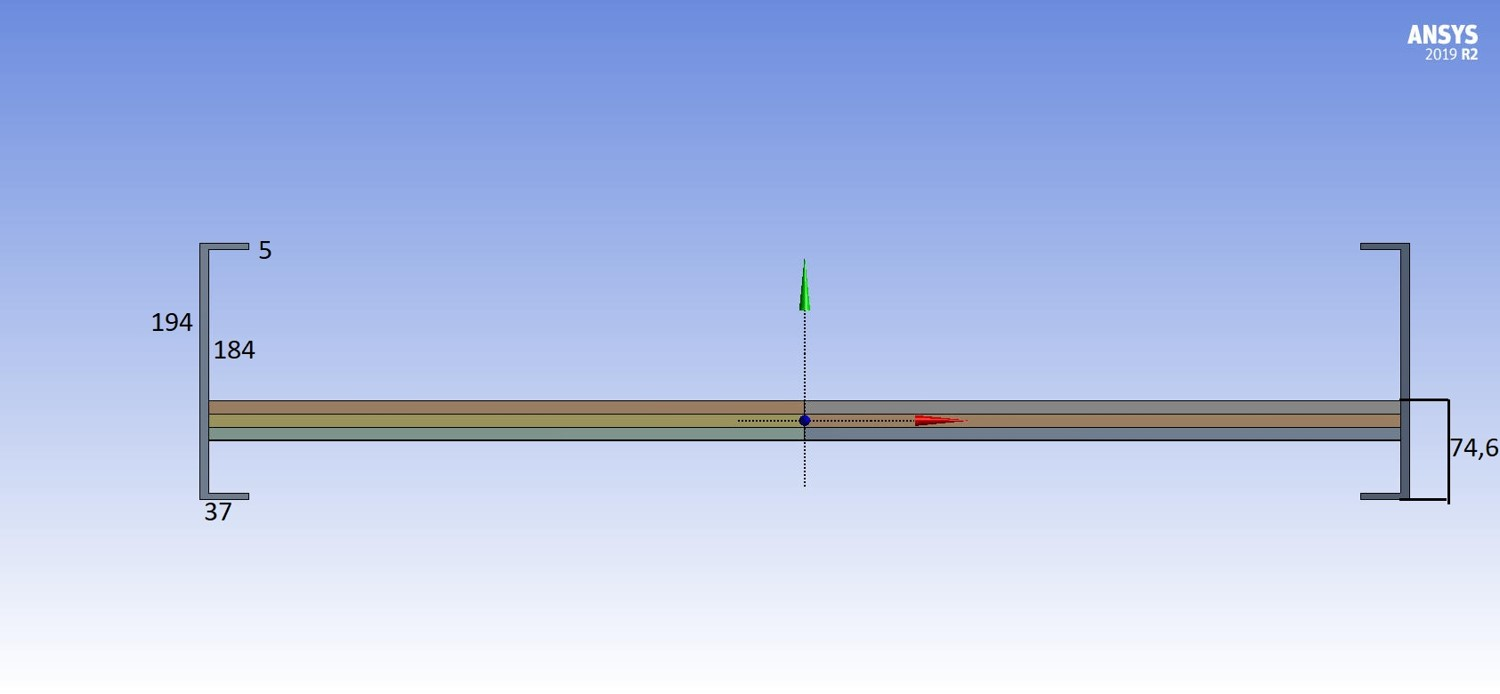}
\end{center}
\caption{Schematic side view of one RPC unit showing details of its mechanical structure with the two ``C''-shaped profiles. All the values displayed in the figure are in mm.}
\label{fig:MechStruct3}
\end{figure}

In order to fulfill the requirements of the 9$\times$9 m$^2$ super-modulus, 4 chambers are assembled in the same mechanical structure of 9$\times$0.9 m$^2$ size, as shown in Fig.~ \ref{fig:MechStruct1}. The gas is serially distributed from one gap to the next. The structure supporting the $4$ chambers is made of two “C”-shaped aluminum profiles fixed at the sides of the chambers (as shown in Fig.~ \ref{fig:MechStruct3}). This structure behaves as a stiff beam supported only at the ends, with a sagitta of about 1-cm. The thickness and the height of the profiles need to be optimized taking into account the weight of the structure. In the final assembly of a $9\, \times \, 9$ m$^2$ super-modulus, two “C”-shaped profiles in contact with one another are screwed together in a single “I”-shaped profile.

\par

\par

\subsection{The RPC front-end electronics}

\begin{figure}[hbtp!]
\begin{center}
\includegraphics[width=1.0\textwidth]{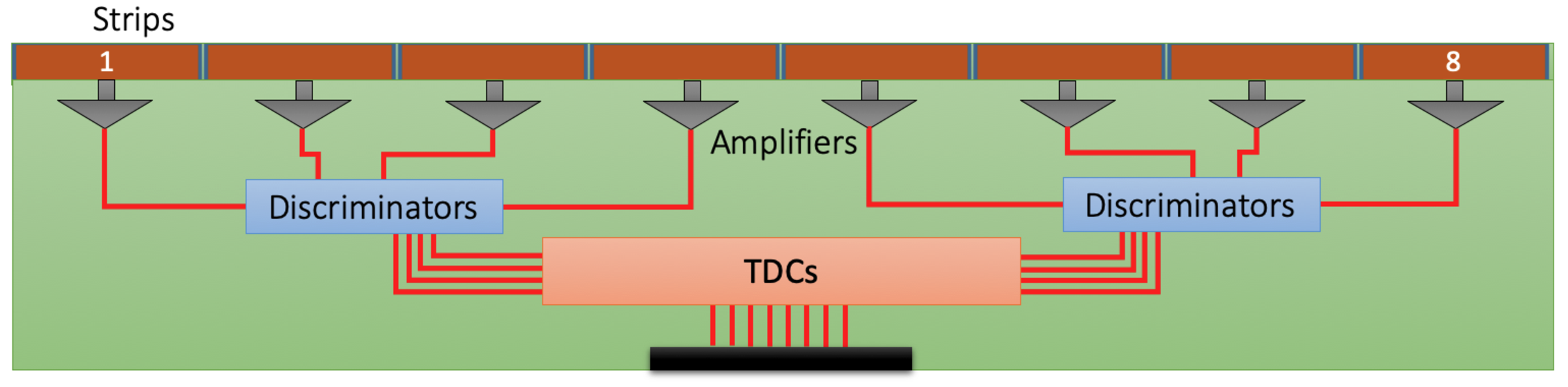}
\end{center}
\caption{Logical layout of an 8-channel RPC front-end circuit board.}
\label{fig:FE1}
\end{figure}

The front-end circuit was developed at the INFN laboratory of Rome Tor Vergata for the RPC detectors of the ATLAS LHC phase-II upgrade \cite{ATLAS-TDR-026}. The circuit consists of a charge preamplifier, a discriminator, and a Low Voltage Differential Signal (LVDS) driver, shown schematically in Fig.~ \ref{fig:FE1}. The amplifier and the driver are built as discrete component circuits. The discriminator is a full-custom circuit, conceived with the purpose of optimizing performance and allowing for large mass production at low cost


\par

\subsubsection{Preamplifier}
\par
\begin{figure}[hbtp!]
\begin{center}
\includegraphics[width=1.0\textwidth]{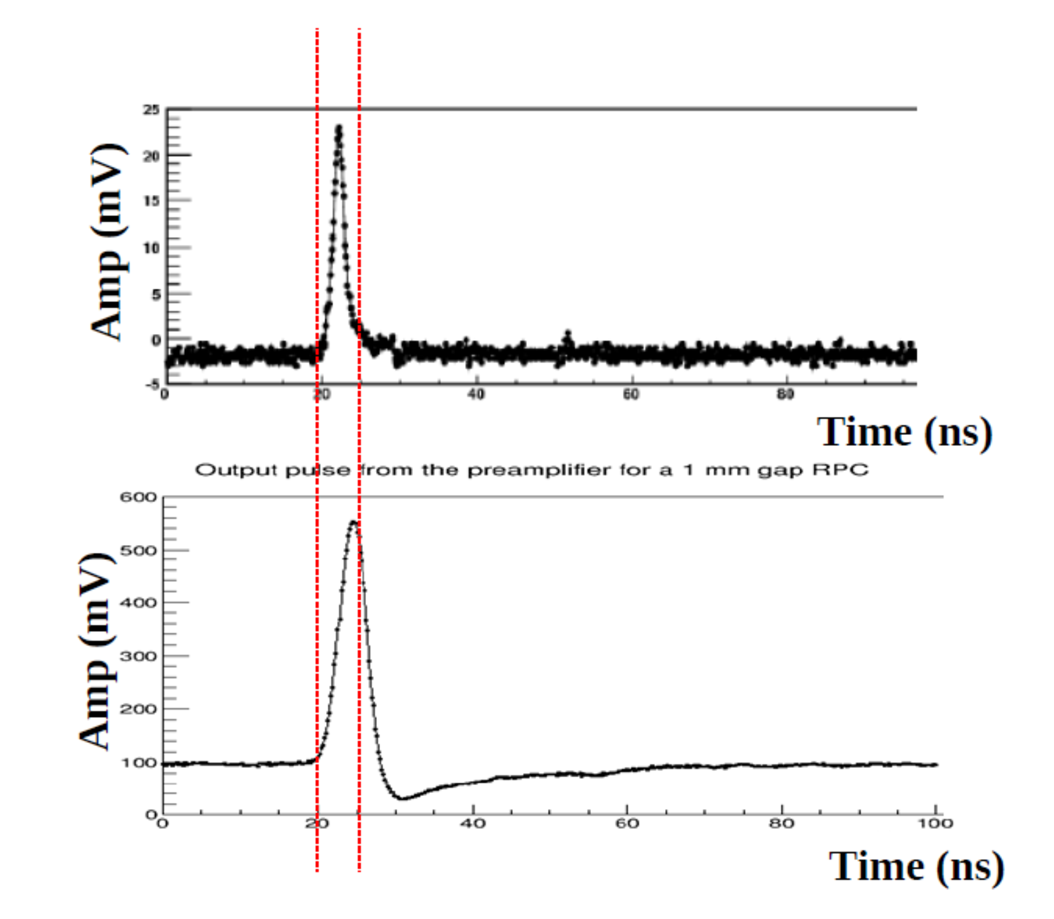}
\end{center}
\caption{Comparison between a bare RPC signal (upper plot) and the corresponding amplified signal (lower plot).}
\label{fig:FE2}
\end{figure}

The preamplifier uses Silicon Bipolar Junction Transistor (BJT) technology. The main feature of this new kind of preamplifier is fast charge integration with the possibility of matching the input impedance to a transmission line. Its working principle is shown in Fig.~ \ref{fig:FE2}, which explains how the injected signal is integrated.

\par

The specifications of the Silicon BJT technology planned for the charge preamplifier are listed in Table \ref{frontend}.

\begin{table}[hbtp!]
    \centering
    \caption{Specifications for the preamplifier in the RPC front-end circuit.}
    \begin{tabular}{l l}
    \hline
 Voltage supply &  $3-5$ V \\
 Sensitivity    &  $0.2-0.4$ mV/fC \\
 Noise (independent of the detector) & $1000$ e$^{-}$ RMS \\
 Input impedance & $50-100$ $\Omega$ \\
 Bandwidth & $10-100$ MHz \\
 Power consumption &  $5$ mW/ch \\
 Rise time for $\delta$(t) input & $300-600$ ps \\
    \hline
    \end{tabular}
    \label{frontend}
\end{table}

\par
\subsubsection{Discriminator}
\par
The requirements for the MATHUSLA cosmic-ray RPC discriminator are the following: a) very low threshold, of the order of $6$-$7$ mV, and b) efficient operation with very fast pulses like the ones produced by RPCs. The very low threshold for cosmic shower detection is not required to improve the rate capability, which is very modest in this case, but rather to improve the linearity range in the measurement of the hit density inside the shower core. The core of the showers can have hit densities exceeding $10^6$/m$^2$, due to the very high energy of the primaries. In this case, the linearity range requires that the charge released in the gas by a single hit is very modest. 
The above requirements are met with a new type of discriminator based on the saturated amplifier: a signal above the discrimination threshold is amplified by multiple stages up to saturation. The output is a square wave of fixed amplitude independent of the input signal. Its duration is the time over the threshold of the input signal, which allows measuring the input amplitude, once the input shape is known. This type of measurement uses a logarithmic scale. This is a crucial characteristic when the input charge can vary over a wide range. In the case of MATHUSLA, the number of particles in a single front-end channel may range over several orders of magnitude, depending on the shower energy and the read-out strip area. 

\begin{figure}[hbtp!]
\centering
\includegraphics[width=1.0\textwidth]{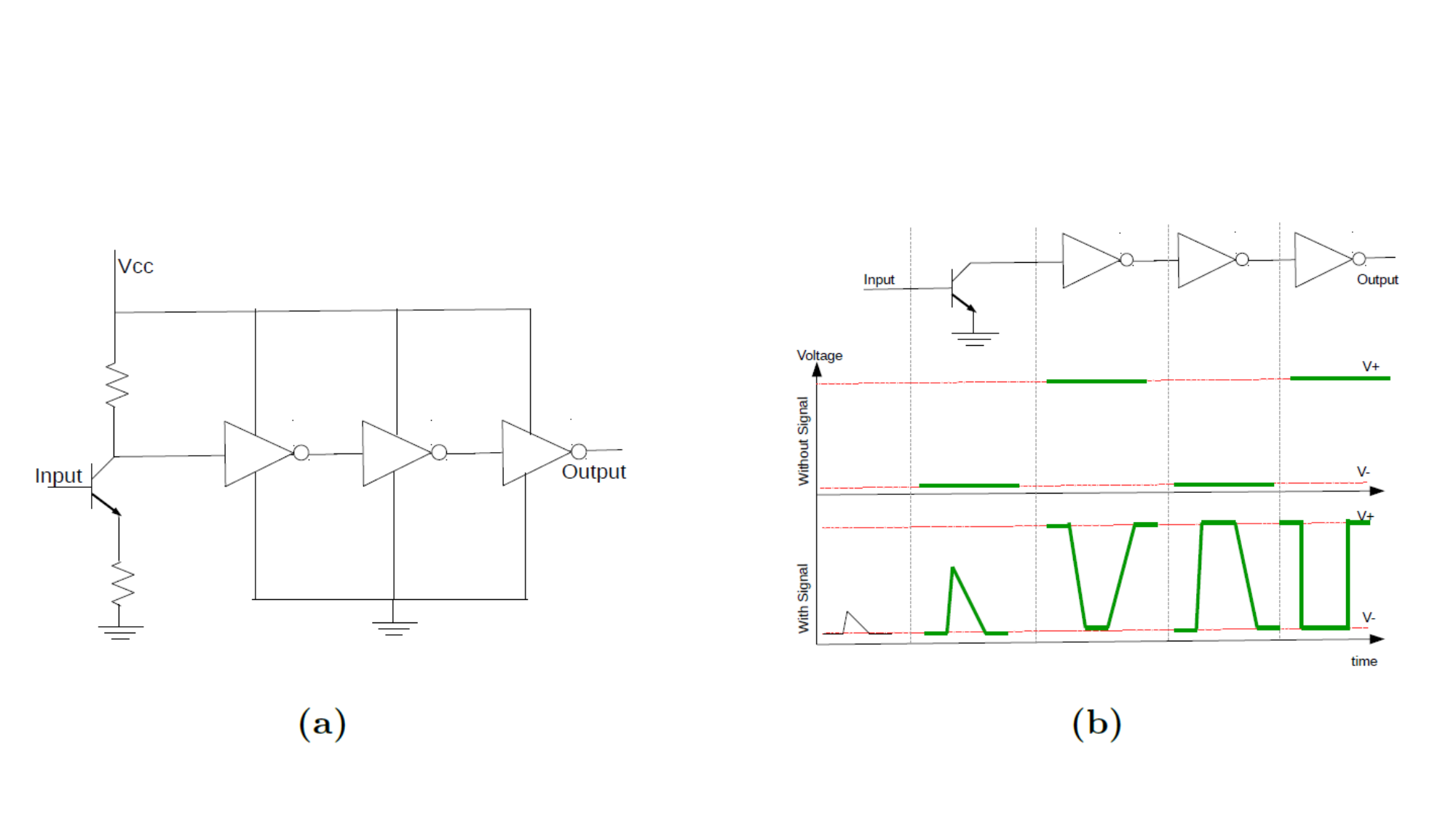}
\caption{(a) Schematic of the discriminator circuit. (b) Signal-processing scheme for the discriminator circuit.}
\label{fig:discr}
\end{figure}

The discriminator circuit is schematically shown in Fig.~ \ref{fig:discr} (a). The first stage is a simple amplifier in the BJT common-emitter configuration. The following three stages can be represented as logic “NOT” gates and have the task of discriminating (first logic stage) and saturating the input signal. As shown in Fig.~ \ref{fig:discr} (b), for each logic stage the output shows sharper edges and the final output is a rectangular wave of inverse polarity with respect to the analogue input.
The discriminator features are summarised in Table \ref{discriminator}.

\begin{table}[hbtp!]
    \centering
    \caption{Specifications for the discriminator in the RPC front-end circuit.}
    \begin{tabular}{l l}
    \hline
 Technology &  $130$ nm Si-Ge BiCMOS \\
 Voltage supply & $1-2.5$ V  \\
 Minimum threshold & $0.3$ $\mu$V \\
 Minimum input-pulse width for threshold linearity & $0.5$ ns \\
 Bandwidth & $10-100$ MHz \\
 Power consumption &  10 mW/ch \\
 Rise time for $\delta$(t) input & 300 ps \\
 Input impedance & $100$ $\Omega$ \\
 Double-pulse separation & 1 ns \\
     \hline
   \end{tabular}
   \label{discriminator}
\end{table}

\par
\subsubsection{TDC}
\par
In a system with a very large number of channels, where time measurements are required, time digitization is one of the most relevant problems due to cabling complexity and TDC cost. For the system proposed for MATHUSLA cosmic-ray measurements, the TDC is embedded in the front-end electronics and is included in the full-custom circuit under development. This is possible thanks to new technologies, $120$-nm SiGe and CMOS, which make possible oscillators and counting scalers that operate at $2-3$ GHz. For the detection of a large-size cosmic shower, the intrinsic single-hit time fluctuations are in the nanosecond range, and therefore a TDC of comparable accuracy would be suitable for this purpose. This makes possible a drastic simplification for the TDC. Indeed, at a frequency of $2-3$ GHz, the simple counting of the periods covering the time interval under measurement results in an uncertainty of $0.5-0.3$ ns that is adequate for cosmic-shower timing, and no interpolation between two consecutive clock signals is required.

\par
\subsubsection{Trigger and data acquisition}
\par
This section describes the MATHUSLA cosmic-ray trigger and its implementation in the electronic
circuits that produce the trigger signal and give the Data Acquisition the start signal. The trigger logic presented here is fully based on the signals supplied by the RPC-layer detector, consequently, the cosmic ray trigger is independent of the main long-lived particle trigger of the experiment.

\par
The MATHUSLA trigger of an EAS generated in the atmosphere by a cosmic primary particle is based on the multiplicity of the shower hits detected by the RPC layer. This multiplicity is approximately proportional to the EAS shower energy on a logarithmic scale. Thus, a multiplicity threshold is equivalent to an energy threshold, even within the large fluctuations of the energy vs. multiplicity correlation. The main task of a multiplicity trigger is to select the simultaneous hits due to an EAS from the accidental rate due to the combinatorial background of the counting rate over a large-area detector. Simultaneous hits due to a cosmic shower can be selected by a proper coincidence among several chambers. The main challenge of this trigger is that the time resolution of the coincidence must take into account the systematic delay among
different hits due to the inclination of the shower front with respect to the vertical direction. For showers propagating almost horizontally this delay is of the order of the linear detector dimension over the speed of light, $L$/$c$. For a $(100\times 100) \, \mbox{m}^2$ detector like MATHUSLA, this delay can be as
large as $(100\sqrt{2} \, \mbox{m}$)/$c$, namely about $450 \, \mbox{ns}$, assuming the diagonal as the maximum distance between two hits. This large delay does not allow for a coincidence time, which is the main parameter for
reducing the accidental rate. This problem can be solved by dividing the detector into subdetectors of gradually increasing areas where the coincidence time resolution can be improved according to the smaller subdetector size. In this section, we give an example of how this trigger concept can be implemented inside the MATHUSLA cosmic-ray detector and how the accidental rate can be evaluated. A more precise trigger study, including the efficiency vs. energy correlation as well as the trigger counting rate, for each multiplicity threshold, requires a full simulation of the process starting from the characteristics of the shower to be triggered.

\begin{table}[hbtp!]
\centering
\caption{Specifications for the relevant RPC trigger parameters.}
\begin{tabular}{ l  l }
\hline
Effective detection area for each submodule   &  $(9\times 9) \, \mbox{m}^2$ \\
Assumed counting rate per unit area  &  $1 \, \mbox{kHz}/\mbox{m}^2$ \\
Multiplicity threshold for EAS detection  &  $50 \, \mbox{hits}$ \\
Number of RPC sub-modules  &  $100$  \\
Number of chambers in one RPC sub-module  &  $40$  \\
Trigger majority threshold in one sub-module  &  $2 \, \mbox{hits}$ \\
Time resolution for a submodule majority  &  $50 \, \mbox{ns}$ \\
Maximum time delay for inclined showers in one sub-module  &  $42 \, \mbox{ns}$ \\
RPC read-out strip length  &  $2.2 \, \mbox{m}$  \\
Inverse signal propagation speed  &  $5 \, \mbox{ns}/\mbox{m}$ \\
\hline
\end{tabular}
\label{trigger_parameters}
\end{table}

\par
In order to evaluate the efficiency of rejecting the accidental background rate and to support it
with some numerical calculations we assume the following working conditions:
\begin{itemize} 
\item	Multiplicity threshold of $50$ hits on the whole RPC detector. The corresponding energy
threshold should be at least an order of magnitude higher than that of ARGO-YBJ if renormalized to the much lower altitude of MATHUSLA.

\item	Intrinsic detector counting rate equal to $1$ kHz/m$^2$. This is an overestimation of the expected
rate, in order to obtain a very conservative evaluation of the accidental background rate.

\item	The hits due to the shower are randomly distributed over the detector. This is also a conservative hypothesis for the accidental background calculation because the shower hits tend to be clustered thus giving a higher probability of several hits in the same sub-detector, which is a crucial point in the following trigger project.
\end{itemize}
\par
The relevant parameters for the MATHUSLA RPC trigger are listed in Table \ref{trigger_parameters} and will be used in the calculations reported below.
The smallest subdetector where the shower multiplicity is considered for trigger purposes is an
RPC sub-module of about $(9\times 9)$ m$^2$. The MATHUSLA detector, according to Fig.~\ref{fig:geometry}, is subdivided into $100$ such units. Each unit contains an RPC plane of the same area, consisting of $40$ chambers arranged in groups of 4 chambers in the same mechanical support as shown in Fig.~\ref{fig:MechStruct1}. The first step of the proposed trigger requires selecting all submodules with at least $2$ hits due to the shower. For each sub-module, the analog sum of all output signals from the front-end circuits is sent to a discriminator with a threshold requiring at least two simultaneous signals.
The signal shaping time is $25$ ns, resulting in a time resolution of about $50$ ns of this majority coincidence. This covers the uncertainties due to both the shower inclination, $42$ ns for a $(9\times 9)$ m$^2$
sub-detector, and the timing propagation of the signal along the readout strips (namely $2.2$ m $\times 5$ ns$/$m $ = 11$ ns)
combined in quadrature. The sub-module counting rate is $81$ kHz so that the accidental
coincidence rate is $(81\times 10^3$ Hz$)^{2} \times (50\times 10^{-9}$ s$) = 328$ Hz for each of the $100$ submodules comprising the RPC detector.

The probability of missing the trigger, because all shower hits fall in different sub-detectors, is
extremely small. For a $50$ hits shower, the average number of hits/subdetectors is $0.5$ and the
expected number of towers showing a coincidence of at least $2$ hits is $9.02$, according to
Poisson statistics.

The next step in the trigger logic is to require the coincidence with a proper multiplicity on the
whole detector. Repeating the procedure followed for the single units, the analog sum of the
$100$ tower outputs is sent to a discriminator with a selectable threshold requiring the
coincidence of two or more towers on the whole detector. The total rate of the analog sum is
$328$ Hz $\times 100=32.8$ kHz. The coincidence time, in this case, must be $450$ ns to take into
consideration the maximum time uncertainty due to the shower inclination over the whole
detector. For a multiplicity threshold of $2$ subdetectors out of a hundred, the accidental rate is:
$(32.8 \times 10^3$ Hz$)^{2} \times  (450 \times 10^{-9}$ s$) = 427$ Hz which is a modest trigger rate.

The probability of missing the trigger for a shower of $50$ hits is negligible. Indeed, the
probability of less than two sub-detectors firing for an expected number of $9.02$ is $1.23\times 10^{-3}$.
If needed, the trigger rate can further be decreased by increasing the multiplicity threshold
above $2$. For a threshold of $3$, the accidental rate is $(32.8 \times 10^3$ Hz$)^{3} \times  (450 \times 10^{-9}$ s$)^2 = 7.1$ Hz.

\par
\subsubsection{Time schedule}
\par

The front-end circuit described here is not yet fully developed. The preamplifier and the discriminator are operational and are included in the RPCs currently being installed for the ATLAS Phase-I upgrade. The integration of the TDC in the circuit is in progress. It requires $1$-$2$ further foundry runs in the next $1$-$2$ years. The current plan is to be ready for the LHC Phase-II upgrade. 

\section{Simulations}
\label{s.simulations}

   In order to explore the performance of the RPC layer and the scintillating planes of MATHUSLA as air-shower cosmic-ray detectors, we have prepared a  set of toy MC simulations that include the production of EAS by the primary cosmic rays, the evolution of the particle cascade until its arrival at the Earth's surface and the physical response of the detectors to the passage of the shower particles. In the following subsections, we describe each of the above simulation steps that are used to produce our MC data.

   \subsection{Air shower simulations}
  \label{mcEAS}
  
  For the generation and development of EAS in the atmosphere, we followed the procedure described in \cite{Alpigiani:2018fgd}. In particular, we  employed the CORSIKA v7.71 package without thinning  \cite{Heck:1998vt} together with the FLUKA 2020 \cite{Ferrari:2005zk,BOHLEN2014211} and the QGSJET-II-04 \cite{qgsjet,qgsjet2} hadronic interaction models.  FLUKA was used to simulate collisions with hadron energies $E_h < 200 \; \gev$, while QGSJET-II-04 was employed to describe hadronic interactions with higher energies. CORSIKA simulates the evolution of the particle cascade through the atmosphere from the production point to the detector level. The output provides  information about the spatial, time, and momentum distributions of the shower particles at the observation point. Without thinning, CORSIKA follows the history of each particle in the EAS until its energy falls below a certain threshold, which is specified at the beginning of the simulation chain. Here, we kept the CORSIKA default values of $3 \, \mev$ for shower electrons and gamma-rays, and  $100 \, \mev$ for shower muons and hadrons.  
  
  We chose the U.S. standard atmosphere model in our simulations. In addition, we selected the curved atmosphere option, as we will work also with very inclined showers. Moreover, the observation level was set at $436 \, \mbox{m a.s.l.}$  ($X \sim 1000 \, \mbox{g}/\mbox{cm}^2$) and the local geomagnetic field was chosen as $B_{x} = 22.2 \, \mu T$ and $B_{z} = 41.9  \, \mu T$ in the simulations. Such values\footnote{The data was taken from the NOAA (National Centers for Environmental Information) webpage at \mbox{https://www.ngdc.noaa.gov/geomag-web/}.} correspond to the characteristics of the site above the ATLAS interaction point, which was initially considered as one of the potential places of the MATHUSLA experiment. The differences with the location of MATHUSLA at the CMS site are small (the altitude is about $374 \, \mbox{m a.s.l.}$ and the corresponding magnitudes of the geomagnetic field are $B_{x} = 22.2 \, \mu T$, $B_{z} = 42.1  \, \mu T$). Therefore, our conclusions regarding the performance of the MATHUSLA detector are not expected to be affected by such differences.

  The EAS were produced with a $\sin \theta \cos \theta$ distribution for two zenith angle intervals: $\theta = 0^\circ - 20^\circ$ (vertical EAS) and $\theta = 70^\circ - 80^\circ$ (inclined EAS), and the shower cores were scattered randomly on the MATHUSLA area. A flat geometry was employed. We produced MC simulations for H and Fe primary nuclei with energies between $1 \, \tev$ and $1000 \, \pev$ assuming an energy spectrum of the form $E^{-2}$. In order to improve the statistics at the highest energies, we generated  MC simulations by energy decade as shown in Table \ref{numCSKsimulations}. Each event was weighted to achieve a unique $E^{-2}$ energy spectrum for the whole energy range.
  
  \begin{table}[ht]
      \centering
       \caption{Distribution of the total number of CORSIKA simulations generated by energy decade. CORSIKA version 7.71 was used together with the FLUKA and QGSJET-II-04 hadronic interaction models. }
     \begin{tabular}{ p{2.5cm}  p{2.5cm} p{2.5cm} }
 \hline
  &  \multicolumn{2}{c}{v7.71} \\
 \hline
Energy decade & $0^\circ - 20^\circ$ & $70^\circ - 80^\circ$\\ 
 \hline
 1-10 TeV   &  $2 \times 10^4$ H  & $2 \times 10^4$ H \\
            &  $2 \times 10^4$ Fe  & $2 \times 10^4$ Fe \\
 10-100 TeV &  $2 \times 10^4$ H  & $2 \times 10^4$ H \\
            &  $2 \times 10^4$ Fe  & $2 \times 10^4$ Fe \\    
 100-1000 TeV & $1 \times 10^4$ H  & $1 \times 10^4$ H \\
            &   $1 \times 10^4$ Fe  & $1 \times 10^4$ Fe \\
 1-10 PeV   &   $7.5 \times 10^2$ H  & $1 \times 10^3$ H \\
            &   $7.5 \times 10^2$ Fe  & $1 \times 10^3$ Fe \\
 10-100 PeV &   $1 \times 10^2$ H  & $2.5 \times 10^2$ H \\
            &   $1 \times 10^2$ Fe  & $2.5 \times 10^2$ Fe \\
 100-1000 PeV &  $3 \times 10^1$ H  & $2.1 \times 10^2$ H \\
            &   $2 \times 10^1$ Fe & $1.5 \times 10^2$ Fe \\
\hline
\end{tabular}
      \label{numCSKsimulations}
  \end{table}

 \subsection{Simulation of the MATHUSLA detector}
 \label{MatSim}
 
  For our study, we built a toy MC model of MATHUSLA using a program written in ROOT CERN \cite{rootcern}. In particular, we  considered a simple geometry consisting of seven extruded scintillating detector layers and the RPC layer as shown in Fig.~\ref{fig:layout_v4}. The present work did not incorporate the last design of MATHUSLA, but a previous one. However, we do not foresee substantial modifications to the main results on the performance of the experiment for cosmic-ray detection. In our simulations, the two scintillating detector layers near the bottom of the instrument are ignored. On the other hand, the RPC chamber plane is located  $2 \, \mbox{m}$ above the intermediate scintillating detector layer that is found in the middle between the top and intermediate groups of scintillating detector planes. Hence, we have five detector layers above the RPC and two of them below it (see Fig.~ \ref{fig:geometry}). The size of the detector was set to $100 \, \mbox{m} \times 100 \, \mbox{m}$. The coordinate system is oriented so that the $X$ and $Y$ axis are parallel to the horizontal dimensions of the detector. In this coordinate frame the detector plane extends from $x, y = 0 \, \mbox{m}$ to $100 \, \mbox{m}$, and from $z = 0 \, \mbox{m}$ to  $9 \, \mbox{m}$. The height is measured from the bottom of the intermediate tracking detector layer. 
 
  The RPC geometry used assumes Big Pads with a conservative size of $1 \, \mbox{m} \times 1 \, \mbox{m}$. These elements provide not only spatial and time data of the shower particles but also local particle density measurements. We output the coordinates of the central positions of the Big Pads that were hit, the arrival times of the first shower particles in each module, and the induced signals. Space-time measurements of the EAS can be made  with individual RPC strips, but to keep the simulation as simple as possible, we have focused our studies on the data from the Big Pads. The amplitudes of the RPC-pad signals are assumed to be proportional to the density of charged particles as found by the ARGO-YBJ RPC measurements \cite{Aielli2012}. 
  We must point out that in a real simulation, it must be considered that the  Big Pads can only be used to measure hit multiplicities with high densities due to their high capacitance. For small density measurements and individual hit counting, the  RPC strips will be used.

For the simulations, the scintillating bars have a length of $5 \, \, \mbox{m}$ and a width of $4 \, \, \mbox{cm}$. They are oriented along the same direction on each plane (parallel to one of the $X$ or $Y$ axis), but they are rotated by $90^\circ$ when passing from one scintillating plane to another one. In the MC simulations, we have considered that the scintillating detector layers are sensitive only to the passage of charged particles and that the output saturates for more than one hit. Our assumption is  conservative.  In practice, the scintillating detector bars will saturate on average at one hit per bar. Consequently, the simulation assumes that the scintillating bars provide information of the impact point and arrival time only for hits of single-charged particles. In case of more than one hit per bar, we only save the central coordinates of the corresponding bar. For more reliable estimations of the performance of the scintillating detector layers, detailed simulations are needed. This task will be done in a future study.  Finally, in our simulations, we have considered that the RPC and the scintillating detector planes have a $100 \, \%$ efficiency for the detection of charged particles.

   To initiate the simulation, the cores of the CORSIKA air showers were randomly distributed in the instrumented area of the detector. Then we injected the shower particles that reach the detector level at the top layer of MATHUSLA. We followed their trajectories through the tracking planes and recorded the spatial coordinates of the hit modules and the corresponding hit times, the latter only in case of one hit per bar. For the Big Pads of the RPC, we also saved the output signals, which are proportional to the local number of incident particles. For the simulations of the detector performance, we only used $e^{\pm}$'s, $\mu^{\pm}$'s, $p^{\pm}$'s, charged pions and kaons. We allowed unstable particles to decay and removed their decay products from the simulations. The lifetimes of the decaying particles were taken from \cite{PDG}. In addition, we neglected the energy losses of the EAS particles due to interactions with the detector and the air between the tracking layers.  We assumed that the entire surface of the detector layers of MATHUSLA are active areas, with no gaps between contiguous scintillating bars or Big Pads.

\begin{figure}[hbtp!]
 \centering
  \includegraphics[width=4.0 in]{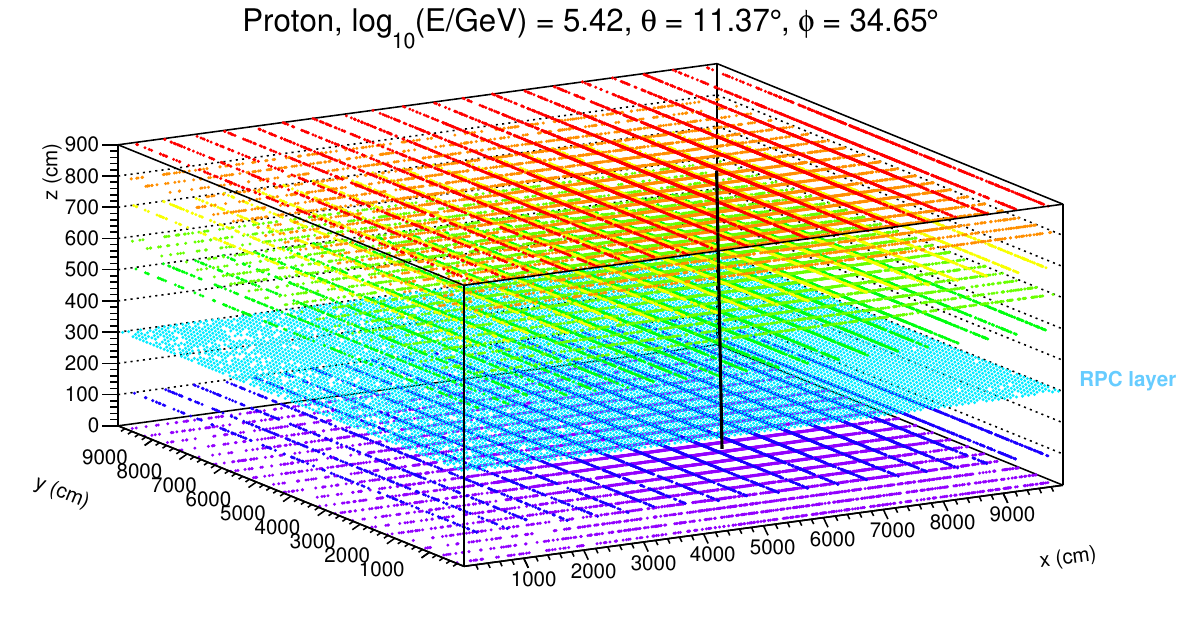}
  \includegraphics[width=2.5 in]{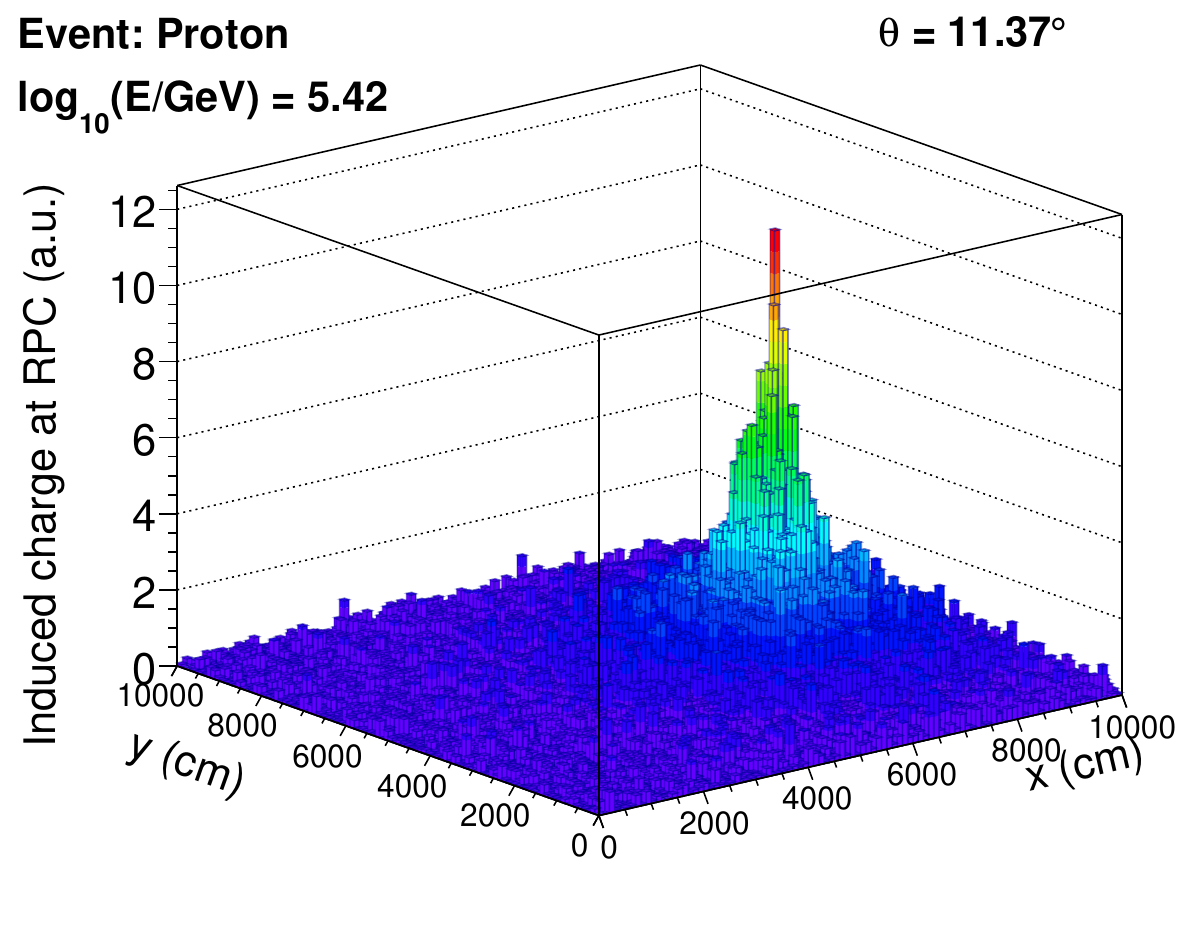}
  \includegraphics[width=2.5 in]{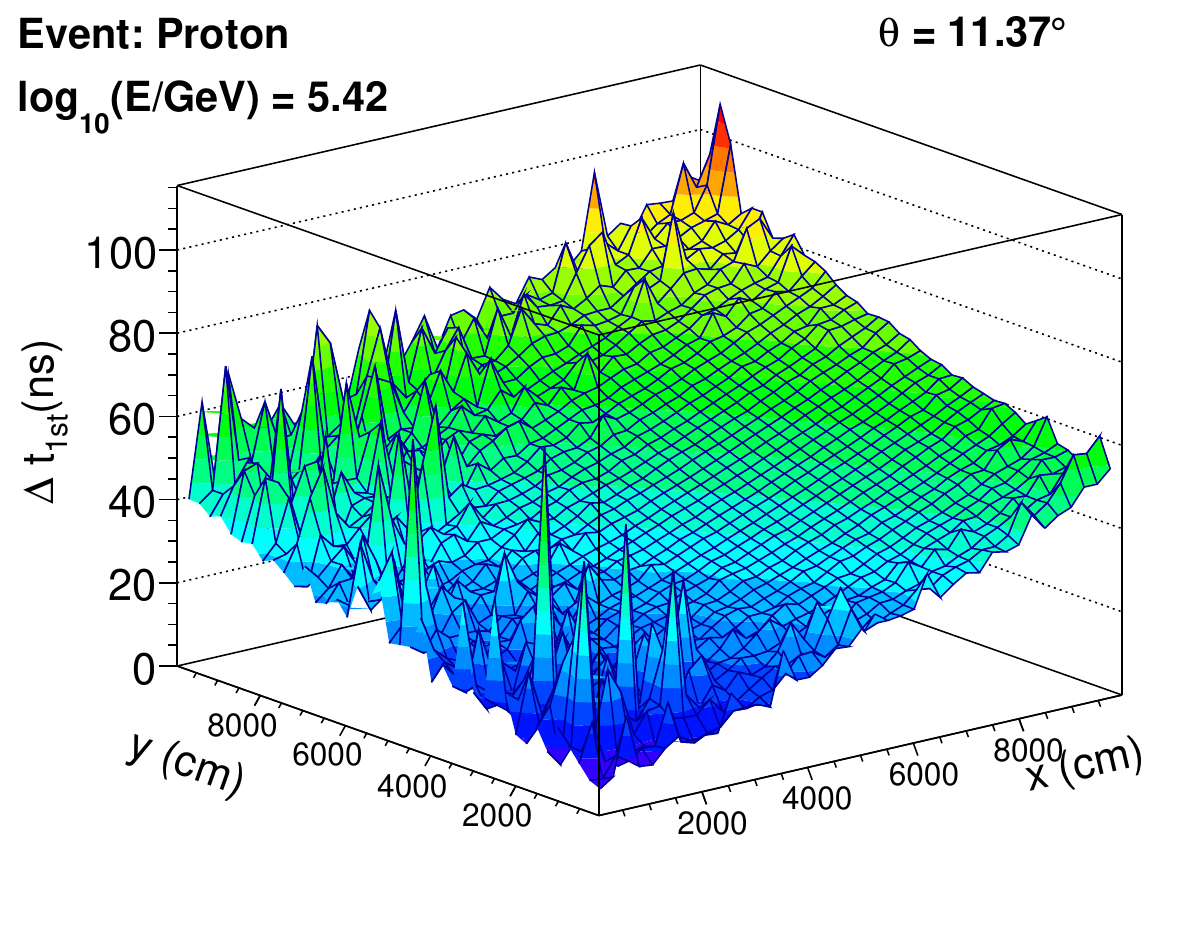}
  \caption{Example of a simulated vertical EAS impinging on the MATHUSLA detector. The event was induced by a primary proton  with an energy of $\log_{10}(E/\gev) = 5.42$ and arrival direction $(\theta, \phi) = (11.37^\circ, 34.65^\circ) $. The azimuth is measured counterclockwise from the $X$ direction to the horizontal component of the momentum vector of the primary cosmic ray. The zenith is the angle between the  momentum vector of the incident nucleus and the vertical direction.  The core of the shower is located at  coordinates $(x_c, y_c) = (67  \, \mbox{m}, 43  \, \mbox{m})$ on the top layer. Top: Central coordinates of the hit scintillating bars and Big Pads of MATHUSLA during the event. Each plane corresponds to a different tracking layer. The third layer of points, from the bottom, are associated to the RPC detector. The axis of the EAS is shown with a continuous black line. Left bottom:  Bi-dimensional distribution of the induced signal (in arbitrary units) at the Big Pads of the RPC. Right bottom: Arrival times of the 1st shower particles intersecting the Big Pads of MATHUSLA RPCs. The timing information is given with respect to the arrival time of the first charged particle of the EAS that hit the RPC plane.}
 \label{3dMCevents}
\end{figure}

\begin{figure}[hbtp!]
 \centering
  \includegraphics[width=4.0 in]{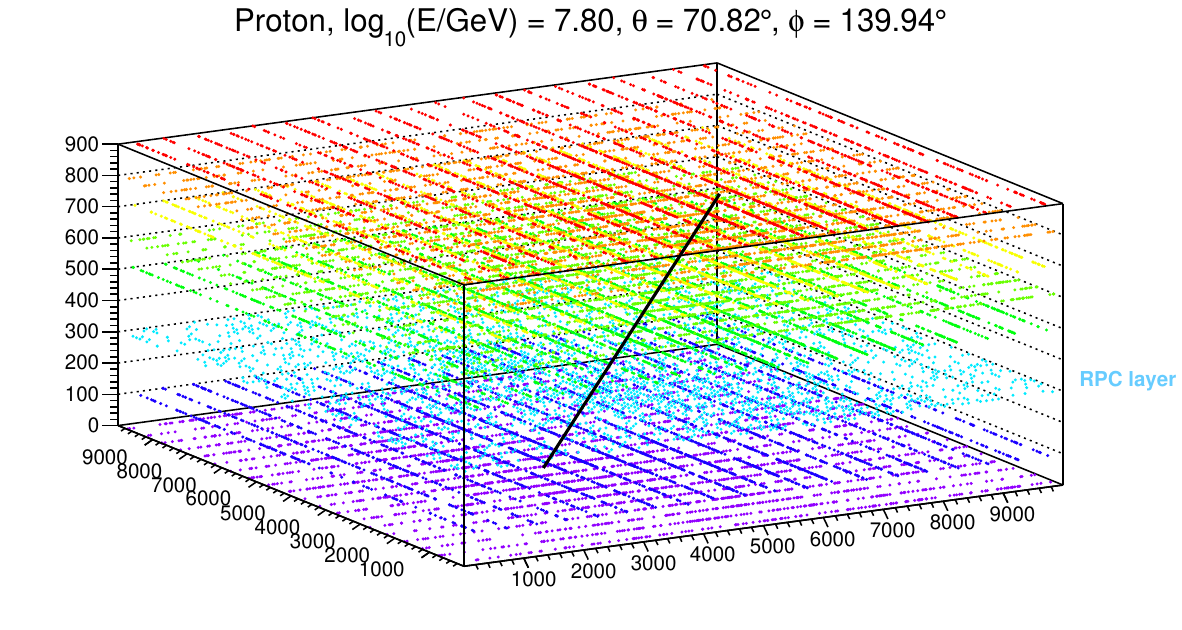}
  \includegraphics[width=2.5 in]{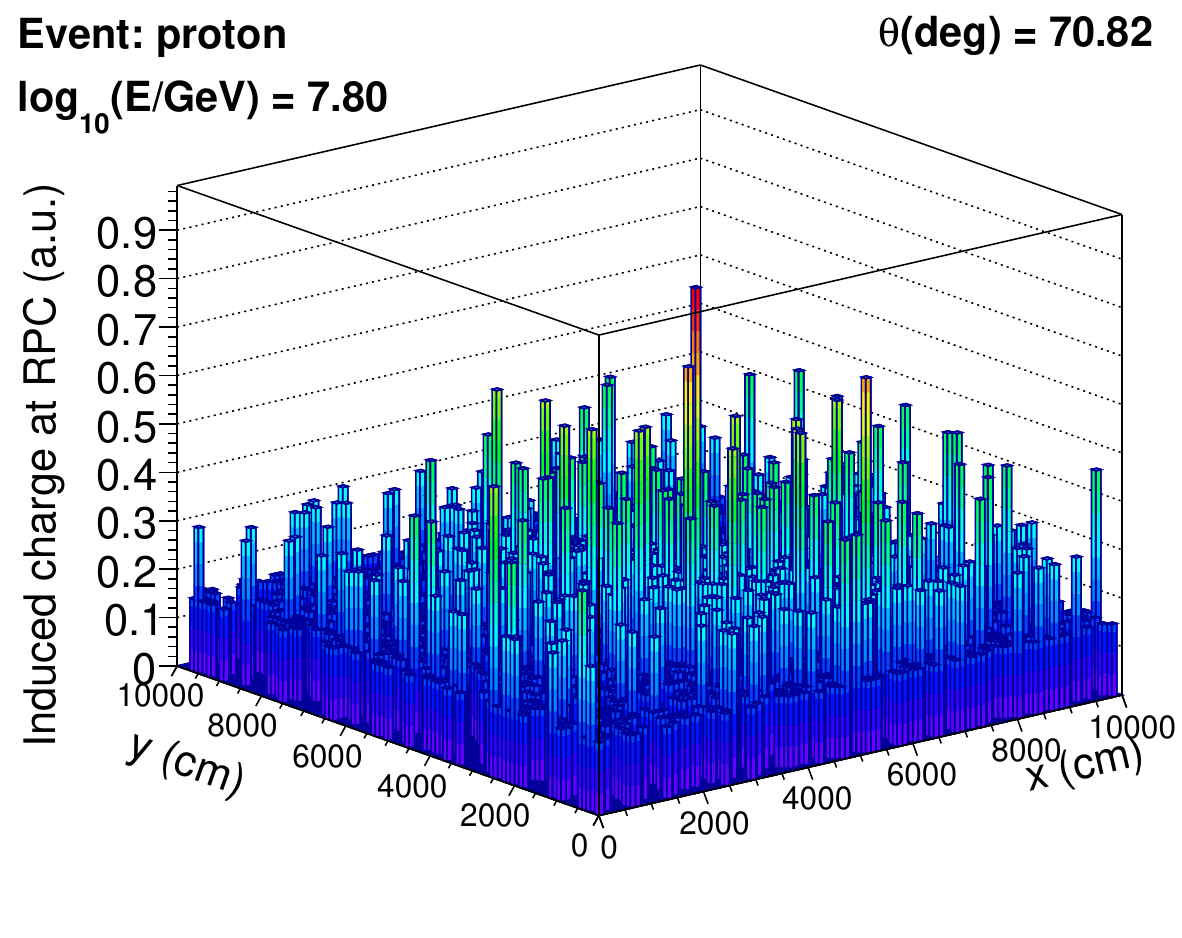}
  \includegraphics[width=2.5 in]{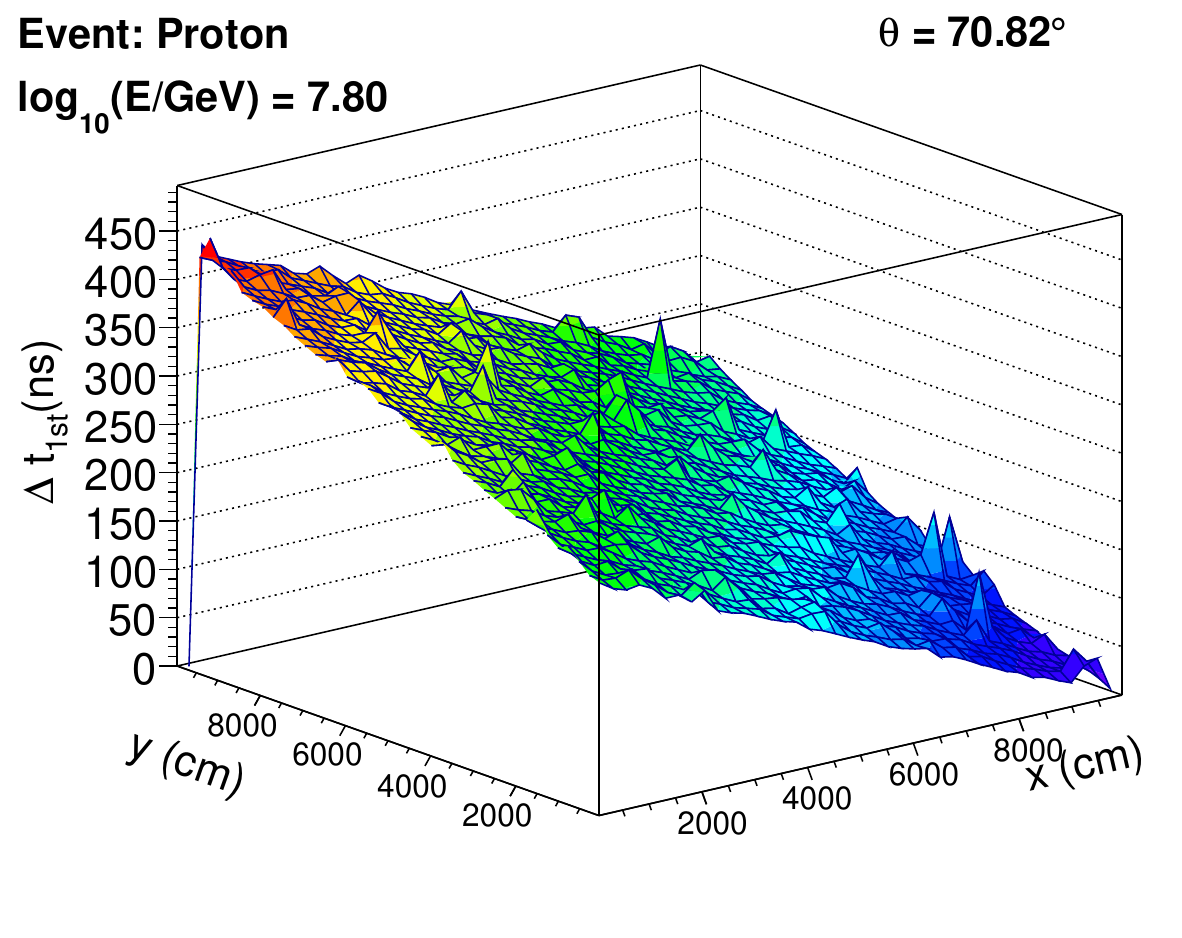}
  \caption{Example of a simulated inclined EAS impinging on the MATHUSLA detector. The event was induced by a primary proton  with an energy of $\log_{10}(E/\gev) = 7.8$ and arrival direction $(\theta, \phi) = (70.82^\circ, 139.94^\circ)$. The azimuth is measured counterclockwise from the $X$ direction to the horizontal component of the momentum vector of the primary cosmic ray.  The zenith is the angle between the  momentum vector of the incident nucleus and the vertical direction. The core of the shower is located at the coordinates $(x_c, y_c) = (60  \, \mbox{m}, 30  \, \mbox{m})$ on the top layer. Top: Central coordinates of the  scintillating bars and Big Pads that are hit.  The points in each horizontal plane belongs to a different tracking layer. The third layer of points, from the bottom, are associated to the RPC detector. The axis of the EAS is shown with a continuous black line. Left bottom:  Bi-dimensional distribution of the induced signal (in arbitrary units) at the Big Pads of the RPC layer. Right bottom: Arrival times of the 1st shower particles intersecting the Big Pads of MATHUSLA RPCs. The timing information is given with respect to the arrival time of the first charged particle of the EAS that hit the RPC plane.}
 \label{3dMCeventsinclined}
\end{figure}

To illustrate the output of the simulation, in Figs.~\ref{3dMCevents} and \ref{3dMCeventsinclined}, we present two examples: one for  a vertical shower and another one for an inclined EAS, respectively. In these figures, we display the 3D patterns of hits at the MATHUSLA detector, the corresponding generated signals at the RPC layer, and the arrival times of the 1st particles hitting each Big Pad of the RPC. 

\section{Event reconstruction}
  \label{EASreco}

  \subsection{Shower core position}
  \label{EASrecocore}
  
  The reconstruction procedure of the EAS events in our analysis goes through different phases. In the first stage, the shower core position and the arrival direction of the EAS are estimated. To find the core position a simple algorithm is applied. For each tracking plane, as a first approximation to the shower core positions, we used the barycenter of the respective signal distributions in the $XY$ space, which are calculated using the following expressions:
  \begin{equation}
   x_c  = \frac{\sum_{i = 1}^{n_{hit}} x_i Q_i}{\sum_{i = 1}^{n_{hit}} Q_i}, \hspace{1cm}
   y_c  = \frac{\sum_{i = 1}^{n_{hit}} y_i Q_i}{\sum_{i = 1}^{n_{hit}} Q_i}.  
     \label{eqcore0} 
  \end{equation}
  Here, $i$ runs over all $n_{hit}$ modules with signal in the corresponding detector plane, $(x_i, y_i)$ is the $XY$ coordinate of the hit $i$-th detector element and $Q_i$ is the corresponding amplitude of the signal of the Big Pad, where for the scintillating bars, $Q_i$ is set to $1$.
  
   We improved the estimates of the EAS core positions by applying a second algorithm that employs the previous core approximations as initial values. In this procedure, for each detector layer, we projected the $XY$ signal distributions on the $X$ and $Y$ axes. The resulting distributions immediately reveal the position of the core, which is located close to the maximum of the projected $X$ and $Y$ histograms. 
   We determine the maximums by fitting the X and Y distributions with lateral functions based on exponential distributions
\begin{equation}
     N_{\mathrm{hits}, x}(x_i) = \alpha_1 e^{-\beta_1 \cdot |x_i - x_{c}|}, \hspace{1cm}
     N_{\mathrm{hits}, y}(y_i) = \alpha_2 e^{-\beta_2 \cdot |y_i - y_{c}|},
     \label{eqcore}
  \end{equation}
  respectively, using a $\chi^2$ minimisation procedure. In the above equations, $\alpha_{1} (\alpha_{2}) >0$ and $\beta_{1} (\beta_{2}) > 0$ are the amplitude and the slope of the projected $X (Y)$ distribution and, as before,  $(x_i, y_i)$ is the coordinate of the $i$-th detector module in a given detector plane. Finally,  $x_c$ and $y_c$ are the positions of the maximums of the projected $X$ and $Y$ histograms, which become our improved approximations to the shower core coordinates. The advantage of this procedure is that it is fast and can be applied for both high and low-energy EAS.
  
  To reduce the effect of fluctuations during the fit, different bin widths are used depending on the fraction of detector elements that are hit, $f_{hit}$. For the RPC histograms of vertical EAS ($\theta \leq 20^\circ$) we used bin widths of $10 \, \mbox{m}$ and $2 \, \mbox{m}$ for fractions of Big Pads $f_{hit} \leq 0.2$ and  $f_{hit} > 0.2$, respectively. Meanwhile, for $\theta > 70^\circ$, we used only $2 \, \mbox{m}$ bins for the RPC distributions.

  For the projected histograms obtained from the scintillating detector layers a bin  width of $10 \, \mbox{m}$ was used for inclined EAS. This bin size was also employed for vertical EAS that induce a fraction of hit scintillating bars $f_{hit} \leq 0.02$. However, for larger  $f_{hit}$ values associated with events with the same zenith angle range, the bin width used was $2 \, \mbox{m}$ ($5 \, \mbox{m}$), if the shortest (longest) dimension of the scintillating bars coincide with the axis 
  of projection used to obtain the histogram.
  
  The bin width selections defined above were chosen to reduce the uncertainties on the core position. We note that a more detailed study would be needed if further improvement is desired.
  
\begin{figure}[hbtp!]
 \centering
  \includegraphics[width=2.5 in]{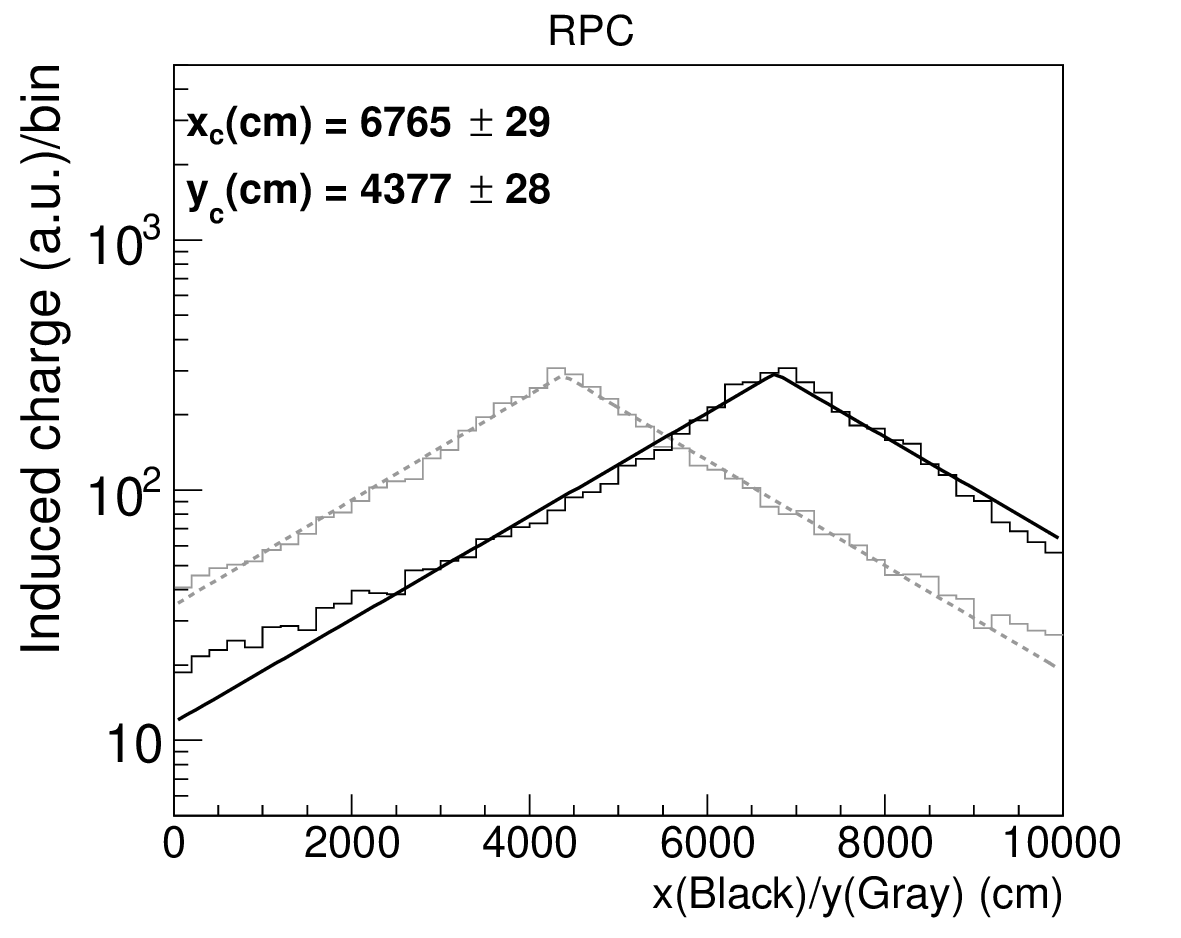}
  \includegraphics[width=2.5 in]{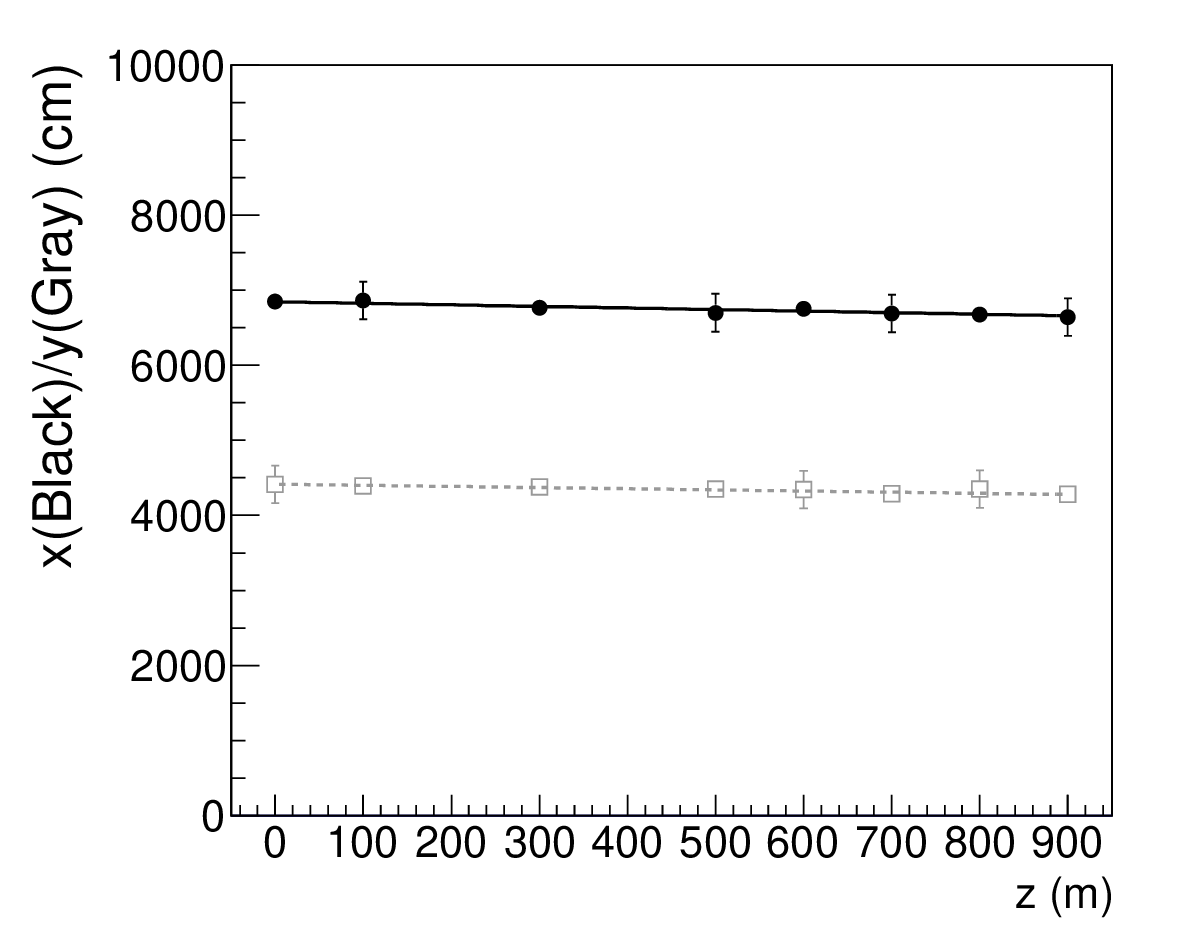}
  \caption{Left: The projected  $X$ (black) and $Y$ (gray) distributions of the number of Big Pads at the RPC layer of MATHUSLA for the vertical MC event of Fig.~\ref{3dMCevents}. The black and gray lines represent the fits with the formula (\ref{eqcore}) to the $X$ and $Y$ projected profiles of the MC shower event, respectively. The estimated coordinates of the EAS core are shown in the upper left part of the figure. Right: Estimated coordinates of the impact point of the shower core of the MC EAS of Fig.~\ref{3dMCevents} at each of the MATHUSLA tracking layers. The open squares in gray show the respective $(z_c, y_c)$ coordinates, and the  black circles, the $(z_c, x_c)$ values of the impact points of the EAS core. The continuous line in black and the dotted line in gray  are the corresponding results of the fits with a linear relation to the $(z_c, x_c)$ and  $(z_c, y_c)$ data points to improve the estimation of the shower core position.}
 \label{2dMCeventsvertical}
\end{figure}

\begin{figure}[b!]
 \centering
  \includegraphics[width=2.5 in]{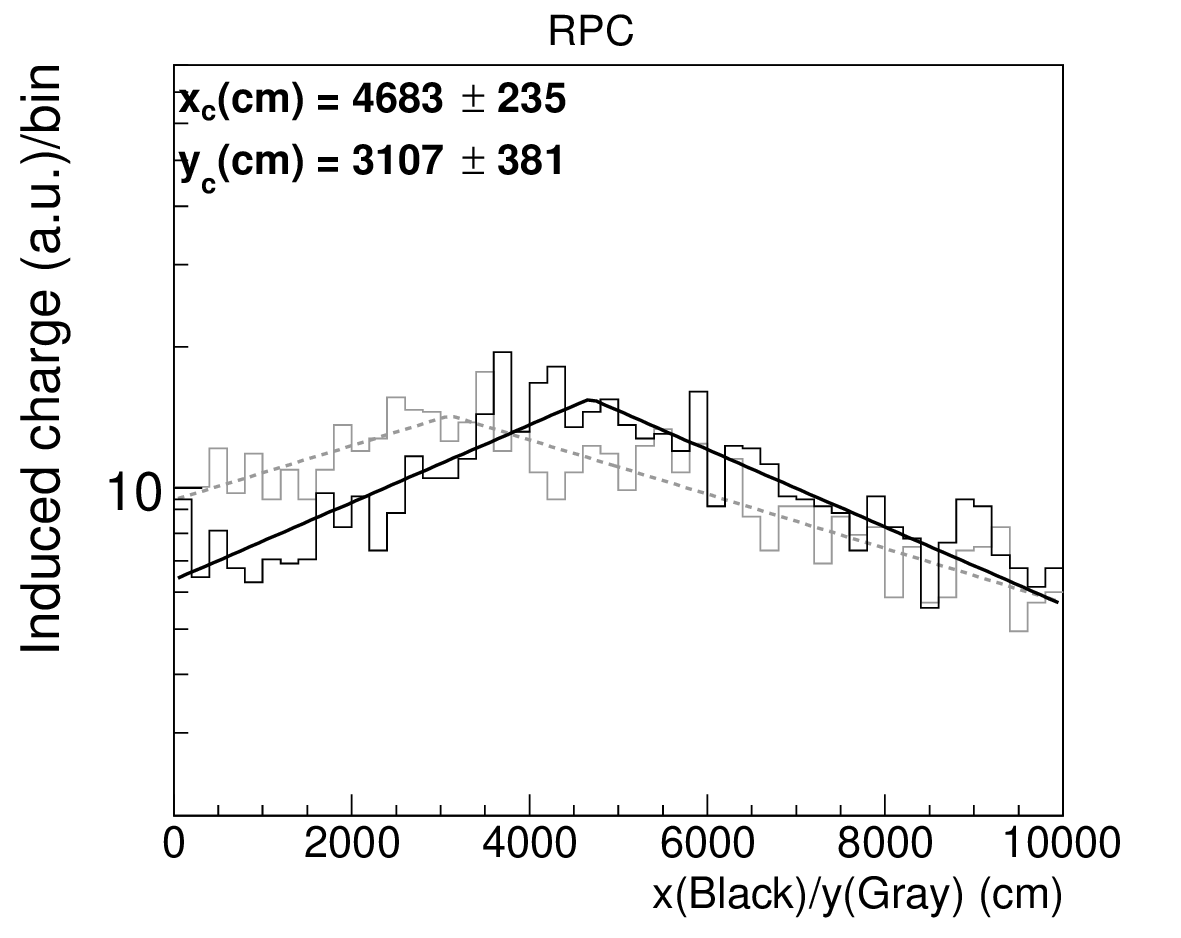}
  \includegraphics[width=2.5 in]{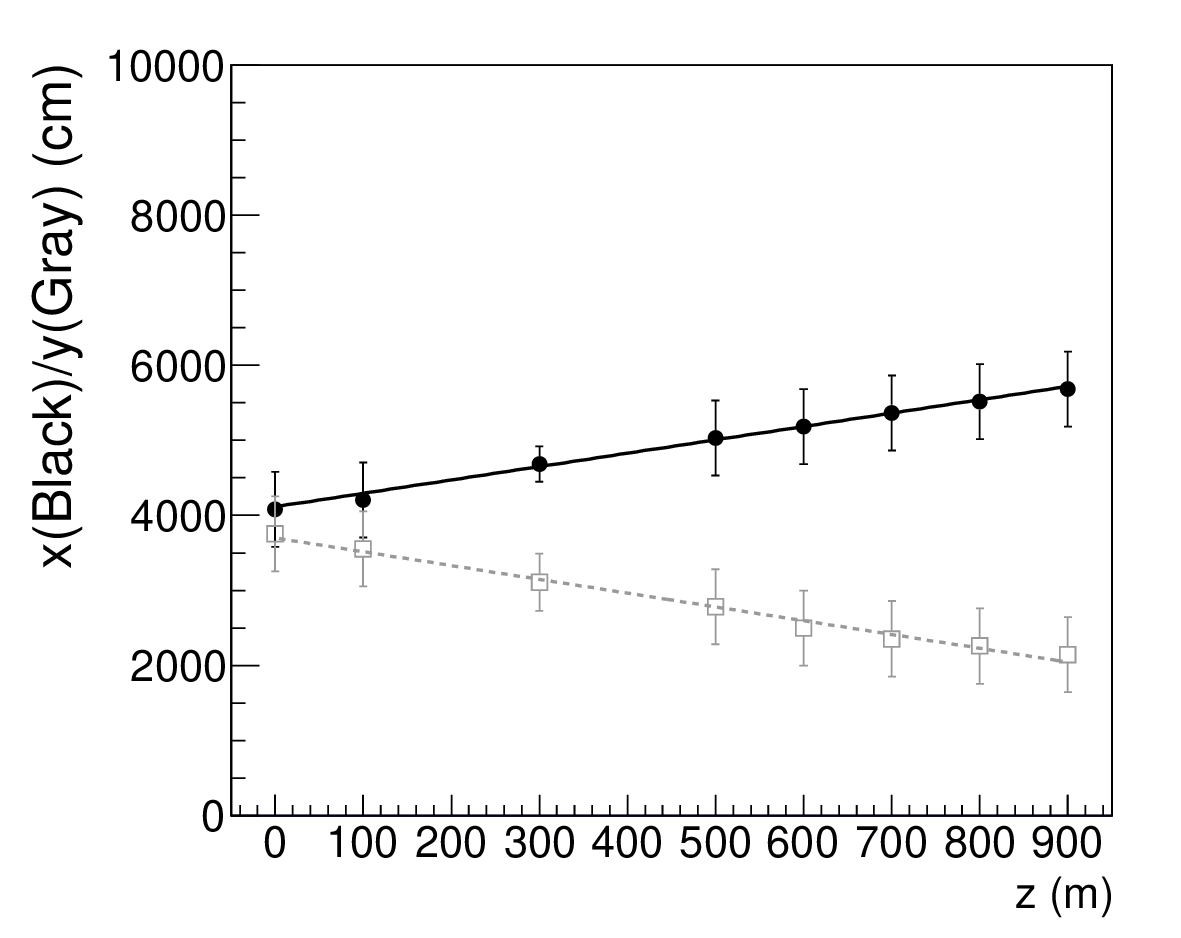}
  \caption{Left: The projected  $X$ (black) and $Y$ (gray) distributions of the number of Big Pads at the RPC layer of MATHUSLA for the inclined MC event of Fig.~\ref{3dMCeventsinclined}. The black and gray lines represent the fits with the formula (\ref{eqcore}) to the $X$ and $Y$ projected profiles of the MC shower event, respectively. Right: Estimated coordinates of the impact point of the shower core of the MC EAS of Fig.~\ref{3dMCeventsinclined} at each of the MATHUSLA tracking layers. The open squares in gray show the respective $(z_c, y_c)$ coordinates, and the  black circles, the $(z_c, x_c)$ values of the impact points of the EAS core. The continuous line in black and the dotted line in gray  are the corresponding results of the fits with a linear relation to the $(z_c, x_c)$ and  $(z_c, y_c)$ data points to improve the estimation of the shower core position.}
 \label{2dMCeventsinclined}
\end{figure}

  To illustrate the  core reconstruction procedure, we present on the left-hand-side of Figs.~\ref{2dMCeventsvertical} and \ref{2dMCeventsinclined} the projected histograms of the signals at the RPC layer of the MATHUSLA detector that were used in the last step of our method to find the EAS core of the vertical and inclined MC events presented in Figs.~\ref{3dMCevents} and \ref{3dMCeventsinclined}, respectively. The corresponding fits with the lateral distribution functions of Eqs.~(\ref{eqcore}), which provide the final estimations of the EAS core positions, are also displayed in the left of Figs.~\ref{2dMCeventsvertical} and \ref{2dMCeventsinclined}  together with the best fitting parameters $(x_c, y_c)$ for each event.

From the right-hand-side in Figs.~\ref{2dMCeventsvertical} and \ref{2dMCeventsinclined}, we found that if there are at least three detector layers where the core position was successfully reconstructed, a fit with a straight line to the corresponding $(x_c, y_c , z_c)$ coordinates of the reconstructed shower core can be applied to get improved positions of the EAS core at each MATHUSLA layer. This procedure could also be used to find the impact points of the shower core in the detector planes where this observable was not successfully reconstructed, provided that the EAS core is inside the instrumented area. This possibility would permit exploiting the tracking capabilities of MATHUSLA, as illustrated on the right-hand side of figs.~\ref{2dMCeventsvertical} and \ref{2dMCeventsinclined}, where we show the fits with a straight line to the core positions of the events in figs.~\ref{3dMCevents} and \ref{3dMCeventsinclined} estimated in each MATHUSLA tracking layer. This technique needs to be further investigated, therefore, it will not be applied in the present analysis.

  \subsection{Arrival direction of the air shower}
  \label{EASrecodirection}

  In a second stage, the arrival direction of the EAS is estimated using the timing information from the RPC. As a first approximation the direction of the EAS is found from the fit parameters $(a_1, a_2, a_3)$ of a 3D plane to the corresponding arrival time data of the shower front. These parameters are determined by solving the following set of linear equations:
  \begin{equation}
      a_1 \cdot x_i  + a_2 \cdot y_i + a_3 = c t_{1st, i}, 
      \label{eqtimes}
  \end{equation}
  using the Moore-Penrose pseudoinverse \cite{Panos:2007}. In the above equation, $i = 1,...,n_{hit}$ runs over each detector element with signal in the layer, where $n_{hit}$ is the total number of hit detector modules. The arrival time of the first EAS particle to the $i$-th detector element is represented by $t_{1st, i}$ and the respective spatial coordinates are denoted by $(x_i, y_i)$, while $c$ is the velocity of the light in vacuum. The fit parameters and the arrival direction of the cosmic ray are obtained from the expressions
  \begin{eqnarray}
      \phi &=&  \arctan{(a_2/a_1)}, \label{eqanglea} \\
      \theta &=& \arcsin{\sqrt{a_1^2 + a_2^2}}.
      \label{eqangleb}
  \end{eqnarray}  
  Here, we are using the CORSIKA definitions for the zenith ($\theta$) and the azimuth ($\phi$) angles of the shower axis \cite{Heck:1998vt}:
 $\theta$ is the measured angle between the  momentum vector of the incident nucleus and the vertical direction, while $\phi$ is defined as the angle between the horizontal projection of the momentum vector of the primary cosmic ray and the $X$ axis. The angle $\phi$ is defined as positive for counterclockwise rotations from the $X$ direction.  In this case, we need to add $\pi$ to the result of equation (\ref{eqanglea}) if $a_1 < 0$ and $a_2 > 0$, and  $-\pi$ when both  $a_1$  and $a_2$ are negative.

To improve the estimation of the arrival direction of the EAS events, we fit the measured time data with a flat EAS front using the minimum $\chi^2$ method and the previous estimations of $\theta$ and $\phi$ as initial values. We minimised the following expression:
  \begin{equation}
      \chi^2 = \sum_{j=1}^{m} \left[ \frac{ t_{1st, j} -  T_0 + (x_j - x_c) n_x/c +  (y_j - y_c) n_y/c  }{\sigma_t}\right]^2,
      \label{eqchi2_direction}
  \end{equation}
  where $T_0$, $\theta$ and $\phi$ are the fitting parameters. Here, $T_0$ is the arrival time of the shower core to the detector layer under consideration, and $n_x$ and $n_y$ are the $X$ and $Y$ components of the normalised vector $\hat{n}$ that is anti-parallel to the momentum of the cosmic ray,
  \begin{eqnarray}
    \hat{n} &=& (\sin \theta \cos(\phi -\pi), \, \sin \theta \sin(\phi -\pi), \, \cos \theta),
    \label{eqnvector}
  \end{eqnarray}  
  $(x_c, y_c)$ is the respective position of the EAS core (estimated with the procedure described in section \ref{EASrecocore}) and $\sigma_t$, the estimated uncertainty in the time measurements, which  we conservatively set to $0.5 \, \mbox{ns}$. The index $j$ runs only over the $m$ detector elements that had a signal within a time window of $10^5 \, \mbox{ns}$ measured from the first hit recorded in the layer. For the minimization procedure, we used MINUIT \cite{James:1994vla} from ROOT.

  Final  approximations of $\theta$ and $\phi$ are obtained by using a cone instead of a plane in the $\chi^2$ fit of the arrival times of the EAS front, which corrects the timing information for curvature effects. In this case, we need to minimise the equation below:
  \begin{eqnarray}
      \chi^2 &=& \sum_{k=1}^{m^\prime} \left[ \frac{t_{1st, k} -  T_0 + (x_k - x_c) n_x/c 
                 + (y_k - y_c)n_y/c   - r_{k} \tan(\rho)/c }{\sigma_t} \right]^2.
      \label{eqchi2_direction2}
  \end{eqnarray} 
  In this relation, $\rho$ is the fit parameter that controls the slope of the cone with respect to the plane perpendicular to the shower axis, and $r_{k}$ is the radial distance of the $k$-th detector element with signal to the EAS core in shower disk coordinates. This variable is defined as
  \begin{equation}
      r_{k} = \sqrt{\left[(x_k - x_c) - \Delta_k \times n_x \right]^2
      +  \left[(y_k - y_c) - \Delta_k \times n_y \right]^2
      +  \left[ \Delta_k \times n_z \right]^2},
      \label{eqrdsc}
  \end{equation} 
 with
  \begin{equation}
      \Delta_k      =  (x_k - x_c) n_x + (y_k - y_c) n_y.
      \label{eqrdsc3}
  \end{equation}  
  In equation (\ref{eqchi2_direction2}), the sum runs only over detector modules  with signals at the RPC layer recorded during a time window of $100 \, \mbox{ns}$ and  $600 \, \mbox{ns}$ (taking as a reference the time of the module with the first hit in the detector layer) for vertical and inclined EAS, respectively. The above procedure will be also applied for inclined EAS and timing data from the scintillating detector planes in case of events with charged particle densities below the saturation threshold of the scintillating bars. 

  \subsection{Lateral density distribution of EAS}
\label{LDFdescription}

   A useful observable in EAS observatories to study the main properties of cosmic rays is the lateral distribution of particle densities or the lateral distribution function (LDF) at the shower front. Therefore, in the next stage of the EAS reconstruction, the LDF of the event is built using the data from the RPC. For this task, first, we computed the local density of deposited charge, $\rho(r)$, in each Big Pad of the RPC detector. Then, we estimated the radial distances, $r$, to the EAS core at shower disk coordinates (using the estimated values for the core position and the arrival direction) of the density measurement and, finally, we calculated the mean values of $\rho(r)$ for different bins of $r$. Here a common bin width of $\Delta r = 4 \, \mbox{m}$ was employed. 

\begin{figure}[t!]
 \centering
  \includegraphics[width=2.5 in]{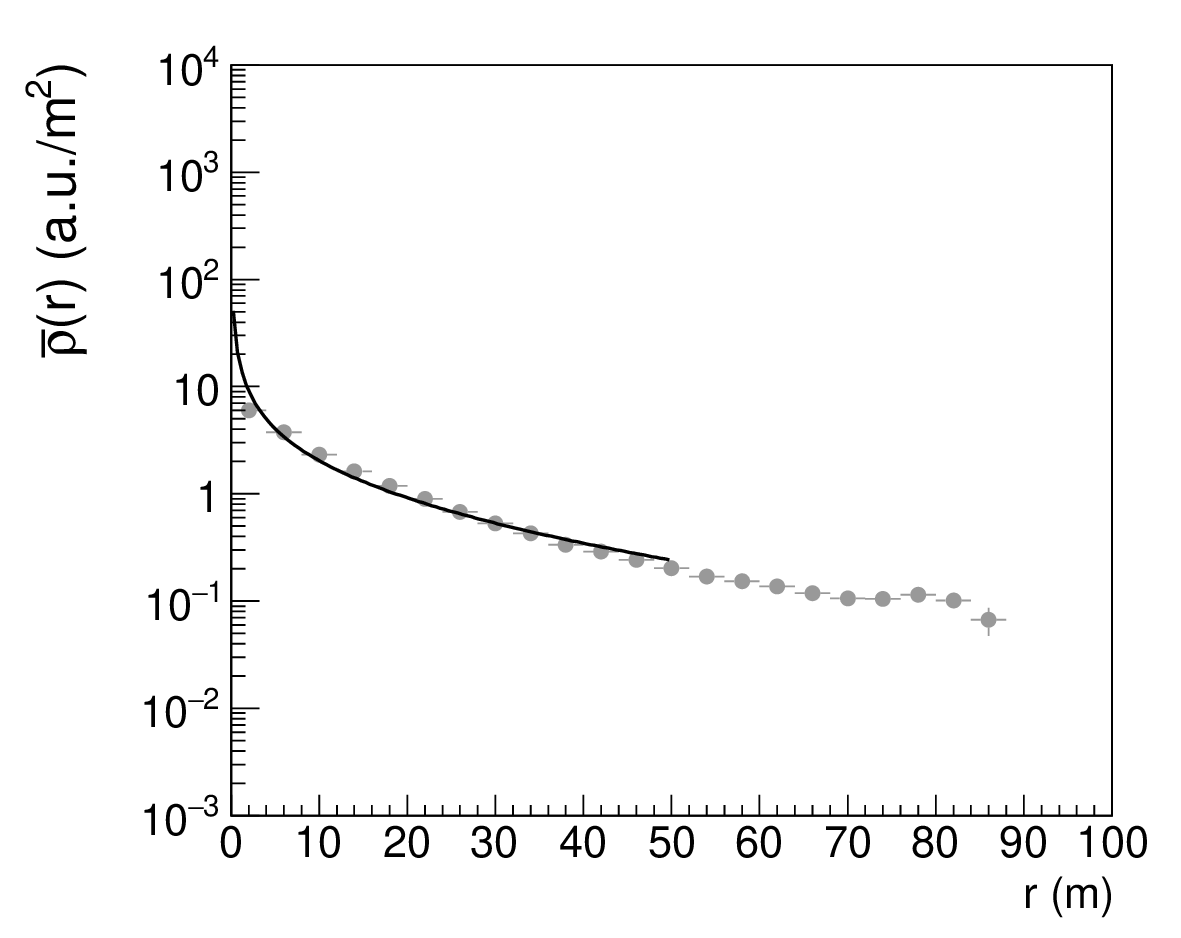}
  \includegraphics[width=2.5 in]{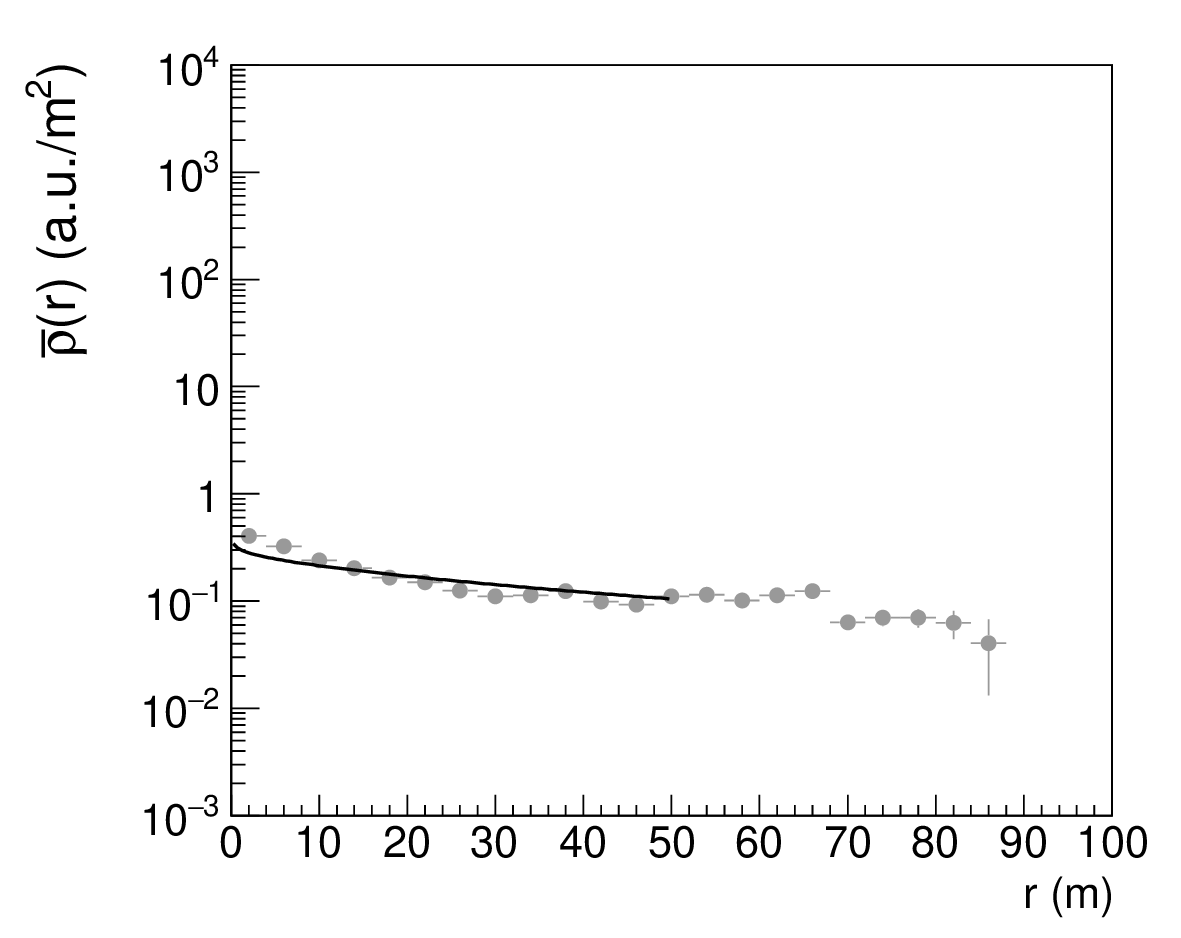}
  \caption{The radial distributions of the mean deposited charge per unit area at the RPC detector of MATHUSLA (at shower disk coordinates) corresponding to the vertical (left) and inclined (right) MC events presented in figs.~\ref{3dMCevents} and \ref{3dMCeventsinclined}, respectively. The LDFs are shown with  data points, which were fitted with the NKG-like function of Eq. (\ref{eqNKG}) (continuous lines).}
 \label{LDFMC}
\end{figure}

The amplitude and the steepness of the radial density distribution of an EAS are considered to be energy and mass composition sensitive parameters of primary cosmic rays, respectively \cite{Apel2006,Bartoli2017b}. Therefore, it is necessary to parameterise the LDFs obtained with the RPC measurements of the EAS to explore the sensitivity to the primary energy and the mass of the cosmic-ray nuclei using this information. The parameterization can be achieved by fitting the LDF data with a Nishimura-Kamata-Greisen (NKG) formula \cite{NKG1,NKG2,NKG3} often used in this context: 
    \begin{equation}
     \rho(r) = A \cdot 
     \left( \frac{r}{r_0}\right)^{s-2}  \left( 1 + \frac{r}{r_0}\right)^{s-3.5},
     \label{eqNKG}
    \end{equation}
   where $A$ is the amplitude of the LDF, $s$ is a parameter (known as the lateral shower age) that determines the flatness of the distribution and $r_0$ is a radial scale, whose value we have chosen as the Moli\`{e}re radius ($\sim 79 \, \mbox{m}$) at sea level.   
   
   Two examples of LDFs obtained from the RPC data for the simulated air showers presented in figs.~\ref{3dMCevents} and \ref{3dMCeventsinclined} are shown in the left and right panels of Fig.~\ref{LDFMC}, respectively. The radial distributions have been fitted with the NKG function (\ref{eqNKG}) for $r \leq 50 \, \mbox{m}$. The fitted functions are also displayed in the above figures. As observed from the plots in Fig.~\ref{LDFMC}, the NKG expression seems to describe reasonably well the LDF of the MC data.

\section{Selection criteria}
\label{selection}

 For this work, we have selected a set of quality cuts to reduce the magnitude of the systematic uncertainties in the reconstructed EAS that are used to study the sensitivity of MATHUSLA to high-energy cosmic rays. 

 We require that the number of elements with signals in each detector layer satisfies the condition $n_{hit} \geq 100$ for vertical events, and $n_{hit} \geq 50$, for inclined EAS. 
 These cuts eliminate  low-energy events with large systematic uncertainties. Note that we have reduced the minimum number of hits for inclined events in order to lower the  energy threshold in our analysis. In addition, for the analysis of the RPC data, we accepted only EAS that passed the core and arrival direction reconstruction procedure and have reconstructed EAS cores within a central area of $90 \; \mbox{m} \times 90 \ \mbox{m}$ at the RPC layer to avoid border effects. Similar cuts were also applied for the study of inclined events recorded with the tracking layers of MATHUSLA. For vertical data from the scintillating layers, we removed the cut that requires the successful reconstruction of the shower direction at the respective layer, since we did not  estimate the shower direction of vertical EAS with these detector elements.

\section{Results}
\label{results}

\subsection{Performance of the RPC}
\subsubsection{Reconstruction efficiency}
\label{effrpc}

\begin{figure}[t!]
 \centering
  \includegraphics[width=2.5 in]{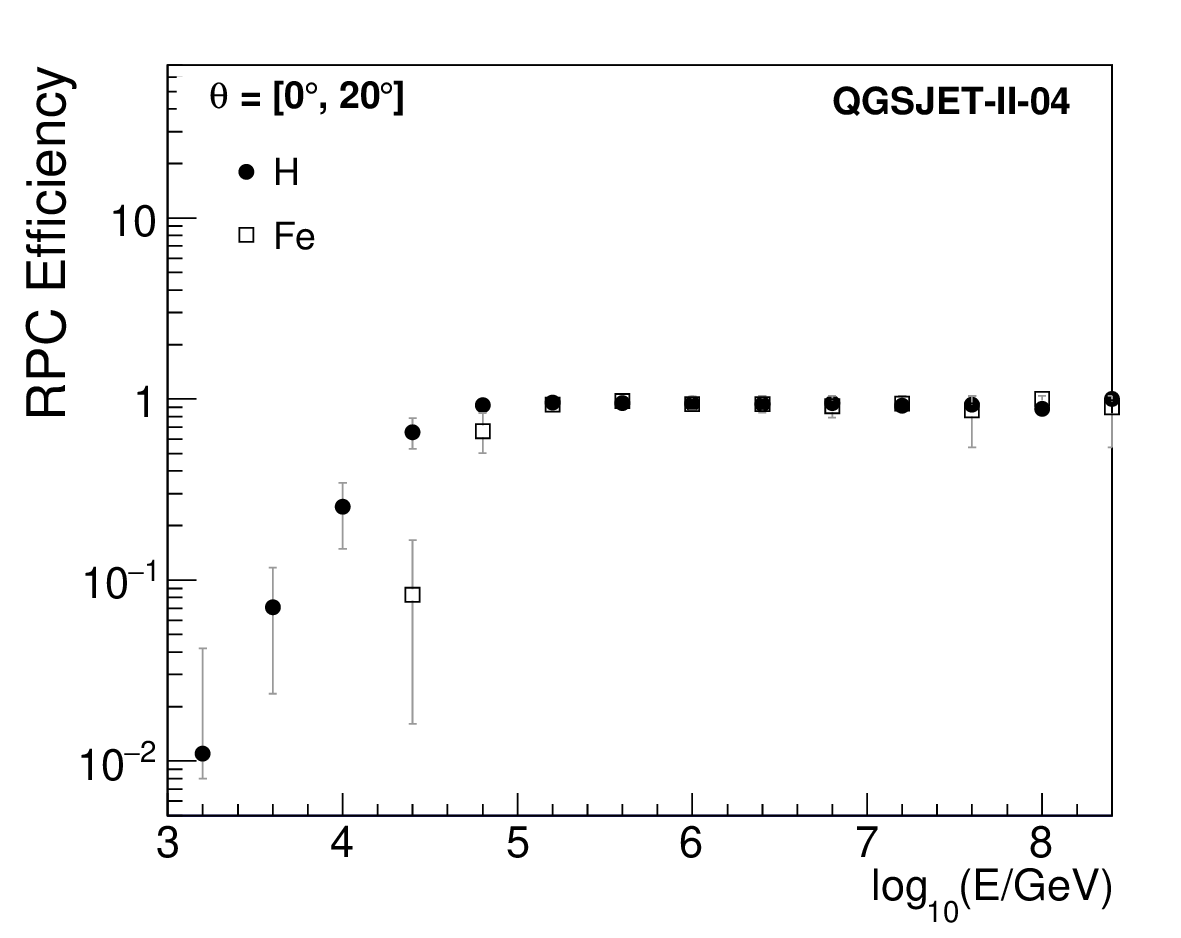}
  \includegraphics[width=2.5 in]{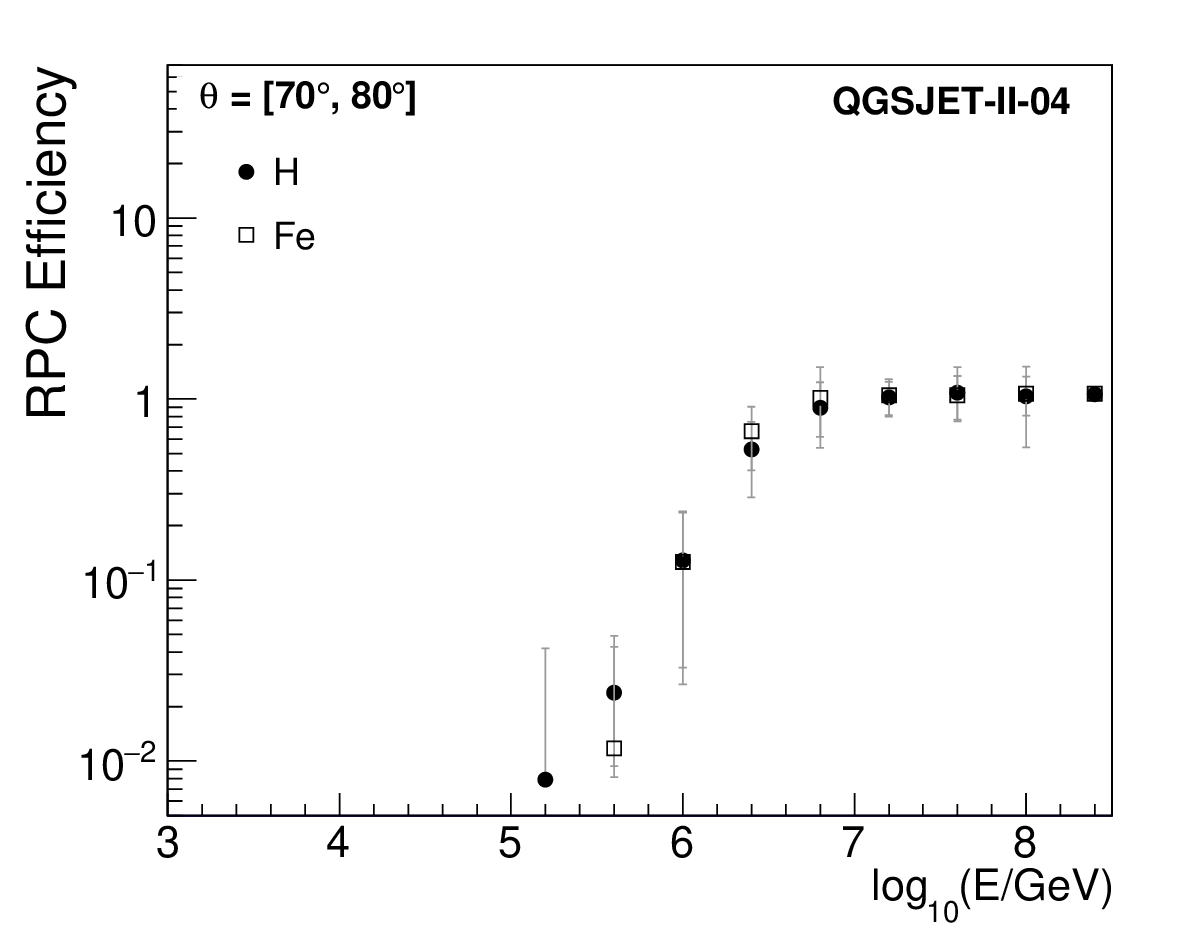}
   \caption{The selection-cut efficiency of the MATHUSLA RPC layer, estimated using the QGSJET-II-04 model, for shower events induced by protons (circles) and iron nuclei (open squares) Left: for vertical EAS with $\theta \leq 20^\circ$ and right: for inclined showers with  $\theta = [70^\circ, 80^\circ]$.}
 \label{Efficiency}
\end{figure}

 The selection-cut efficiency of the proposed RPC layer of MATHUSLA to vertical, $\theta \leq 20^\circ$, and inclined, $\theta = [70^\circ, 80^\circ]$, EAS induced by protons and iron nuclei are presented in Fig.~\ref{Efficiency}. The efficiencies were estimated using our MC simulations generated with QGSJET-II-04. From these graphs, we observe that MATHUSLA  reaches maximum efficiency at $10^{14} \, \ev$ for vertical air showers, and increases to $10^{16} \, \ev$  for larger zenith angles, which is expected due to atmospheric attenuation of the EAS.  We did not simulate a particular trigger condition in our MC simulations. For EAS detection, the RPC is expected to provide the trigger in MATHUSLA, which will be the same for vertical and inclined events. A more detailed EAS analysis with a specific trigger condition for MATHUSLA will be included in a further study. 

\subsubsection{Systematic uncertainties in EAS reconstruction}
\label{systematicuncertainty_rpc}

\begin{figure}[ht!]
 \centering
  \includegraphics[width=2.5 in]{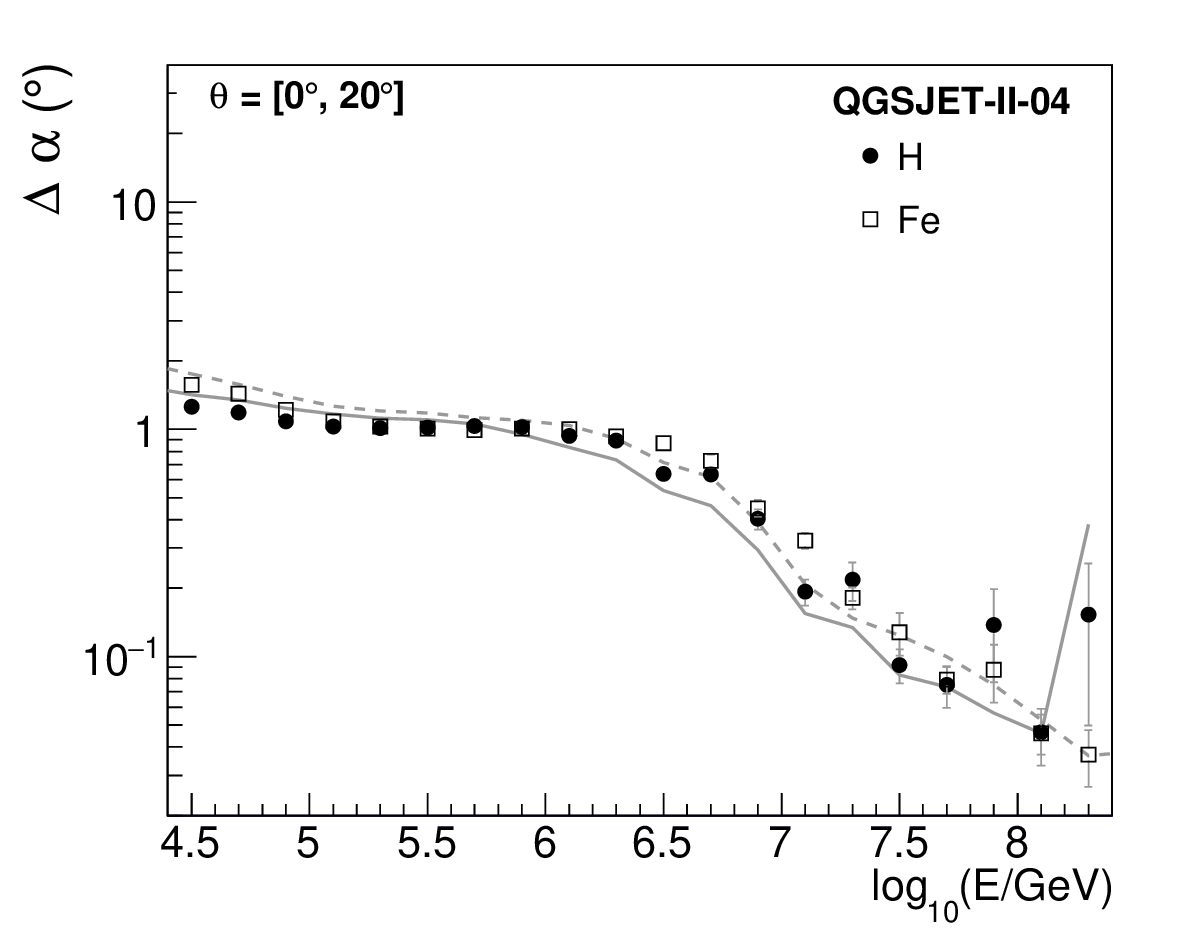}
  \includegraphics[width=2.5 in]{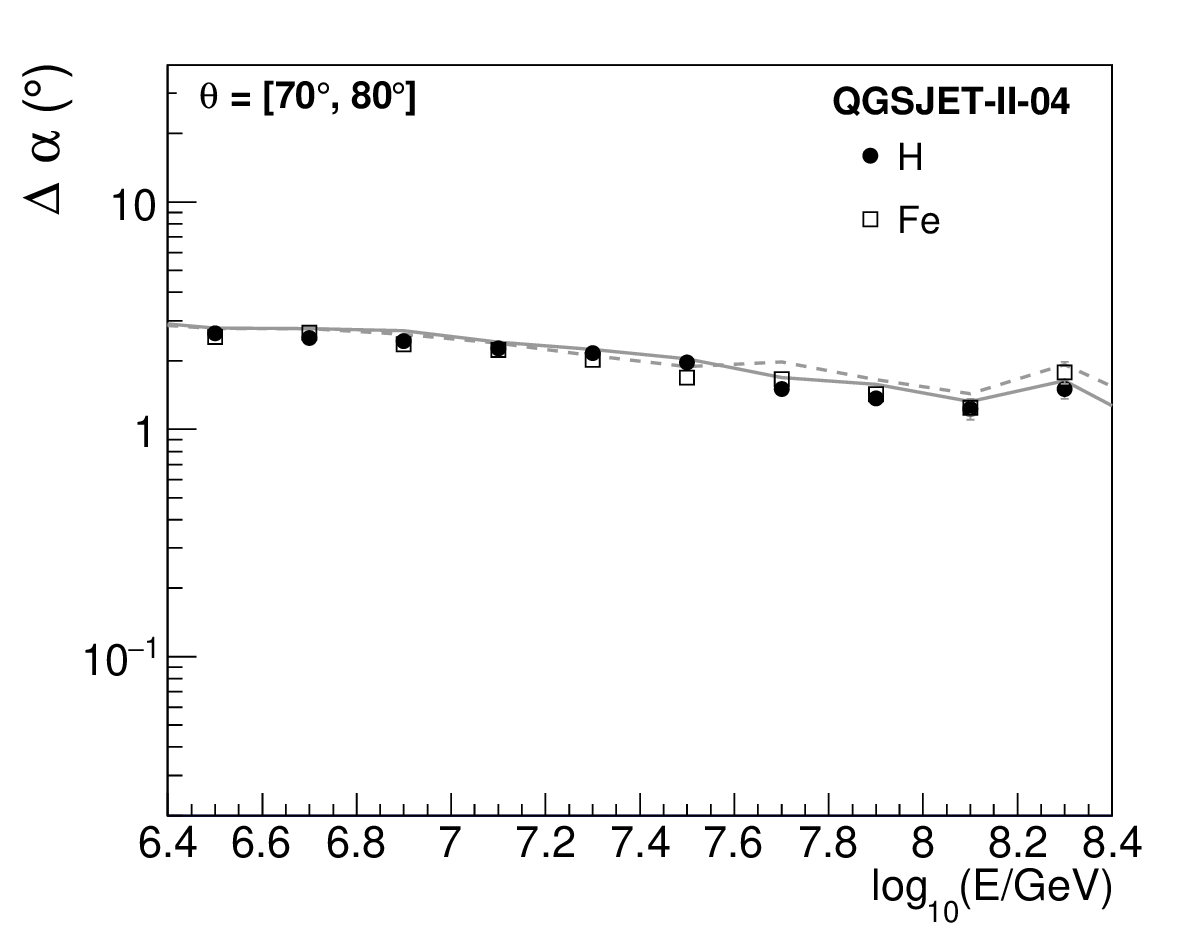}
  \caption{The bias and resolution on the arrival direction of EAS expected for the MATHUSLA RPC layer as estimated from MC events induced by H and Fe nuclei. The simulations were produced with QGSJET-II-04. Quality cuts were applied. Symbols represent the angular bias for protons (circles) and iron nuclei (open squares). The vertical errors are standard uncertainties of the mean. Continuous and dashed lines are used to represent the angular resolutions for H and Fe induced events, respectively. Left: Estimations for vertical EAS with $\theta \leq 20^\circ$. Right: Results for inclined showers within the interval  $\theta = [70^\circ, 80^\circ]$.}
 \label{biasAngleRPC}
\end{figure}

 The performance of the proposed RPC layer as a cosmic-ray detector  was additionally investigated by calculating the systematic uncertainties expected from the EAS reconstruction described in section \ref{EASreco}. First, we have estimated the bias and resolution of the arrival angle of cosmic-ray EAS as a function of the primary energy. For this aim, the data is divided in different energy intervals of width $\Delta \log_{10}(E/\mbox{GeV}) = 0.2$ and for each bin, the distributions of the angular systematic uncertainty are calculated. This systematic is defined as the angle between the reconstructed and true shower axes of the EAS events. Then, the mean and the $68 \%$ containment region of these distributions are obtained, which are defined as the angular bias $\Delta \alpha$  and resolution  $\Delta \alpha_{68}$ of the RPC layer to EAS events, respectively, and are plotted versus the central energy of the corresponding energy bin.  The results  are shown in Fig.~\ref{biasAngleRPC} for vertical (left plot) and inclined events (right graph). The uncertainties were estimated for H and Fe-induced EAS. The left plot of Fig.~\ref{biasAngleRPC} shows that  both $\Delta \alpha$  and $\Delta \alpha_{68}$ diminish with the primary energy from $E = 3.2 \times 10^{13} \, \ev$ to  $2.5 \times 10^{17} \, \ev$. At low energies ($\lesssim 3 \times 10^{16} \, \ev$), the angular bias and the resolution vary between $1^\circ$ and $2^\circ$, while at high energies, both  uncertainties decrease rapidly to values close to $\sim 0.3^\circ$. On the other hand, the right plot of  Fig.~\ref{biasAngleRPC} shows that for inclined events in the interval $E = [3.2 \times 10^{16} \, \ev, 2.5 \times 10^{17} \, \ev]$, the angular resolution is not as good. At the RPC layer, the bias $\Delta \alpha$ is larger for inclined events than for vertical EAS. From MC simulations, at energies between $3.2 \times 10^{16} \, \ev$ and $2.5 \times 10^{17} \, \ev$, we estimated that  $\Delta \alpha = [1.3^\circ, 2.7^\circ]$ and $\Delta \alpha_{68} = [1.3^\circ, 2.9^\circ]$  for inclined showers. In this case, we still observe  that both uncertainties decrease at high energies as for vertical EAS, albeit the reduction is not as pronounced as observed for the case of vertical showers. 

 \begin{figure}[b!]
 \centering
  \includegraphics[width=2.5 in]{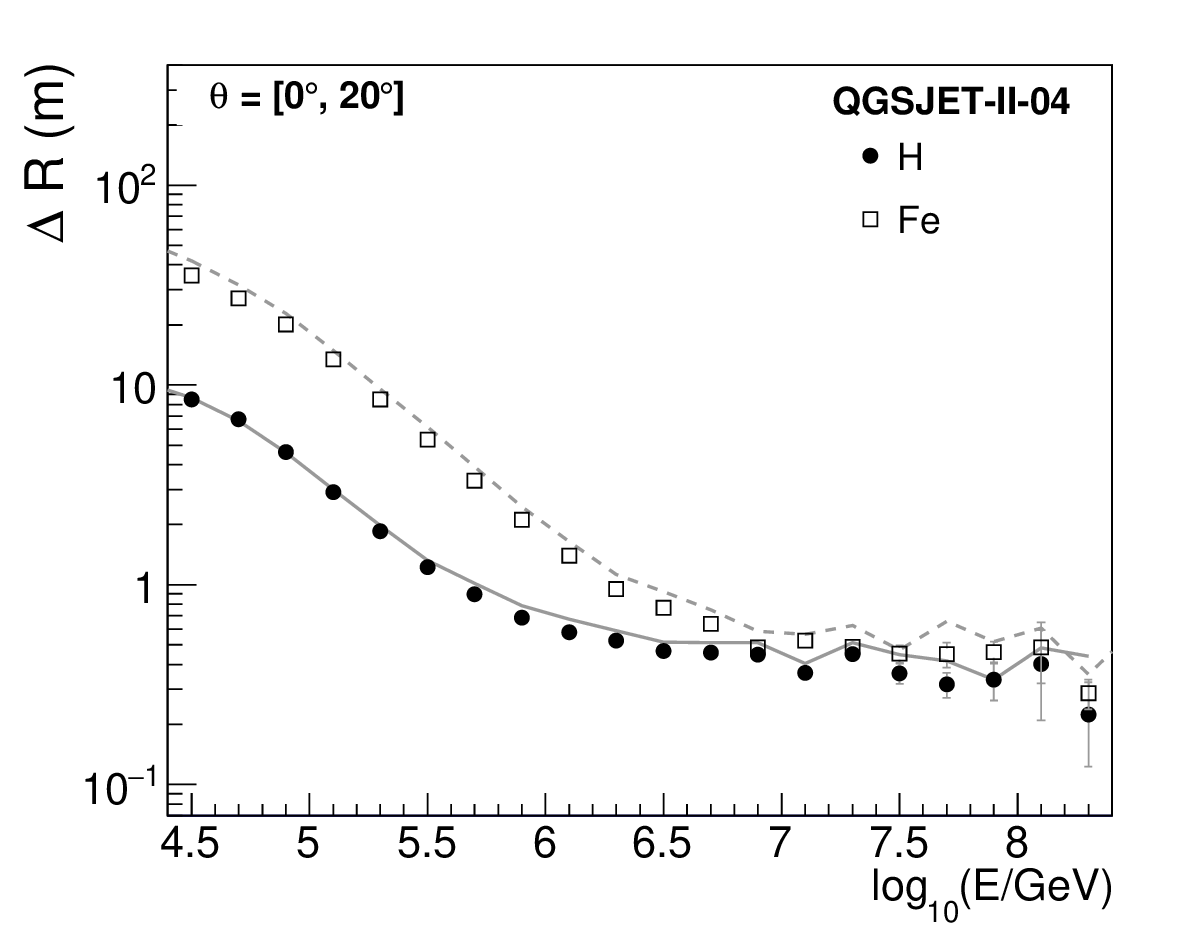}
  \includegraphics[width=2.5 in]{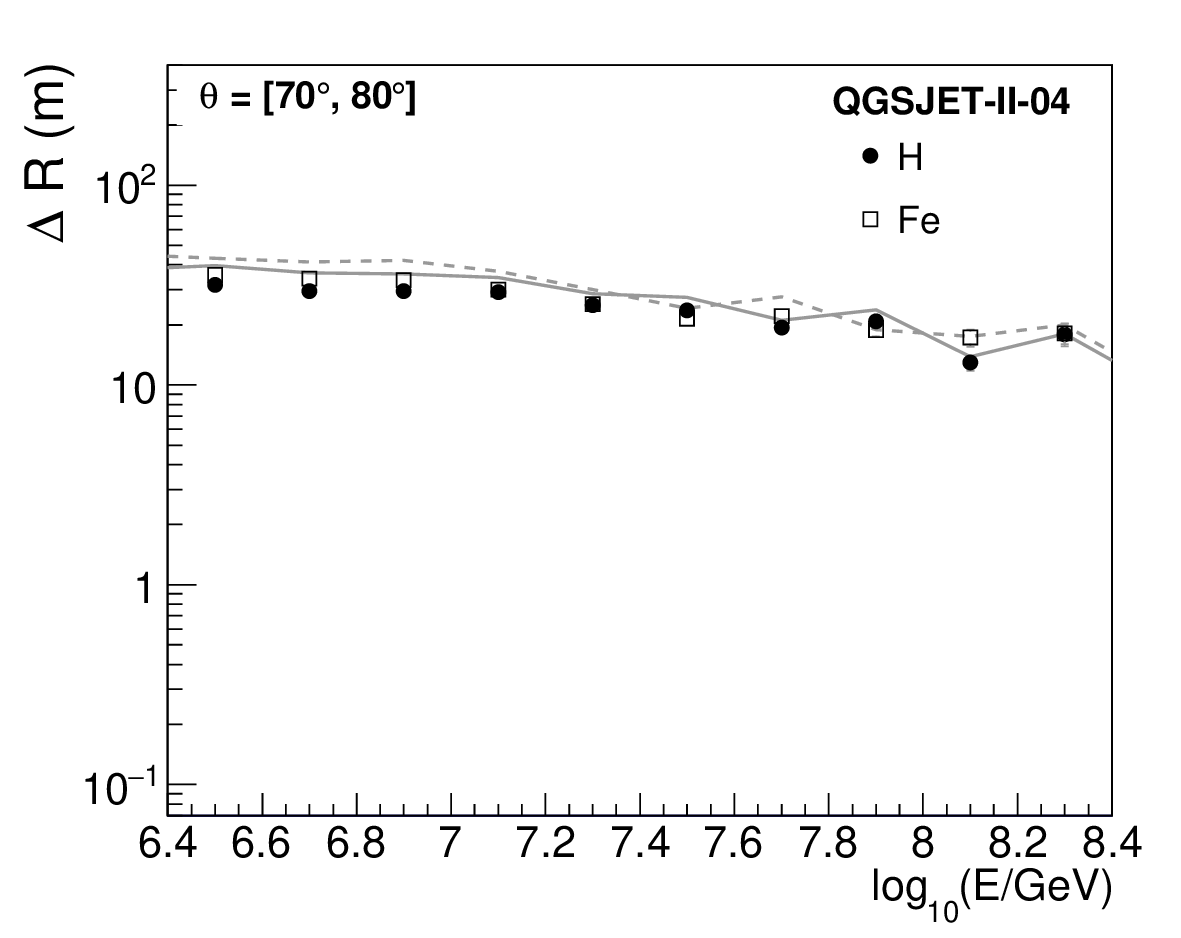}
  \caption{The bias and resolution on the shower core expected for the MATHUSLA RPC layer as estimated from MC events induced by H and Fe nuclei. The simulations were produced with QGSJET-II-04. Quality cuts were applied. Symbols represent the bias on the shower core location for  events produced by protons (circles) and iron nuclei (open squares). The vertical uncertainty errors are standard uncertainties of the mean. Continuous and dashed lines are used to represent the core resolutions for H and Fe-induced events, respectively. Left: Estimations for vertical EAS with $\theta \leq 20^\circ$. Right: Results for inclined showers within the interval  $\theta = [70^\circ, 80^\circ]$.}
 \label{biasCoreRPC}
\end{figure}

We also estimated the bias $\Delta R$ and resolution $\Delta R_{68}$ on the EAS core position expected for the RPC layer at different primary energies. These quantities are defined as the mean and the $68 \%$ containment region of the distribution for the distance between the reconstructed and the true shower core locations, respectively, for each energy bin.  $\Delta R$ and $\Delta R_{68}$ are  presented in Fig.~\ref{biasCoreRPC} for vertical and inclined EAS. We employed a bin width of   $\Delta \log_{10}(E/\mbox{GeV}) = 0.2$ .  The plots were estimated for H and Fe events using QGSJET-II-04. The left plot in Fig.~\ref{biasCoreRPC} shows that the reconstruction uncertainties for the shower core position  of vertical events decrease from a few tens of meters at $3.2 \times 10^{13} \, \ev$ up to a fraction of a meter between $\sim 10^{15} \, \ev$ and $10^{16} \, \ev$, above which the reconstruction uncertainties remain almost constant with the primary energy. In particular, the bias (resolution) of the EAS core position at  $3.2 \times 10^{13} \, \ev$ is  $\Delta R = 8.6 \, \mbox{m}$ ($\Delta R_{68} = 8.6 \, \mbox{m}$) and $35.4 \, \mbox{m}$ ($42.1 \, \mbox{m}$) for H and Fe events, respectively. At energies larger than $10^{15} \, \ev$, $\Delta R$ and  $\Delta R_{68}$ are about  $0.4 \, \mbox{m}$ for vertical showers created by protons and  $0.5 \, \mbox{m}$ above $10^{16} \, \ev$ for iron nuclei with $\theta < 20^\circ$.
 
  On the other hand, from Fig.~\ref{biasCoreRPC}, right, we see that for energies greater than $3.2 \times 10^{15} \, \ev$, the shower-core bias and resolution for inclined data are of the order of some tens of meters based on our shower reconstruction method of MC simulations for the MATHUSLA RPC layer. We also observe that both $\Delta R$ and $\Delta R_{68}$ are reduced at higher energies, however, their corresponding magnitudes in the high-energy domain do not reach the same values that we found for vertical events. The reason is the reduction in the number of hit Big Pads close to the EAS core for inclined events due to attenuation effects in the atmosphere.  For inclined EAS and energies between $3.2 \times 10^{15} \, \ev$ and  $2.5 \times 10^{17} \, \ev$, $\Delta R$ decreases from  $36 \, \mbox{m}$ to $\sim 14 \, \mbox{m}$, while the core resolution is reduced from $44 \, \mbox{m}$  to approximately $16 \, \mbox{m}$.

\subsubsection{Fraction of hit Big Pads at the RPC layer}
 \label{section_fhitRPC}

\begin{figure}[ht!]
 \centering
  \includegraphics[width=2.5 in]{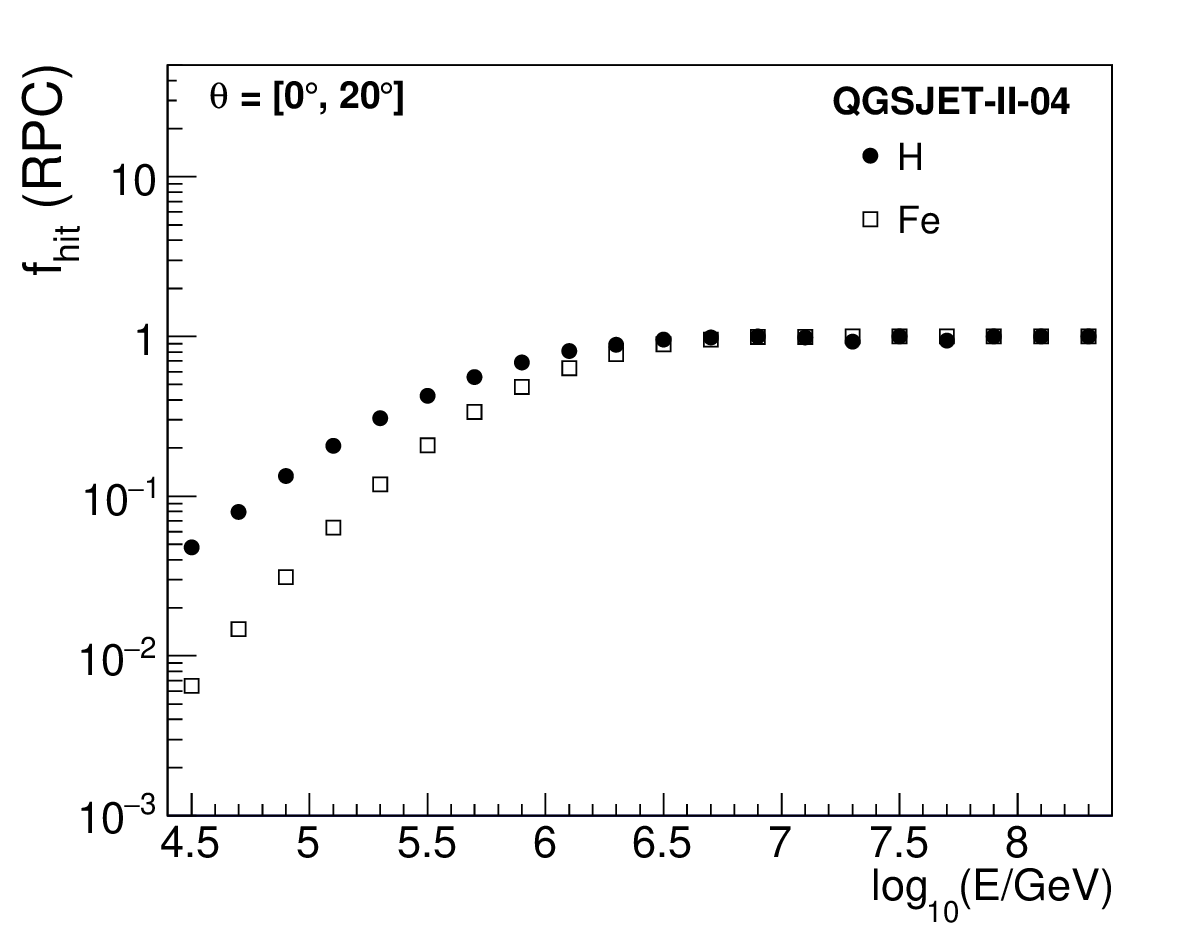}
  \includegraphics[width=2.5 in]{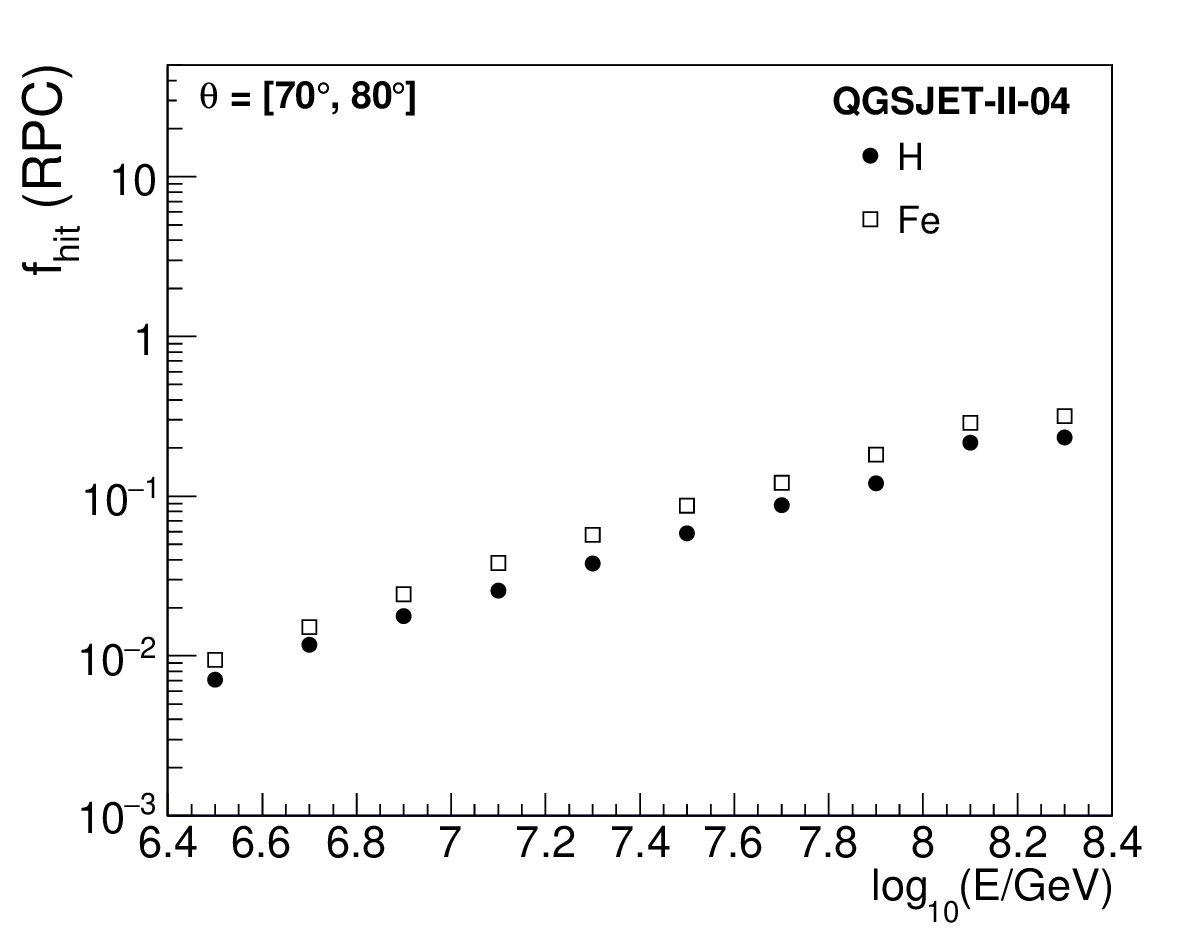}
  \caption{The mean fraction of hits in the Big Pads at the MATHUSLA RPC layer obtained from  MC simulations with QGSJET-II-04 for proton (circles) and iron (open squares) cosmic-ray primaries. The results are presented for two zenith angle intervals: left column: $\theta = [0^\circ, 20^\circ]$ and right column: $\theta = [70^\circ, 80^\circ]$.  Standard uncertainties on the mean were estimated, but they are smaller than the size of the symbols.}
 \label{fhitrpc}
\end{figure}

 We have also investigated the mean fraction of hit Big Pads $f_{hit}$ at the RPC layer of MATHUSLA that is expected from EAS events of different energies, see Fig.~\ref{fhitrpc}.  As before, calculations were performed with our MC simulations for vertical and inclined showers, and for protons and iron nuclei as primaries.  We used energy bins of size  $\Delta \log_{10} (E/\gev) = 0.2$. For the estimations, we computed $f_{hit}$ event-by-event, then, for a given primary energy bin and cosmic-ray nuclei, we determined the  corresponding average values of $f_{hit}$. From these plots, we can see that $f_{hit}$  increases for vertical showers up to an energy threshold of $\sim 8 \times 10^{15} \, \ev$ at which point the mean hit fraction saturates.  For inclined events with $\theta > 70^\circ$, there is no saturation of $f_{hit}$ at least up to energies of $2.5 \times 10^{17} \, \ev$.

According to Fig.~\ref{fhitrpc}, below the energy threshold for the saturation of $f_{hit}$, the expected fraction of hits is different for proton and iron-induced events. For $\theta < 20^\circ$, $f_{hit}$ is greater for light primaries than for heavy nuclei because proton primaries have a larger production of charged shower particles. For events with $\theta > 70^\circ$, where the atmospheric absorption of the electromagnetic component is more important and the fractional abundance of the muon component increases, the fraction of hits associated with iron events tends to be slightly larger than for proton-induced showers. 

 The fraction of hits increases with the primary energy of the EAS below the $f_{hit}$ saturation region for both vertical and inclined events.  For inclined EAS, this dependence is linear in a logarithmic scale. These results suggest that $f_{hit}$ could be used to estimate the cosmic-ray energy below the threshold for $f_{hit}$ saturation.

 \subsubsection{Maximum density of hit particles at the Big Pads due to EAS events}

\begin{figure}[ht!]
 \centering
  \includegraphics[width=2.5 in]{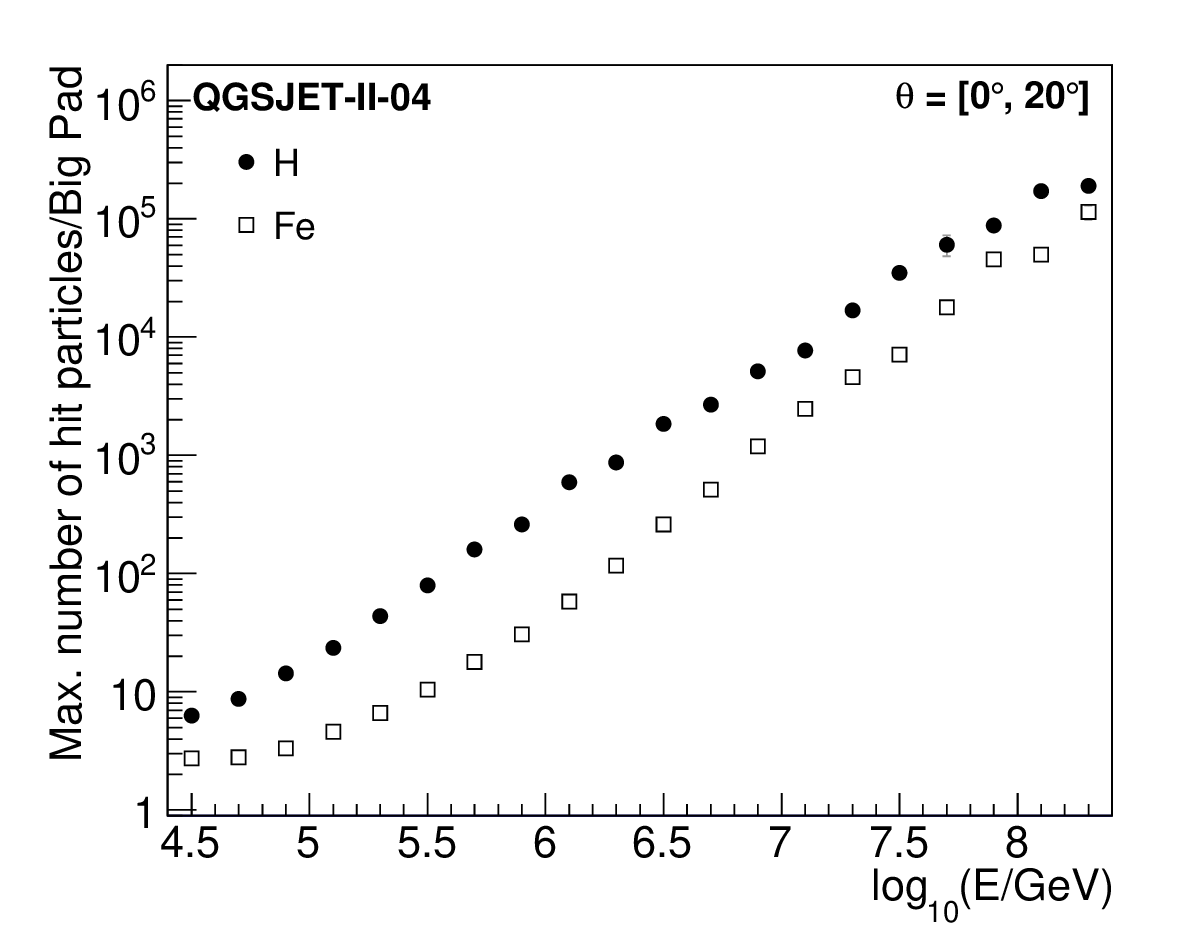}
  \includegraphics[width=2.5 in]{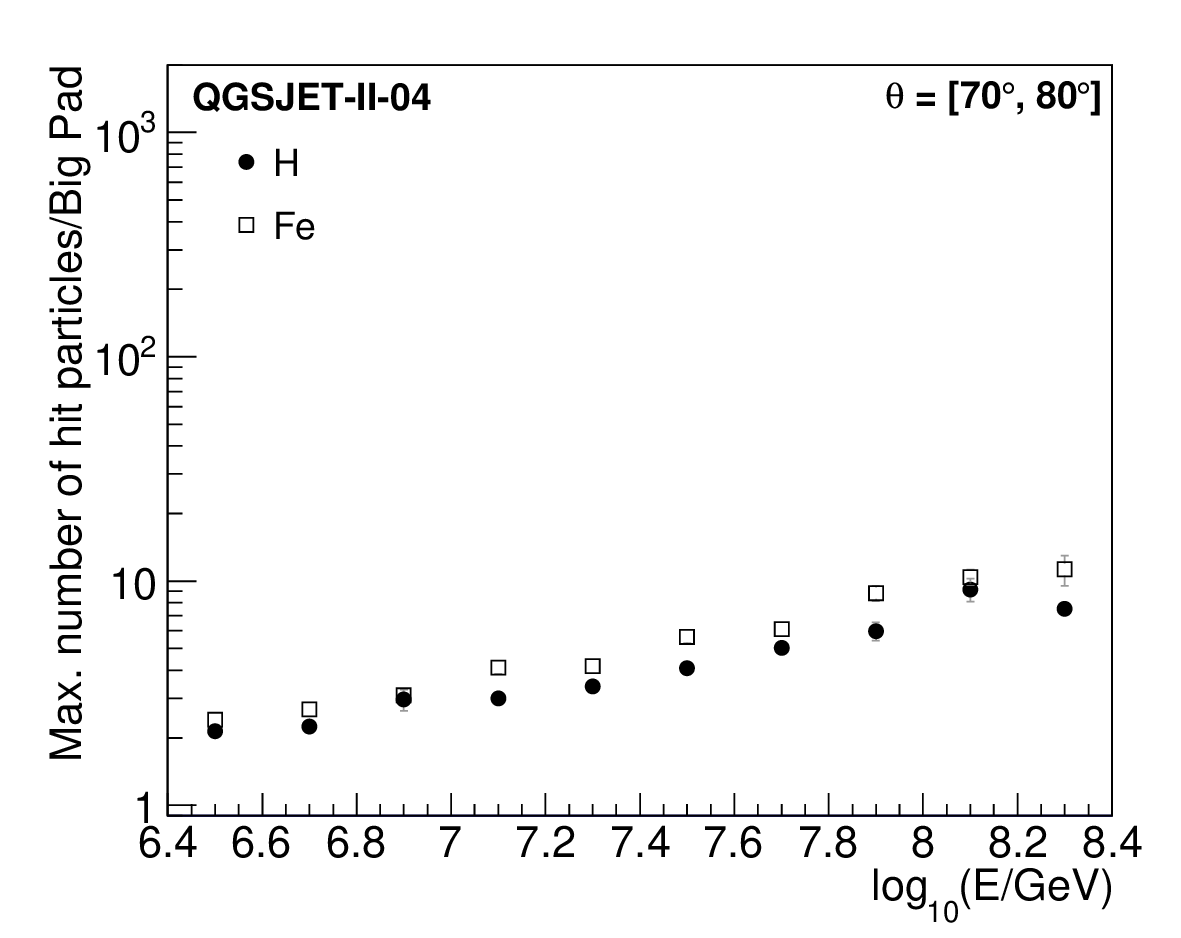}
  \caption{The mean maximum number of charged particles per Big Pad at the RPC layer of MATHUSLA induced by EAS events as a function of the primary energy. The plots were estimated from MC data after applying our selection cuts and employing the QGSJET-II-04 model. Two primary nuclei were simulated: H (circles) and Fe (open squares). MC simulations for vertical EAS ($\theta \leq 20^\circ$) are presented on the left panel and  for inclined showers ($\theta = [70^\circ, 80^\circ]$), on the right one. Vertical error bars correspond to statistical uncertainties of the mean.}
 \label{Maxhit}
\end{figure}

 Individual  saturation at the Big Pads of the RPC of MATHUSLA is expected to appear for more than  $\sim 10^{6}$ hits of charged particles per module. These densities can be found for air showers of very high energies. Therefore, there could be some EAS energy threshold for which, we would begin to see some saturation of signals at the Big Pads due to shower events. In order to investigate this shower energy threshold, we have used our MC simulations to calculate the mean of the maximum number of hit-charged particles per Big Pad for EAS  as a function of the primary energy. The plots for H and Fe-induced showers are presented in Fig.~\ref{Maxhit} for vertical and inclined events. From  the results of Fig.~\ref{Maxhit}, we do not find any potential saturation at the Big Pads due to hadronic-induced EAS in the energy range of interest. 

\subsubsection{Spatial structure of EAS at the RPC layer}
\label{spatialstructure}

\begin{figure}[ht!]
 \centering
  \includegraphics[width=5.2 in]{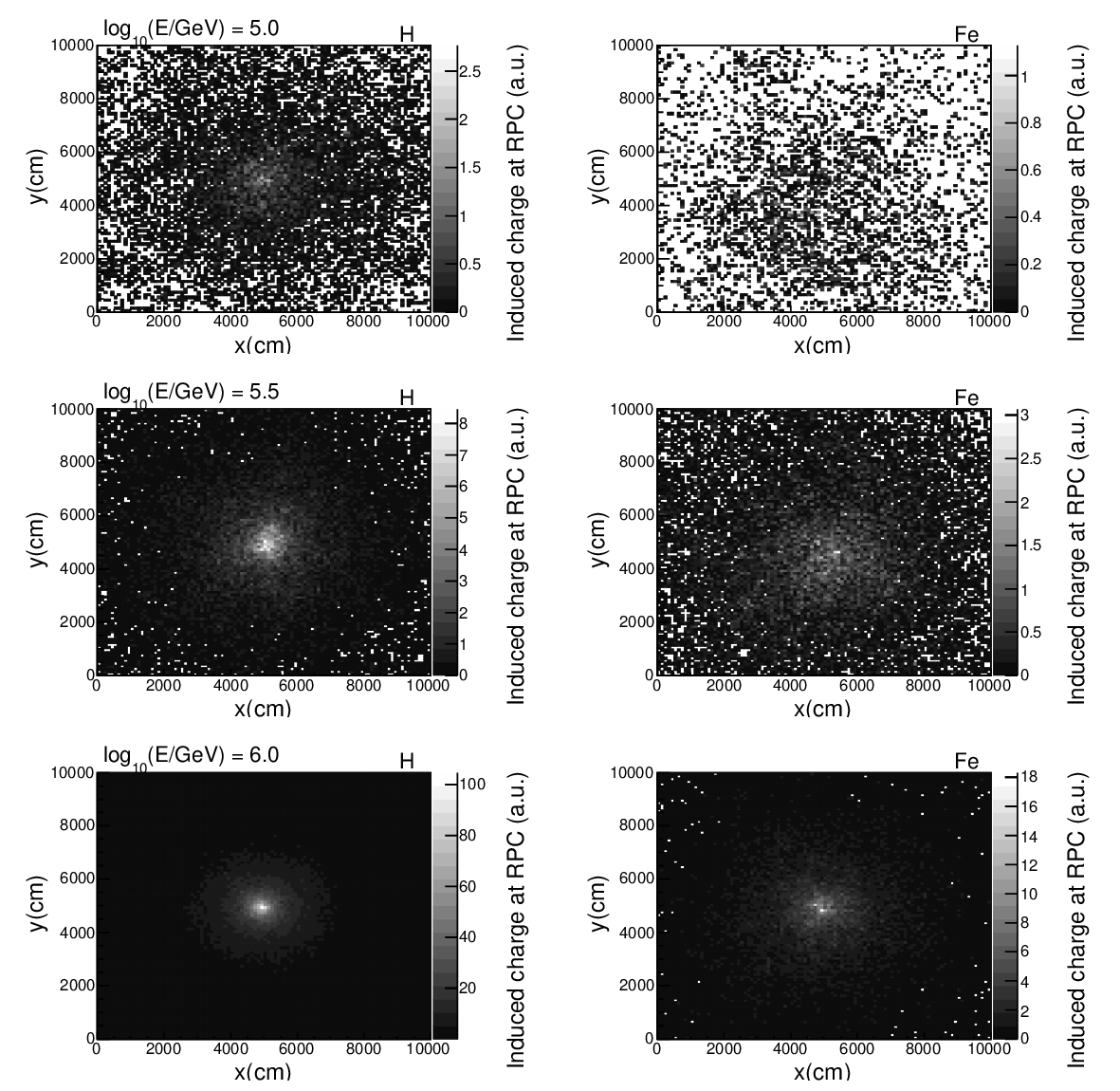}
  \caption{The 2D charge distributions of the RPC MATHUSLA layer for individual MC events with $\theta = 7^\circ$ produced by H and Fe primaries at three different energies. The simulations were produced using QGSJET-II-04. The shower cores hit the center of the instrument on the top scintillation layer.  The column on the left was calculated for H primaries, while the column on the right, for Fe nuclei. The top row contains events with energy $\log_{10}(E/\gev) = 5.0$, the middle row, with  $\log_{10}(E/\gev) = 5.5$ and the bottom row, with  $\log_{10}(E/\gev) = 6.0$. The bins have dimensions $1 \, \mbox{m} \times 1 \, \mbox{m}$.}
 \label{Mor3d}
\end{figure}

 One of the advantages of instrumenting MATHUSLA with an RPC layer is the possibility of studying the individual morphology of EAS in detail.  To illustrate this point, we present in Fig.~\ref{Mor3d} some MC examples that show the bi-dimensional signal distributions at the Big Pads produced by H and Fe vertical events with shower cores at the centre of the detector for three different primary energies: $\log_{10}(E/\gev) = 5.0$, $5.5$ and $6.0$. The observed patterns in each of these plots show noticeable differences depending on the energy and mass of the primary nuclei. For example, the shower core has a sharper edge, larger amplitudes, and fewer fluctuations at high energies than at low energies. From Fig.~ \ref{Mor3d}, we observe that for energies in the interval from $10^{5} \, \gev$ to $10^{6} \, \gev$, the proton-induced EAS are more compact and less clumpy than the showers produced by iron nuclei. The  distinctive morphological features of the EAS measured with the RPC layer are potential observables for distinguishing the nature of the incident cosmic-ray particle and its primary energy. 

 Another advantage of the RPC layer is that it clearly shows the position of the EAS core at very high energies. For showers produced by iron nuclei of low energies (smaller than a few $10^{5} \, \gev$), the EAS core is more difficult to locate from the signal patterns at the RPC because these types of EAS fronts are clumpy, flatter (as we will see later) and are subject to strong fluctuations. This is the reason for the rapid increase of the systematic uncertainty on the shower core position, which we have observed in Fig.~\ref{biasCoreRPC} for Fe-induced events in the low-energy regime.
 
 Further details about the physical differences of the EAS fronts can be obtained with the RPC layer from the corresponding measurements of the lateral density profiles, which are presented in the next subsection.

 \subsubsection{Lateral distribution of EAS from the RPC data}
\label{2dLDF} 

\begin{figure}[ht!]
 \centering
  \includegraphics[width=2.5 in]{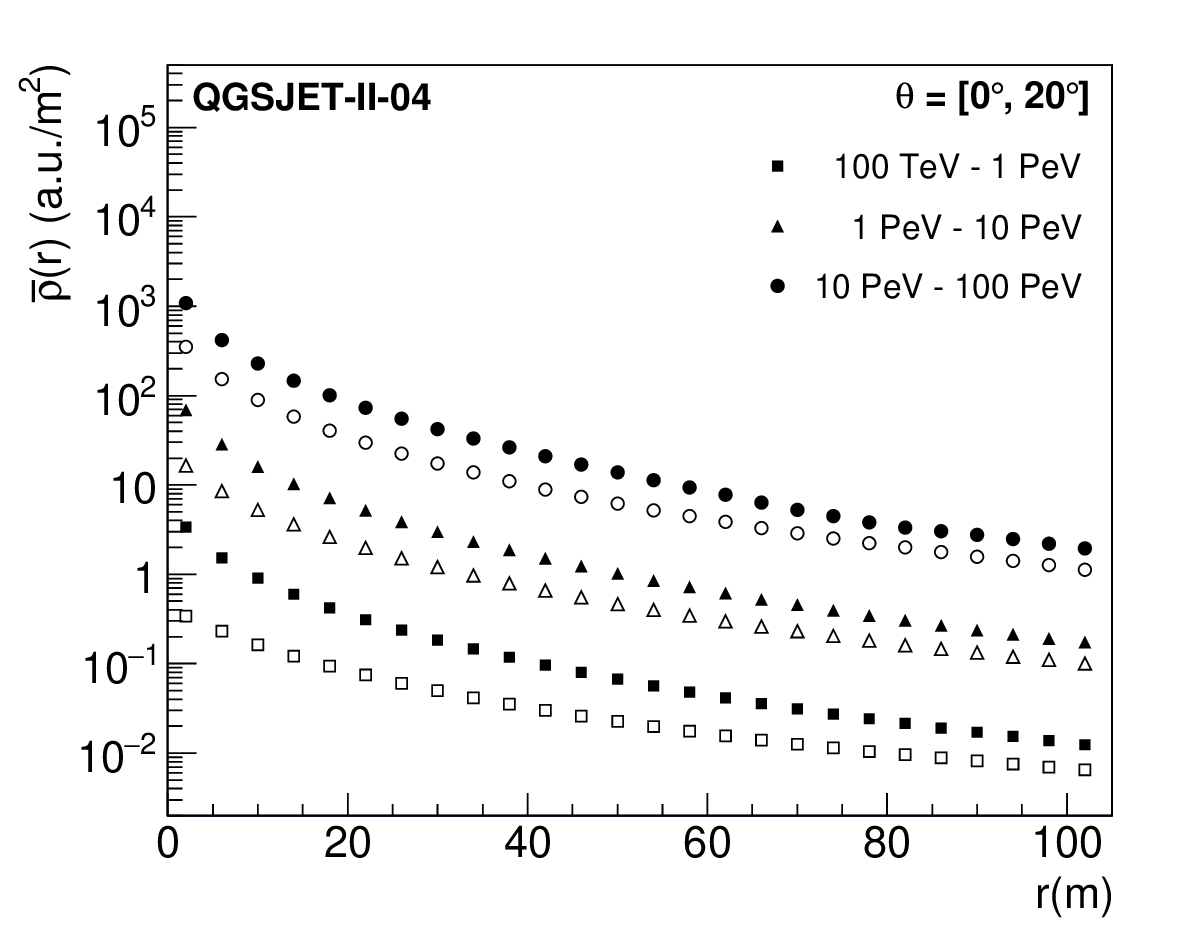}
  \includegraphics[width=2.5 in]{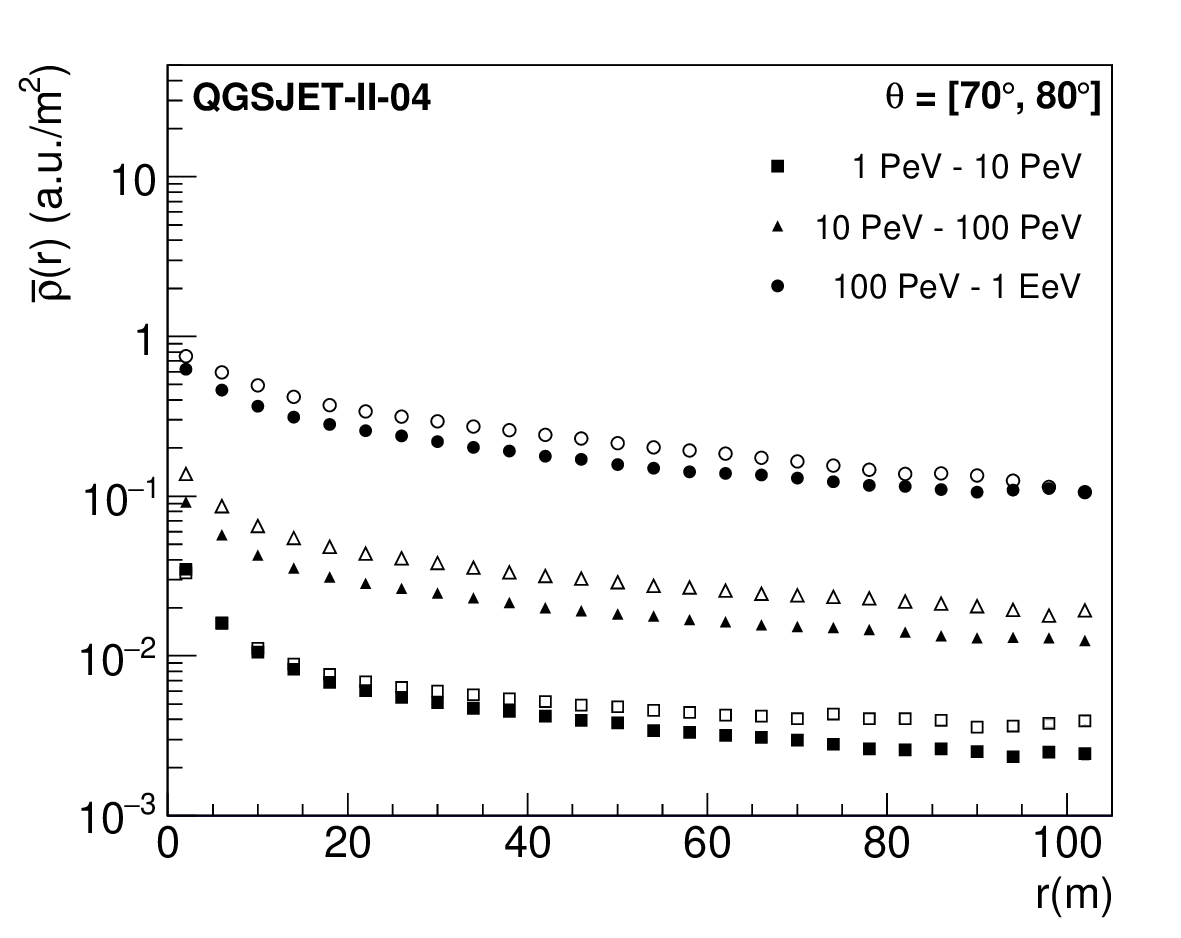}
  \caption{The mean lateral distribution for charged particles of EAS at high energies expected from our simulations of the MATHUSLA RPC detector. On the left, for events with zenith angles in the range $\theta = [0^\circ, 20^\circ]$ for three primary energy intervals: $E = [100 \, \tev, 1 \, \pev]$,  $[1 \, \pev, 10 \, \pev]$,  $[10 \, \pev, 100 \, \pev]$, the latter are represented by  squares,  triangles, and circles, respectively. On the right, for $\theta = [70^\circ, 80^\circ]$ and  $E = [1 \, \pev, 10 \, \pev]$,  $[10 \, \pev, 100 \, \pev]$ and  $[100 \, \pev, 1 \, \eev]$ . The energy intervals are represented by squares, triangles, and circles, respectively. Solid symbols show the results for EAS induced by protons, and hollowed points, for iron nuclei.} 
 \label{LDFmean}
\end{figure}

The radial density profile of EAS as measured with the RPC Big Pads of MATHUSLA could be a useful observable for studying primary particles on an event-by-event basis. In Fig.~\ref{LDFmean}, we show the mean radial density profiles, $\bar{\rho}(r)$, calculated from our RPC simulations for proton and iron primaries at different energy ranges and for both vertical and inclined events. The lateral distributions were estimated at shower disk coordinates. For the calculation,  we divided the shower front plane in concentric disks of width $\Delta r = 4 \, \mbox{m}$ from  $r = 0\, \mbox{m}$ to $200\, \mbox{m}$. For each disk with radius $r_i$, we estimated the average, $\bar{\rho}(r_i)$, of the density measurements of the Big Pads, whose centers were located inside the bin $r_i$. The local density at each Big Pad was estimated by dividing the respective signal by the projected area of the module $A_{BP}(\theta) = 1 \, \mbox{m}^2 \times \cos(\theta)$, where $\theta$ is the zenith angle of the shower axis.

It is clear from the plots in Fig.~\ref{LDFmean} that there are differences among the expected radial density profiles of EAS at the RPC layer depending on the energy, nature and arrival direction of the incident cosmic ray. For example, we observe that the amplitude of the distributions increases with the cosmic-ray energy but decreases with the zenith angle of observation.  In addition, the slopes of the density profiles are flatter for low-energy showers than for high-energy showers, as well as for vertical events produced by heavy cosmic-ray nuclei in comparison with vertical EAS due to light mass primaries. From these results, we expect that the RPC data of the lateral density distributions of EAS will provide information on the primary energy and the composition of high-energy cosmic rays. Hence, we should have some shower observables associated to the radial density profiles that could be employed as cosmic-ray energy and composition estimators. We have looked for some of these observables and will present the results in the next subsections.

 \subsubsection{Potential estimators of the cosmic-ray energy from the RPC data}
 \label{energyestimators}

\begin{figure}[t!]
 \centering
  \includegraphics[width=2.5 in]{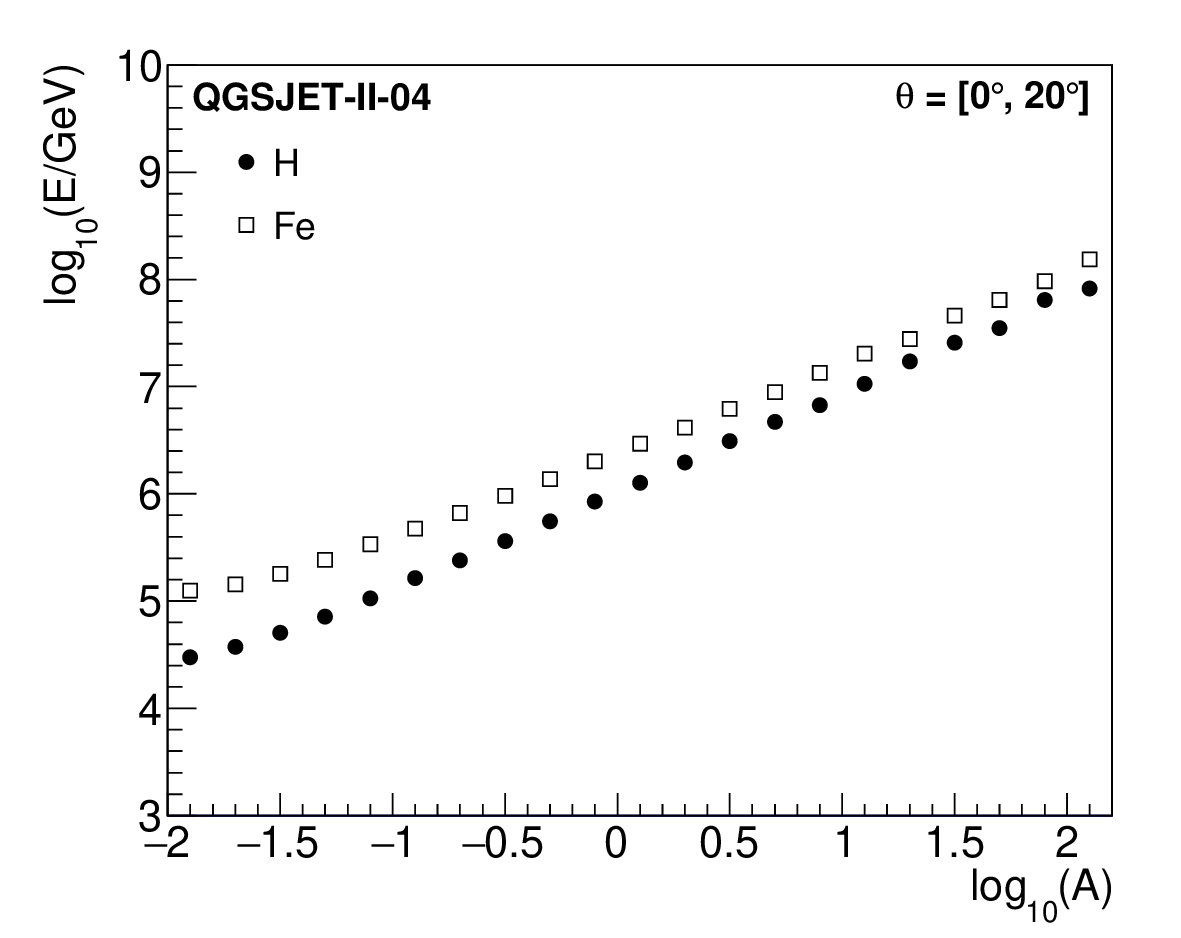}
  \includegraphics[width=2.5 in]{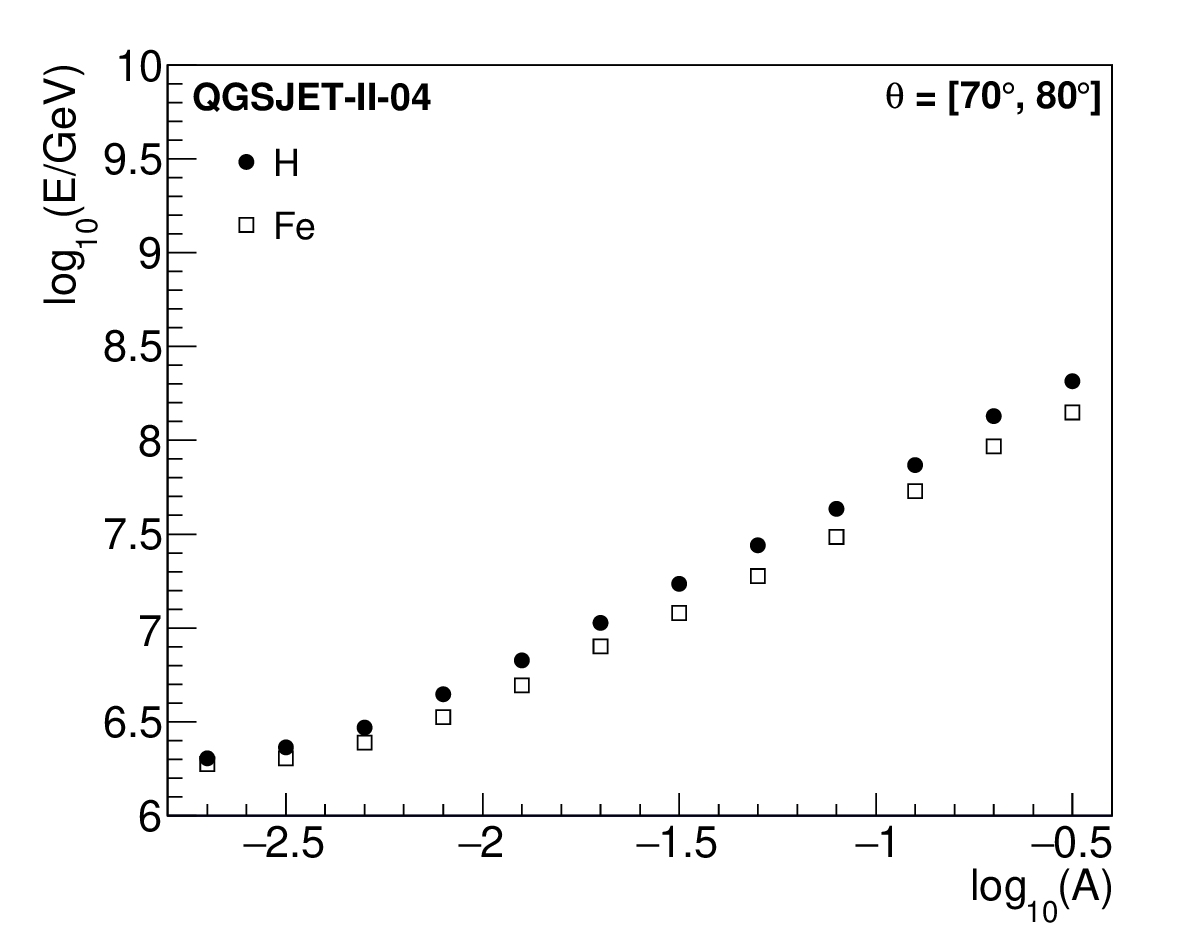}
  \caption{Mean primary energy as a function of the amplitude of the lateral distribution $A$ as defined by equation (\ref{eqNKG}) for H (circles) and Fe nuclei (open squares) simulated with QGSJET-II-04.  The plot on the left was built for EAS with $\theta \leq 20^\circ$, while the plot on the right was estimated for showers within the interval  $\theta = [70^\circ, 80^\circ]$. Vertical error bars represent the uncertainty of the mean. In general, the uncertainties are smaller than the size of the data points.}
 \label{logEvslogA}
\end{figure}

Using our MC simulations, we investigated the amplitude $A$ of the lateral shower density distributions and the value of the charge density at a fixed radial distance from the centre of the shower core as potential estimators of the primary energy of the EAS events. First, we calculated the dependence of the mean logarithm of the primary energy $\log_{10}(E/\gev)$  on $\log_{10}(A)$ for H and Fe events and vertical and inclined EAS. The results are presented in Fig.~\ref{logEvslogA}. We used MC events with $f_{hit} > 0.03$ for vertical data and $f_{hit} > 0.003$ in the case of inclined showers. This selection was chosen to reduce the influence of low-energy events in the results, which have large reconstruction uncertainties. 

  The plots from Fig.~\ref{logEvslogA} show an approximately linear relation between $\log_{10}(E/\gev)$ and  $\log_{10}(A)$ for both proton and iron primaries in the energy region of maximum reconstruction efficiency, which indicates that for vertical and inclined  air shower events, the amplitude $A$ can be used to estimate their primary energy. It is important to point out that, according to  Fig.~\ref{logEvslogA}, left, a mass dependence of the energy calibration function based on the $A$ observable for vertical events is expected to introduce systematic effects when reconstructing the energy spectrum of cosmic rays with this energy scale. Such dependence, however, decreases at large zenith angles, see Fig.~\ref{logEvslogA}, right. 

\begin{figure}[b!]
 \centering
  \includegraphics[width=2.5 in]{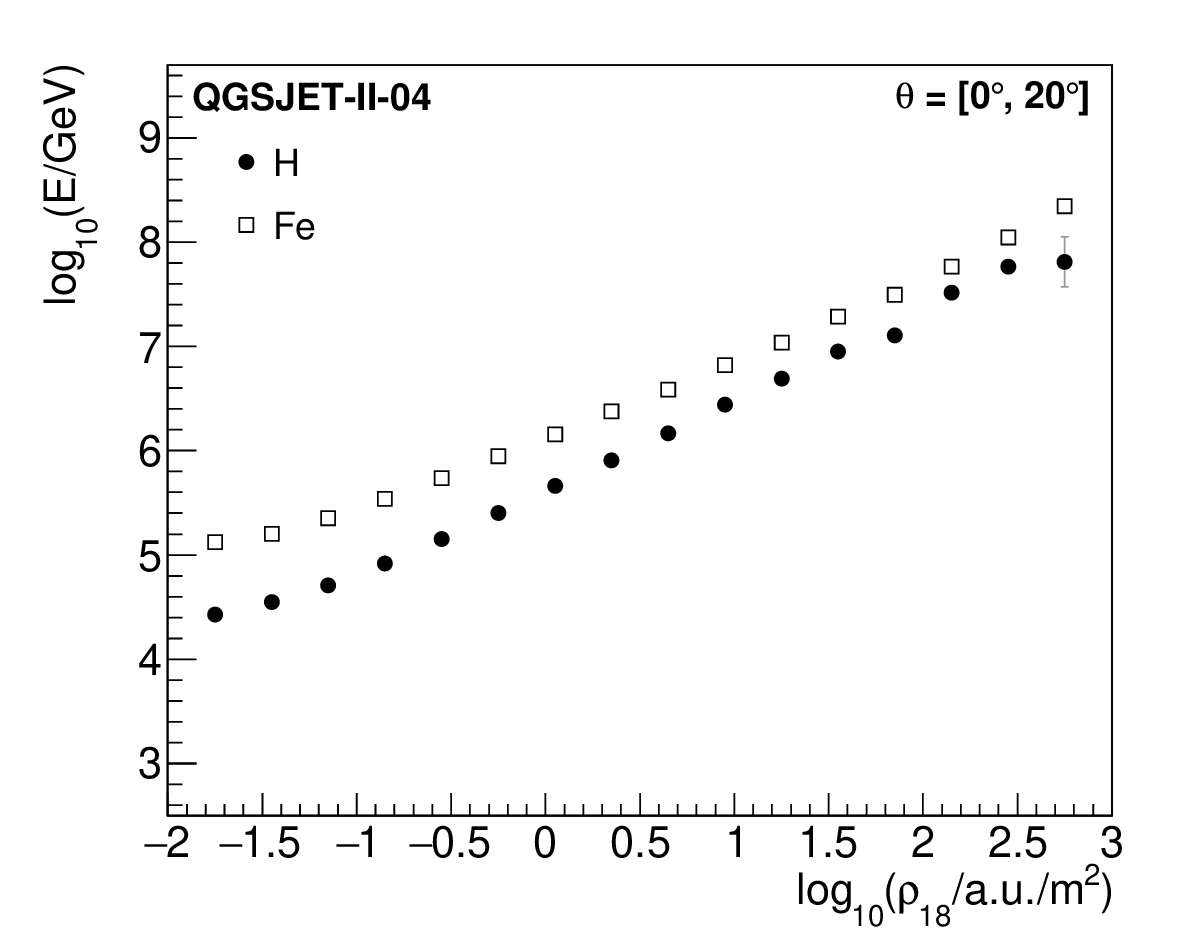}
  \includegraphics[width=2.5 in]{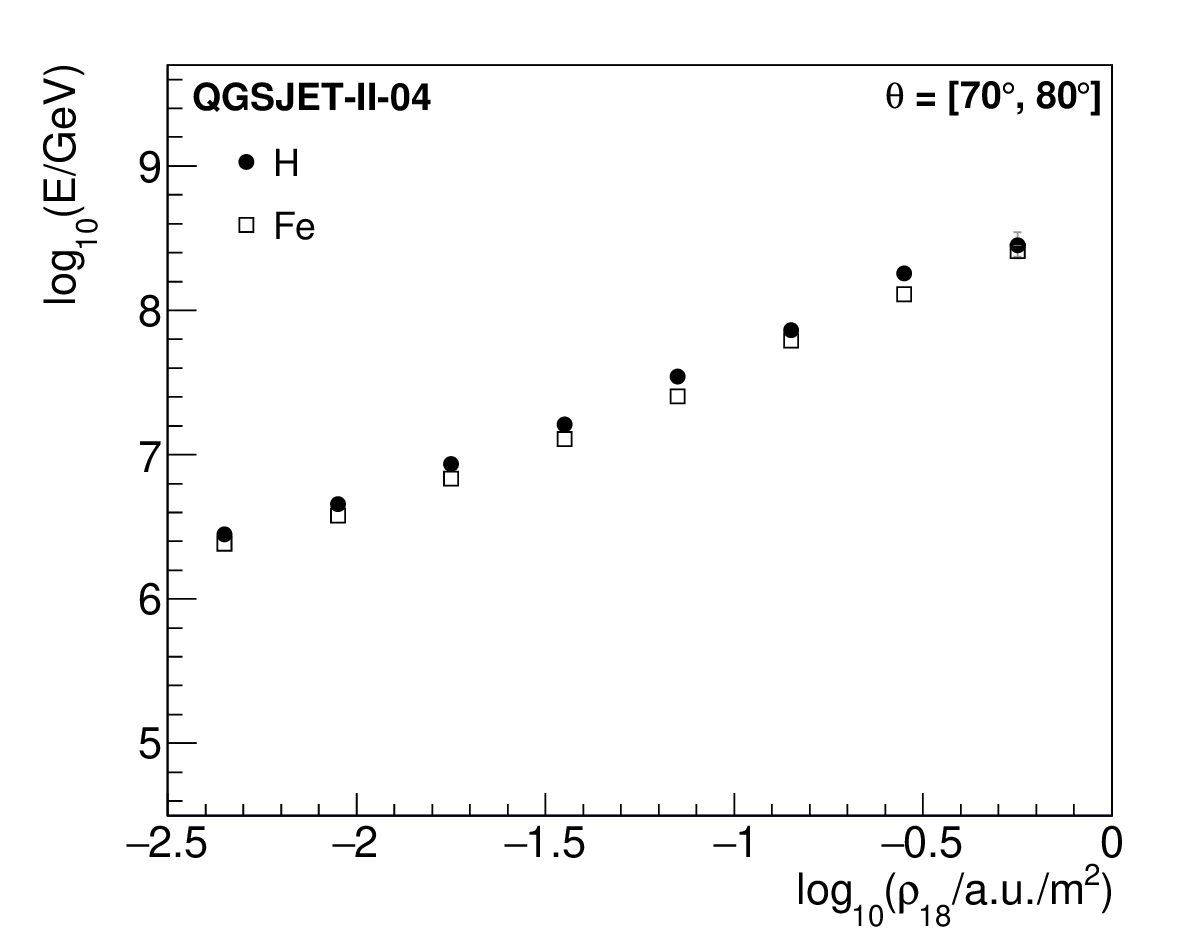}
  \caption{Primary energy  as a function of the measured lateral density at the radial bin centred at $18 \, \mbox{m}$ with the MATHUSLA RPC layer for EAS due to H (circles) and Fe (open squares) nuclei simulated with QGSJET-II-04 and after applying the quality cuts. The plot on the left was obtained for showers with $\theta \leq 20^\circ$, while the plot on the right was estimated for events within the interval  $\theta = [70^\circ, 80^\circ]$. Vertical error bars represent the standard uncertainty of the mean.}
 \label{logEvsrho}
\end{figure}

  As an alternative energy estimator, we also explored the value of the lateral density distribution of the event at a fixed radial distance from the core, in shower disk coordinates. In this case, as an example, we selected the mean density at the radial bin $r = 18\, \mbox{m}$, which we denoted as $\rho_{18} = \rho (r = 18 \, \mbox{m})$. The  mean logarithm of the  primary energy versus $\log_{10}(\rho_{18})$, as computed with MC data, is shown in the plots of Fig.~\ref{logEvsrho} for protons and iron nuclei and within the intervals, $\theta = [0^\circ, 20^\circ]$ and $\theta = [70^\circ, 80^\circ]$. We have applied again the cuts $f_{hit} > 0.03$ and $f_{hit} > 0.003$ for vertical and inclined EAS, respectively.
 In both cases, and independently of the primary mass composition,  we observe from the plots in Fig.~\ref{logEvsrho} that the curves for the average $\log_{10}(E/\gev)$ vs $\log_{10}(\rho_{18})$ dependence is approximately linear in the energy region of maximum reconstruction efficiency, which implies that $\rho_{18}$ could be also employed as a primary energy estimator. Nevertheless, for vertical EAS, we observe a dependence on the mass of the primary nuclei. This dependence is also observed for values of $\rho (r)$ up to distances of at least $r = 48 \, \mbox{m}$ for $\theta < 20^\circ$.
 
\subsubsection{Sensitivity of the lateral shower age to the mass of the cosmic ray}
\label{ageandmass}

 We have previously noted, in section \ref{spatialstructure}, that the shape of the lateral distribution in MATHUSLA could be sensitive to the type of the primary particle. One of the parameters of the lateral distribution of EAS that is sensitive to the primary composition is the lateral shower age, $s$, as defined, for example, in equation (\ref{eqNKG}). This parameter measures the flatness of the lateral distribution of the EAS and it is related to the depth of the maximum development of the air shower in the atmosphere, $X_{max}$, \cite{Apel2006,Bartoli2017a}. It is generally known that, at a given energy, light nuclei create air showers with steeper lateral distributions and larger values of $X_{max}$ than heavy primaries. Based on such differences, we say that the EAS associated with the first group are young and the second set, old. On average, the young EAS have relative $s$ values smaller than the old ones. Thereby the age parameter has the potential to be used in mass composition studies (see, for example, \cite{HAWCCR23}).  
  
 \begin{figure}[b!]
 \centering
  \includegraphics[width=2.5 in]{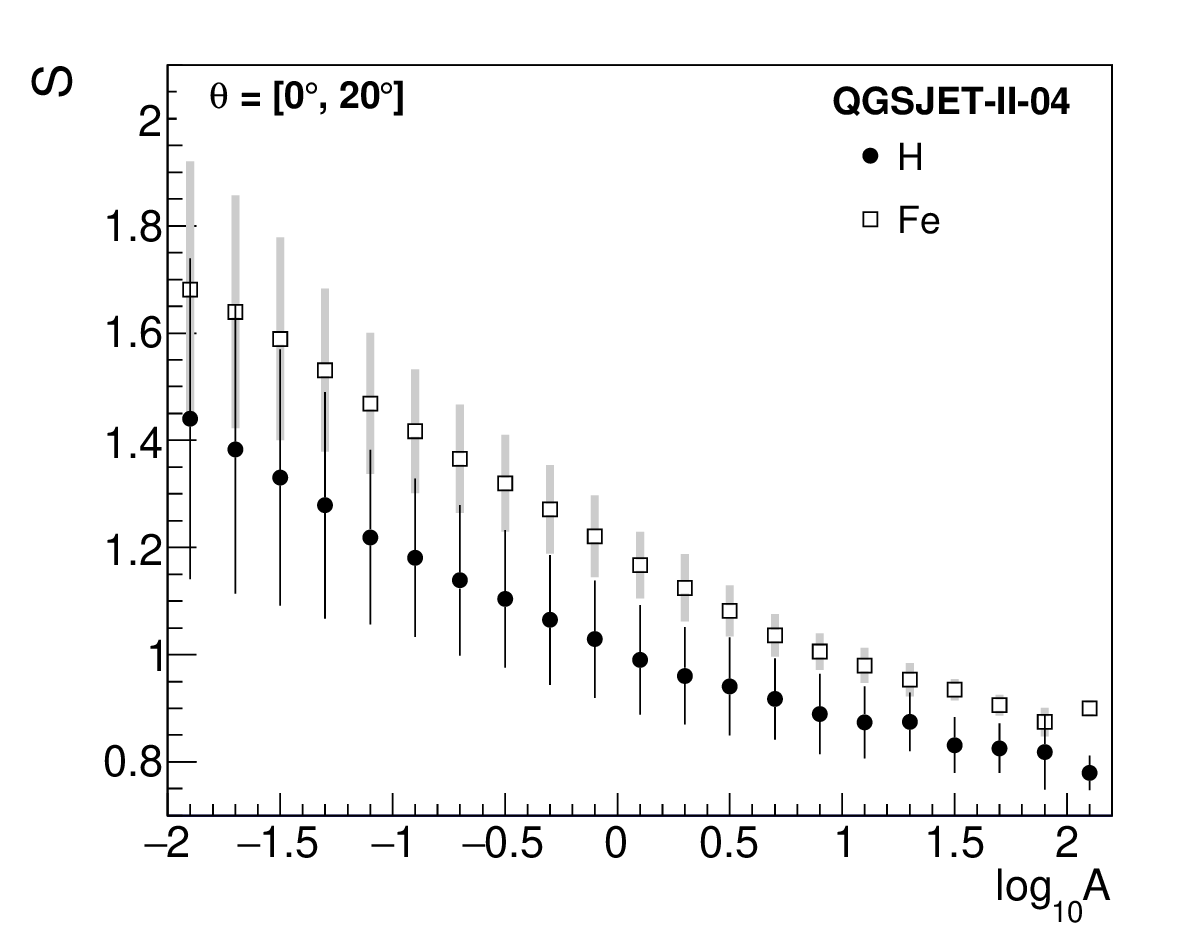}
  \includegraphics[width=2.5 in]{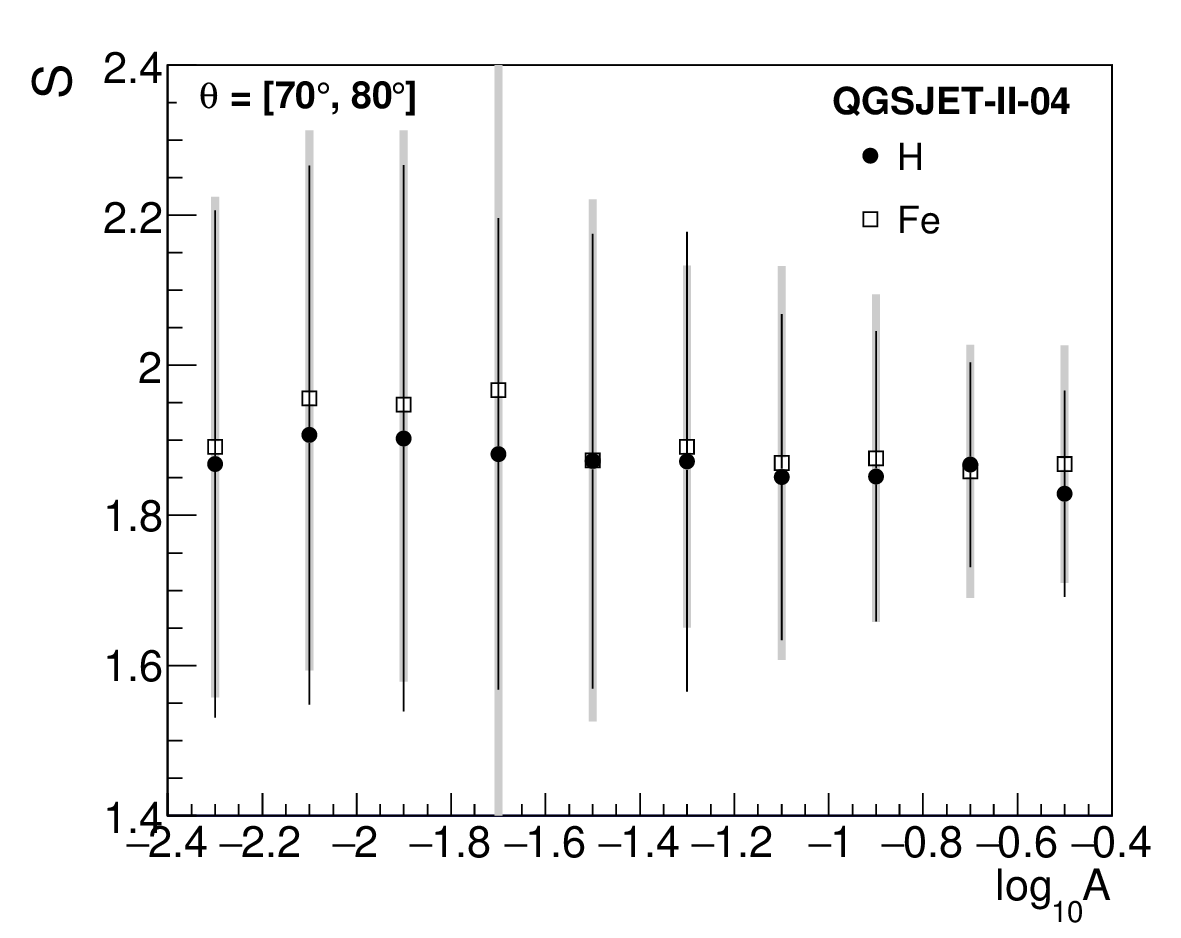}
  \caption{The mean lateral shower age, $s$, defined in eq. (\ref{eqNKG}), against the logarithm of the amplitude $A$ of the lateral density distribution of  EAS expected from measurements with the RPC of MATHUSLA. The graphs were estimated from MC simulations using QGSJET-II-04. The circles and open squares represent the results for protons and iron primaries, respectively. The plots on the left are valid for showers with $\theta \leq 20^\circ$, while  those on the right, for $\theta = [70^\circ, 80^\circ]$. Vertical uncertainties show the standard deviations of the measurements.}
 \label{AgevslogA}
\end{figure} 

To investigate the sensitivity of the $s$ parameter to the primary cosmic-ray nature in MATHUSLA, we plot in Fig.~\ref{AgevslogA} the mean age parameter versus the amplitude $A$ as calculated from our MC simulations for the RPC of the apparatus. The plots were computed with our MC simulations for vertical and inclined EAS using both H and Fe primaries. By comparing the graphs for both nuclei for vertical EAS (c.f.  Fig.~\ref{AgevslogA}, left), we found that there is a mass classification capability in the age parameter above some $100 \, \tev$ in MATHUSLA, as the separation of both curves is larger than one standard deviation in this regime. It is noteworthy that, for vertical events, the mass sensitivity of the age parameter is, in general,  enhanced at high energies, at least up to $10^{17} \, \ev$, because the uncertainty bars shrink.

 For inclined showers, the mass sensitivity of the age parameter in MATHUSLA decrease as seen from the right plot in Fig.~\ref{AgevslogA}, since EAS fluctuations are larger and the  mean shower age values for H and Fe are very similar. This means that for composition analysis of the primary nuclei with $\theta > 70^\circ$, alternative mass-sensitive parameters must be explored.

\subsection{Performance of the scintillating layers}
\label{sciperformance}

  The analyses shown in the previous sections show that the RPC detector layer proposed for MATHUSLA could be well suited for studying high-energy cosmic rays in the experiment. However, the capabilities of the instrument as a cosmic-ray detector could be enhanced by adding  data from the scintillating layers of MATHUSLA. Before exploring some of these possibilities, we need to investigate the performance of the scintillating detectors for EAS detection. For this aim, we carried out additional analyses,  which will be presented in the following subsections.

  \begin{figure}[b!]
 \centering
  \includegraphics[width=2.5 in]{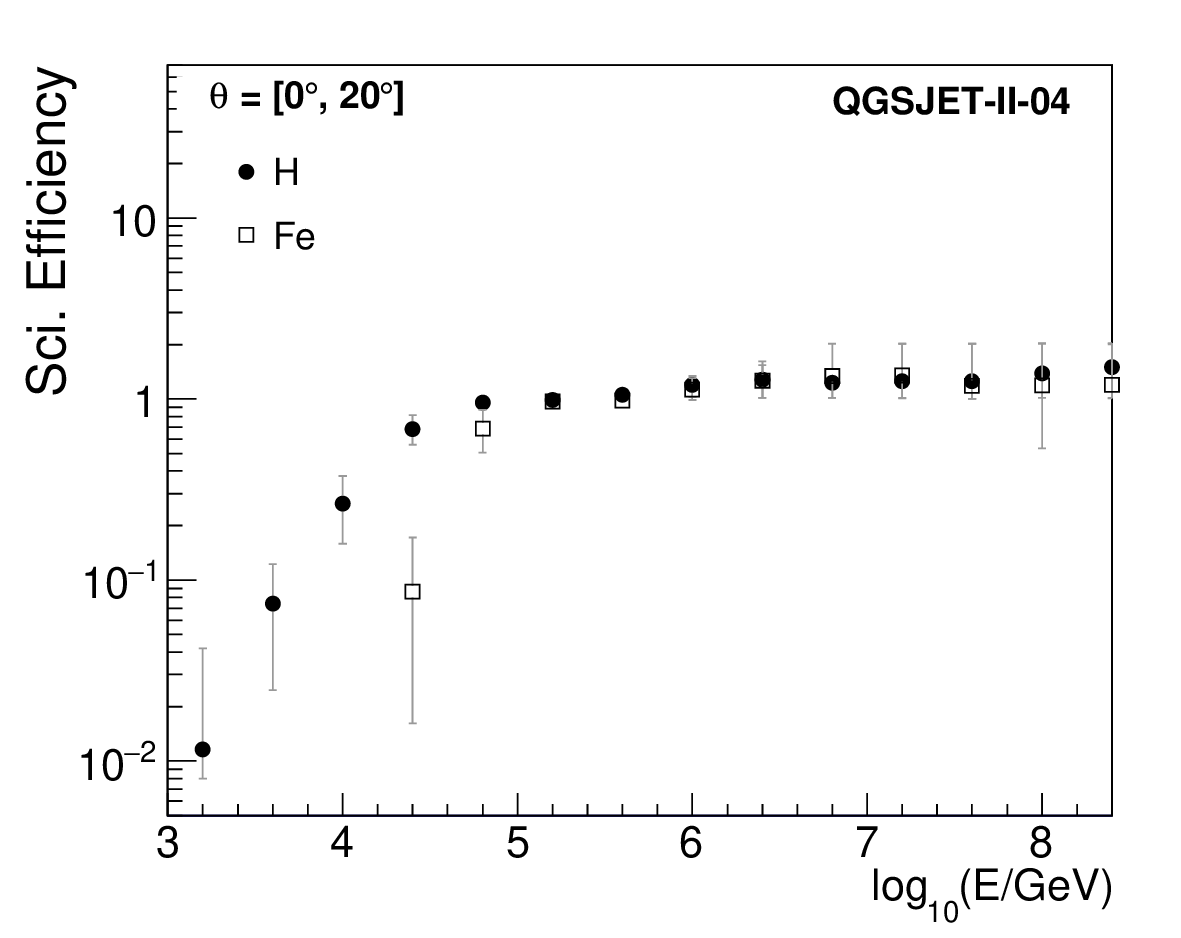}
  \includegraphics[width=2.5 in]{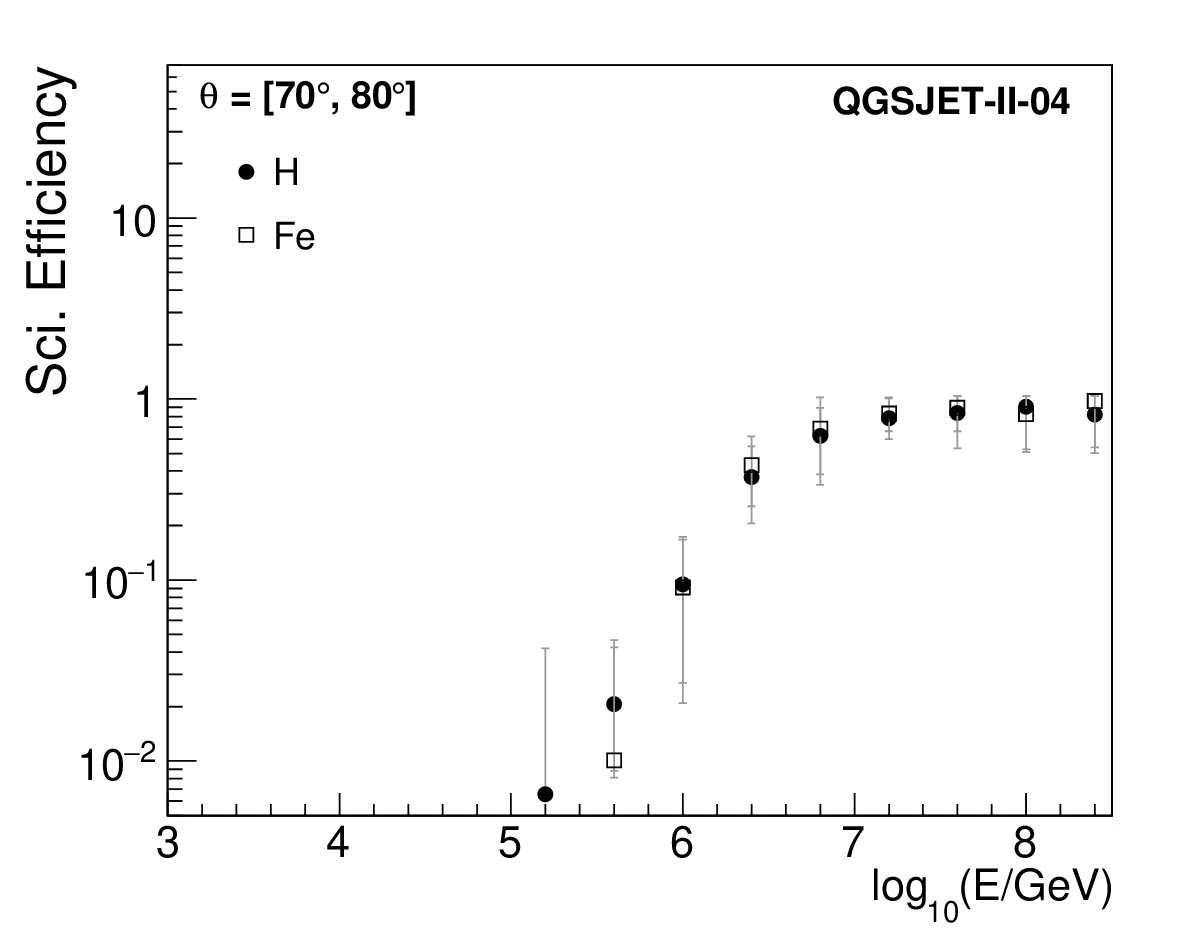}
   \caption{The selection cut efficiency of the top MATHUSLA scintillating layer, estimated using the QGSJET-II-04 model, for shower events induced by protons (black circles) and iron nuclei (open squares) Left: for vertical EAS with $\theta \leq 20^\circ$ and right: for inclined showers with  $\theta = [70^\circ, 80^\circ]$. The uncertainty bars show the statistical uncertainties. The upper and lower limits of the statistical uncertainties show the values of the $68.2 \, \%$ and $31.8 \, \%$ containment regions, respectively, of the efficiency distributions  at a given energy, which were estimated by dividing the MC data into sub-samples according to the energy interval and calculating the selection cut efficiency for each of them.}
 \label{Efficiency_sci}
\end{figure}

\subsubsection{Reconstruction efficiency}
\label{effsci}

   We start the analyses by investigating the energy range of sensitivity of the scintillating layers of MATHUSLA for cosmic-ray observations.  For this task, using MC simulations,  we calculated the selection cut efficiency of the scintillating planes as a function of the primary energy. Plots for vertical and inclined events at the top scintillating layer are shown in Fig.~\ref{Efficiency_sci} to illustrate the results.  By comparing these graphs with those in  Fig.~\ref{Efficiency}, we observe that the energy thresholds for maximum reconstruction efficiency of EAS at the top scintillating layer are similar to those for the RPC detector plane. In particular, for vertical EAS, we found that the energy threshold of the scintillating detector layer is close to $10^{14} \, \ev$, while for inclined events, it is approximately $10^{16} \, \ev$.  Similar conclusions apply to the  other MATHUSLA scintillating detector layers since their  efficiency plots are similar to the ones presented in Fig.~\ref{Efficiency_sci}. 

   \subsubsection{Systematic uncertainties in EAS reconstruction}
\label{systematicuncertainty_sci}
 
\begin{figure}[t!]
 \centering
  \includegraphics[width=3.0 in]{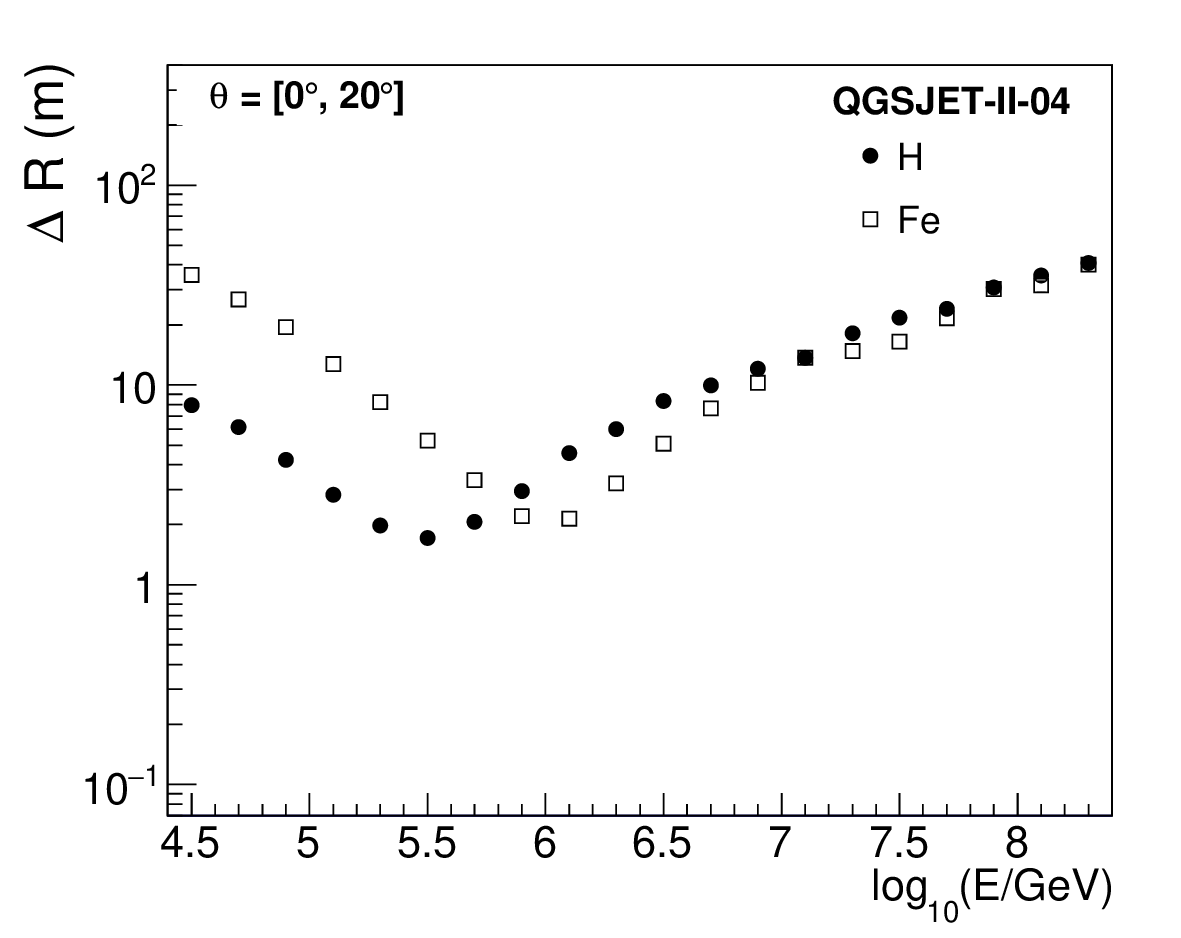}
  \caption{The bias  on the shower core position of vertical EAS with $\theta \leq 20^\circ$ expected for the MATHUSLA scintillating layers,  as estimated from a selected sample of MC events induced by H (circles) and Fe (open squares)  nuclei.  The simulations were produced with the QGSJET-II-04 model. Quality cuts were applied.  Vertical uncertainty bars represent the statistical uncertainties on the mean. They are smaller than the size of the data points.}
 \label{biasEAS}
\end{figure}

\begin{figure}[b!]
 \centering
  \includegraphics[width=2.5 in]{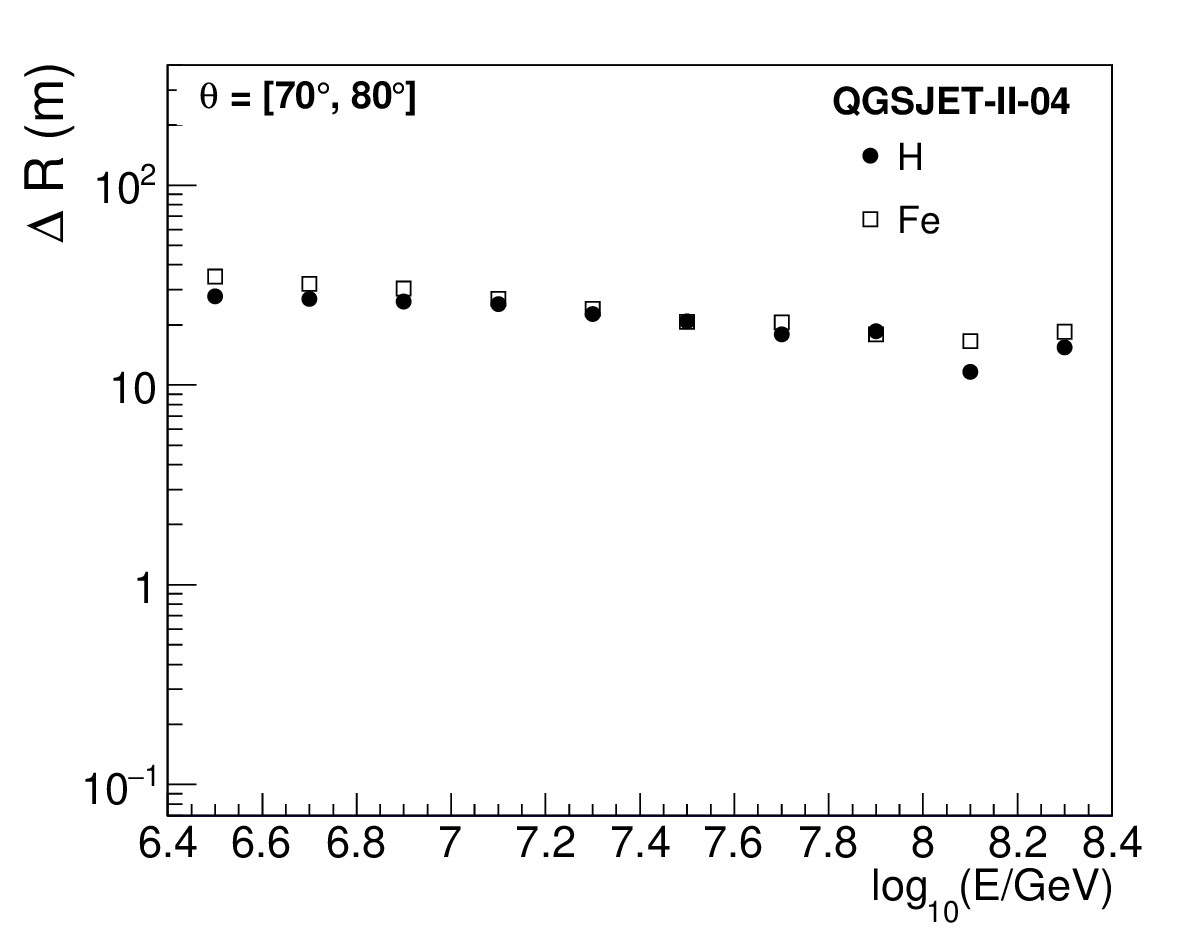}
  \includegraphics[width=2.5 in]{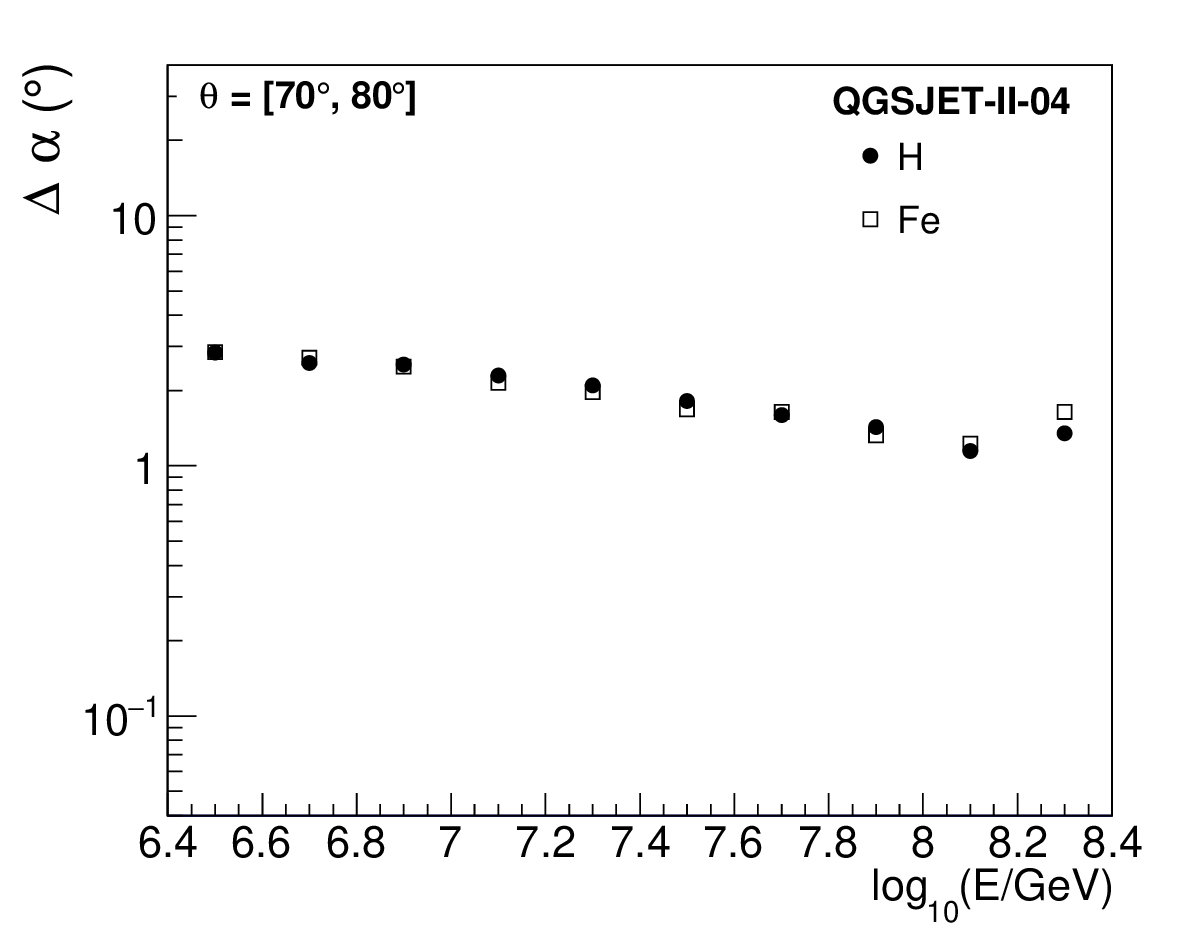}
  \caption{The bias on the EAS core position (left) and arrival direction (right) of inclined showers within the interval  $\theta = [70^\circ, 80^\circ]$ expected for the MATHUSLA scintillating layers. The plots were estimated from a selected sample of MC events using H (circles) and Fe (open squares)  nuclei as primaries. The simulations were produced with QGSJET-II-04 model. Quality cuts were applied.  Vertical uncertainty bars represent the statistical uncertainties on the mean. They are smaller than the size of the data points.}
 \label{biasCoreSci}
\end{figure}

  We continue the study of the performance of the MATHUSLA scintillating detector layers with an estimation of the  systematic uncertainties that could be expected for the reconstruction of EAS events at the scintillating planes using the procedures described in section \ref{EASreco}. To evaluate the systematic uncertainties, we have calculated the bias $\Delta R$ of the shower core position for vertical and inclined events and the bias $\Delta \alpha$ in the arrival direction of inclined showers.  We did not calculate $\Delta \alpha$ for vertical EAS due to saturation problems  at some scintillating bars as explained in section \ref{EASreco}.  In Fig.~\ref{biasEAS}, we present the shower core bias expected for the MATHUSLA scintillating layers vs the primary energy.  On the other hand, in  Fig.~\ref{biasCoreSci}, we show the corresponding bias on the arrival direction $\Delta \alpha$ and the core position $\Delta R$ for inclined EAS. For the calculation of the bias, we divided the primary energy range in bins of $\Delta \log_{10} (E/\gev) = 0.2$ and followed the procedure described in section \ref{systematicuncertainty_rpc}.

\begin{figure}[t!]
 \centering
  \includegraphics[width=2.5 in]{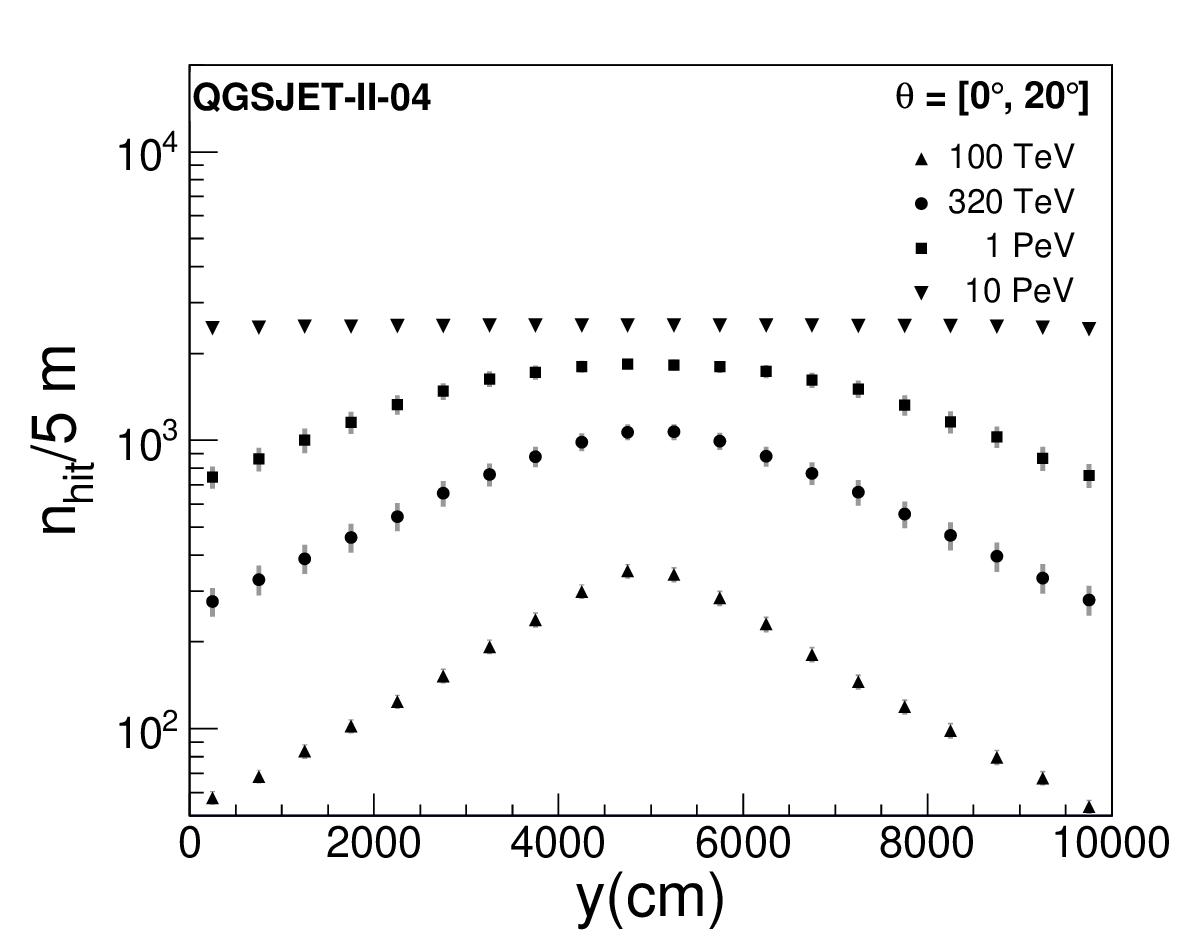}
  \includegraphics[width=2.5 in]{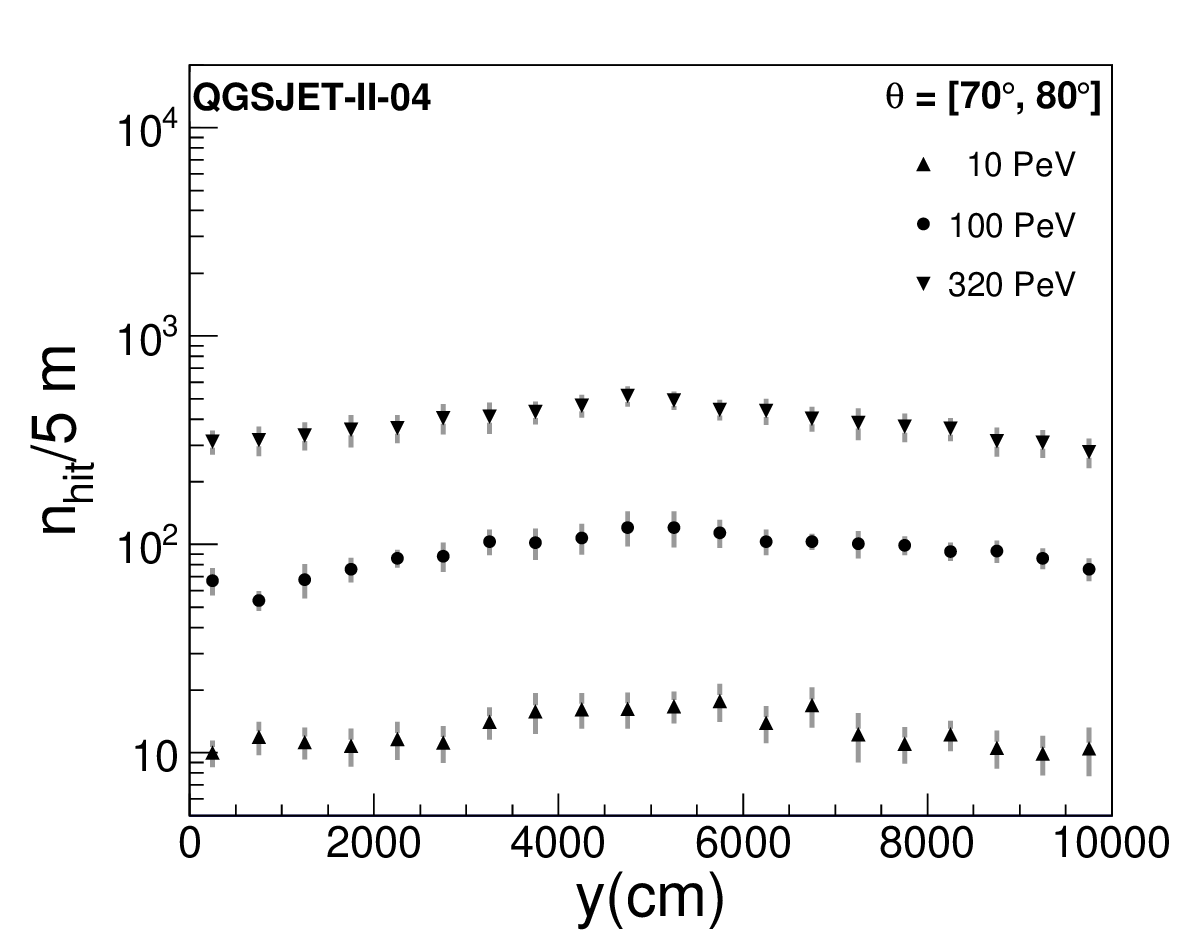}
  \caption{Mean projections on the $Y$ direction of the number of hit bars at the second scintillating detector layer of MATHUSLA (counted from the top of MATHUSLA) expected for proton induced EAS with different energies according to QGSJET-II-04. The shower cores are at the center of the top MATHUSLA layer. Left: Vertical events with energies of  $100 \, \tev$ (upward triangles), $320 \, \tev$ (circles), $1 \, \pev$ (squares) and  $10 \, \pev$ (downward triangles). Right: Inclined EAS with energies  $E = 10 \, \pev$ (upward triangles),  $100 \, \pev$ (circles) and  $320 \, \pev$ (downward triangles).  Projections were build for bins in $y$ of $5 \, \mbox{m}$. Vertical uncertainty bars represent uncertainties on the mean.}
 \label{ProfY_sci}
\end{figure}

In Fig.~\ref{biasEAS}, we note that for vertical events, the bias on the shower core position $\Delta R$ at the scintillating layers is less than $41  \, \mbox{m}$  for H and Fe primaries, respectively, in the region from $ 3.2 \times 10^{13} \, \ev$ to  $2.5 \times 10^{17} \, \ev$. From the same figure, we also find that the average $\Delta R$ decreases in the energy interval from  $E = 3.2 \times  10^{13} \, \ev$ to  $3.2 \times 10^{14} \, \ev$ ($1.3 \times 10^{15} \, \ev$) for proton and iron nuclei, respectively, and then increases at large energies to $41 \, \mbox{m}$. The reduction in $\Delta R$ at $\tev$ energies is due to an increment in the number of hit scintillating bars with the primary energy, while the increase appears  because at very high energies the EAS core covers most of the surface of the scintillating layers with a large number of tracks (see Fig.~\ref{ProfY_sci}, left). When all the scintillating layers are covered by tracks it is difficult to locate with sufficient precision the shower core with our reconstruction method (see subsection \ref{EASrecocore}). It is worth noting that, from Fig.~\ref{biasEAS}, left, the bias $\Delta R$ for EAS produced by iron nuclei is larger than for proton events at $\tev$ energies. This difference is related to the fact that for iron-induced EAS, the shower core is not as well defined as for proton-induced events with the same primary energy. 
 
   For showers with $\theta > 70^\circ$ (see Fig.~\ref{biasCoreSci}, left), saturation effects in the fraction of hit scintillating bars are no longer important for  $E = 3.2 \times 10^{15} \, \ev - 2.5 \times 10^{17} \, \ev$. However, the quality of the reconstruction of the EAS core diminishes mainly because the shower core is elongated due to the large inclination of the shower axis, and since the lateral particle density is flatter than for vertical EAS (see Fig.~\ref{ProfY_sci}, right). In addition, the shower fluctuations tend to bias the reconstruction of the EAS core. From  Fig.~\ref{biasCoreSci}, left, we observe that the bias of the EAS core of inclined events decreases with the primary energy. We also notice that at  $3.2 \times 10^{15} \, \ev$, the systematic uncertainty is equal to $28  \, \mbox{m}$ and $35  \, \mbox{m}$ for proton and iron events, and decreases to approximately $17  \, \mbox{m}$ at $2.5 \times 10^{17} \, \ev$.

   Finally, in Fig.~\ref{biasCoreSci}, right, we notice that the bias $\Delta \alpha$ in the arrival direction of inclined EAS decreases from $2.9^\circ$ at about $E =  3.2 \times 10^{15} \, \ev$ up to  $1.2^\circ$ at $2.5 \times 10^{17} \, \ev$ at the scintillating layers. In this case, the magnitude and energy dependence of  $\Delta \alpha$ is similar for inclined EAS produced by H and Fe primaries.

\subsection{Fraction of detector hits per layer}
\label{fhit}

\begin{figure}[t!]
 \centering
  \includegraphics[width=2.5 in]{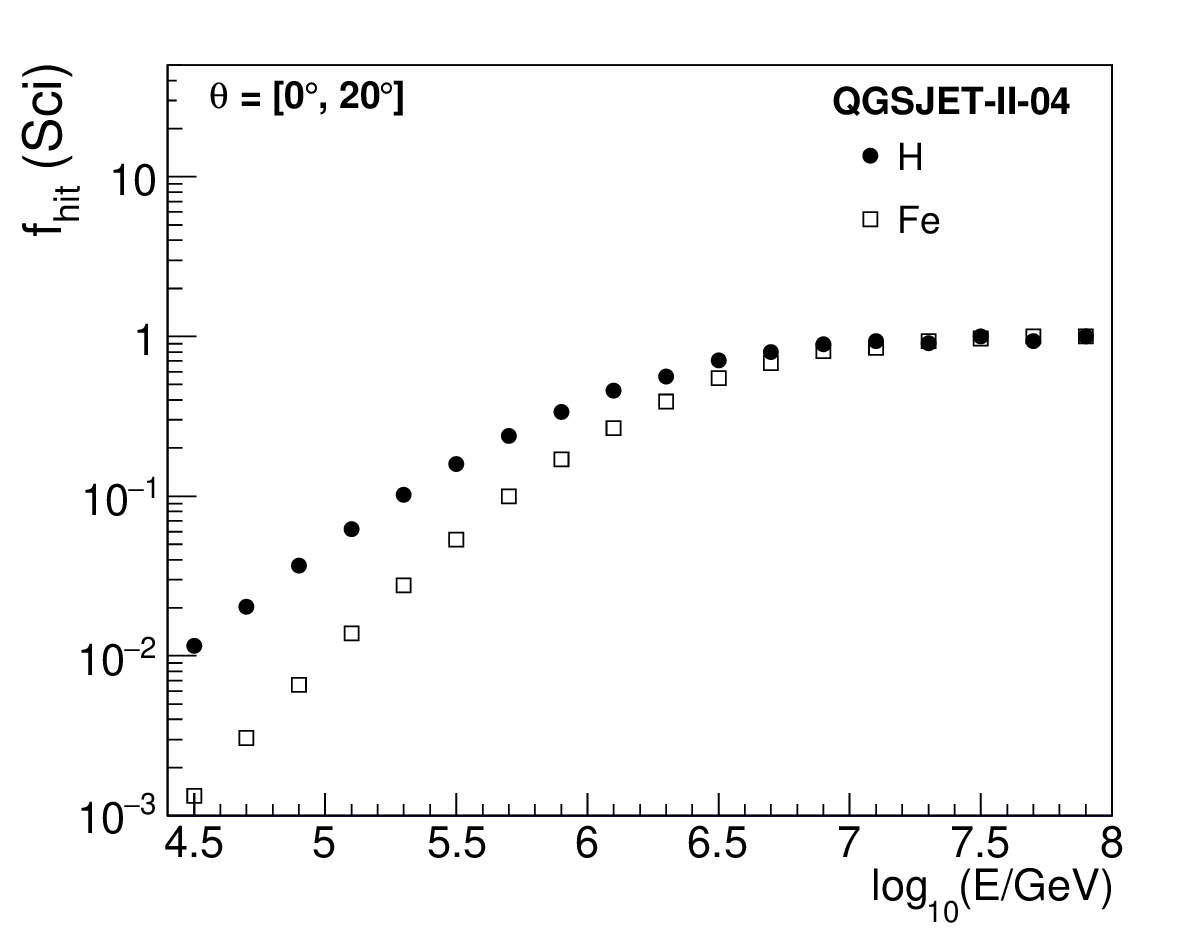}
  \includegraphics[width=2.5 in]{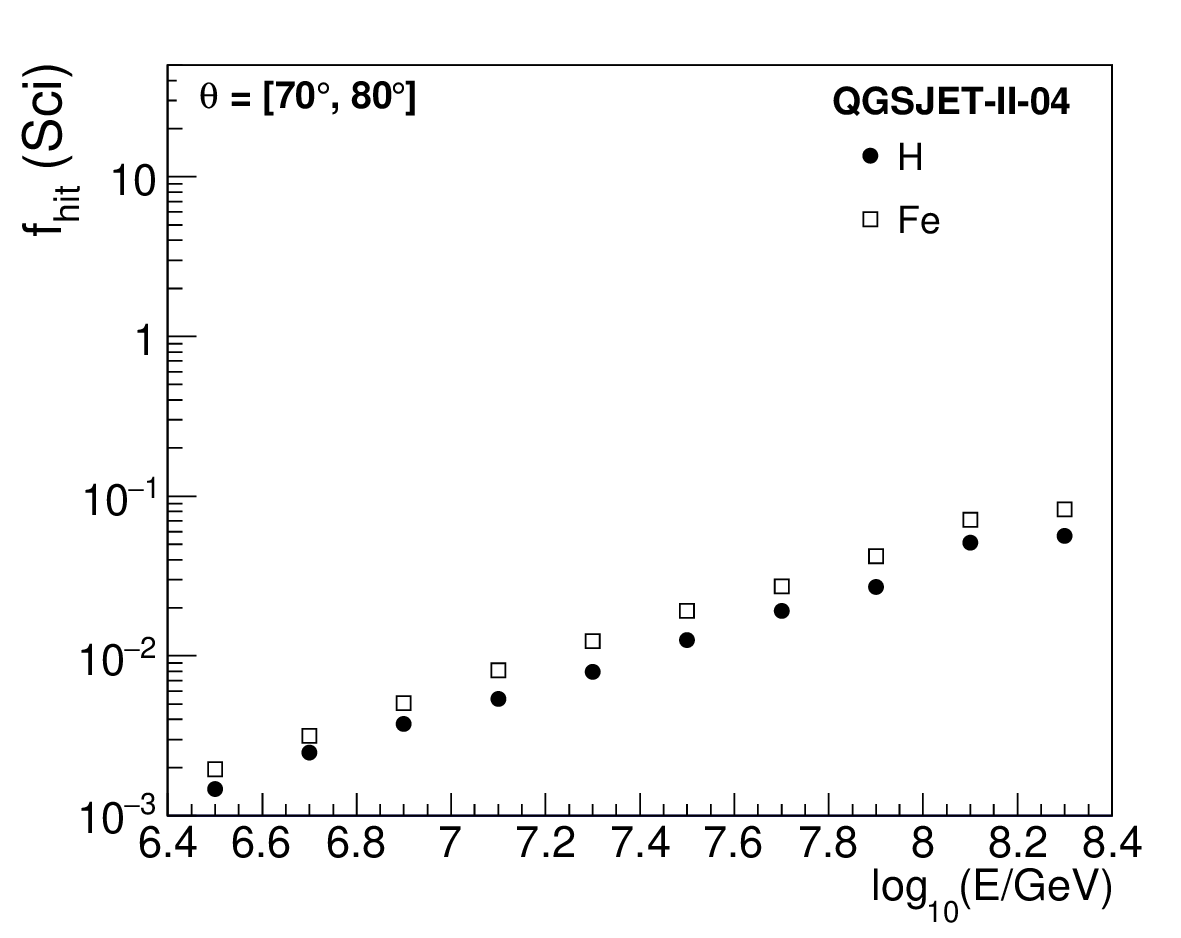}
  \caption{The mean fraction of hit scintillating bars at the MATHUSLA detector layers obtained from the MC simulations with QGSJET-II-04 for proton (circles) and iron (open squares) cosmic-ray primaries. The results are presented for two zenith angle intervals: left: $\theta = [0^\circ, 20^\circ]$ and right: $\theta = [70^\circ, 80^\circ]$. Standard uncertainties on the mean were estimated, but they are smaller than the size of the symbols.}
 \label{fhitsci}
\end{figure}

  To continue with the study of the performance of the scintillating layers of MATHUSLA for cosmic ray detection, we also estimated the mean fraction  $f_{hit}$ of scintillating bars activated in each MATHUSLA detector layer by the EAS at different primary energies using MC data and following the procedure described in section \ref{section_fhitRPC}. We used again energy bins of size  $\Delta \log_{10} (E/\gev) = 0.2$.  The results for $f_{hit}$  are presented in Fig.~\ref{fhitsci}. From here, we observe an energy dependence of $f_{hit}$ with the primary energy at the scintillating layers analogous to that observed at the RPC system (c.f. Fig.~\ref{fhitrpc}). However, for vertical EAS all the detector elements of the scintillating layers are activated at higher EAS energies ($4 \times 10^{16} \, \ev$ ) than for the RPC detector. For inclined EAS, as in the case of the RPC system, we found  that  $f_{hit}$ for the scintillating layers is not saturated below  $2.5 \times 10^{17} \, \ev$ and that $f_{hit}$ increases with the primary energy following a power-law with a small sensitivity to the primary mass.
  
\subsection{Saturation of the scintillation detector layers of MATHUSLA}
 \label{saturation_sci}

  Another key point in the performance of the 
  scintillating detectors of MATHUSLA studied here is the saturation of the individual scintillating bars. Saturation at each bar occurs, on average, for particle densities greater than $\rho_{th} \sim 5\, \mbox{charged part.}/\mbox{m}^2$. Therefore, if the particle density of the EAS exceeds this limit, we will lose information on the distribution of charged particles at the shower front at the scintillation detector layers of MATHUSLA.
 
\begin{figure}[t!]
 \centering
  \includegraphics[width=2.5 in]{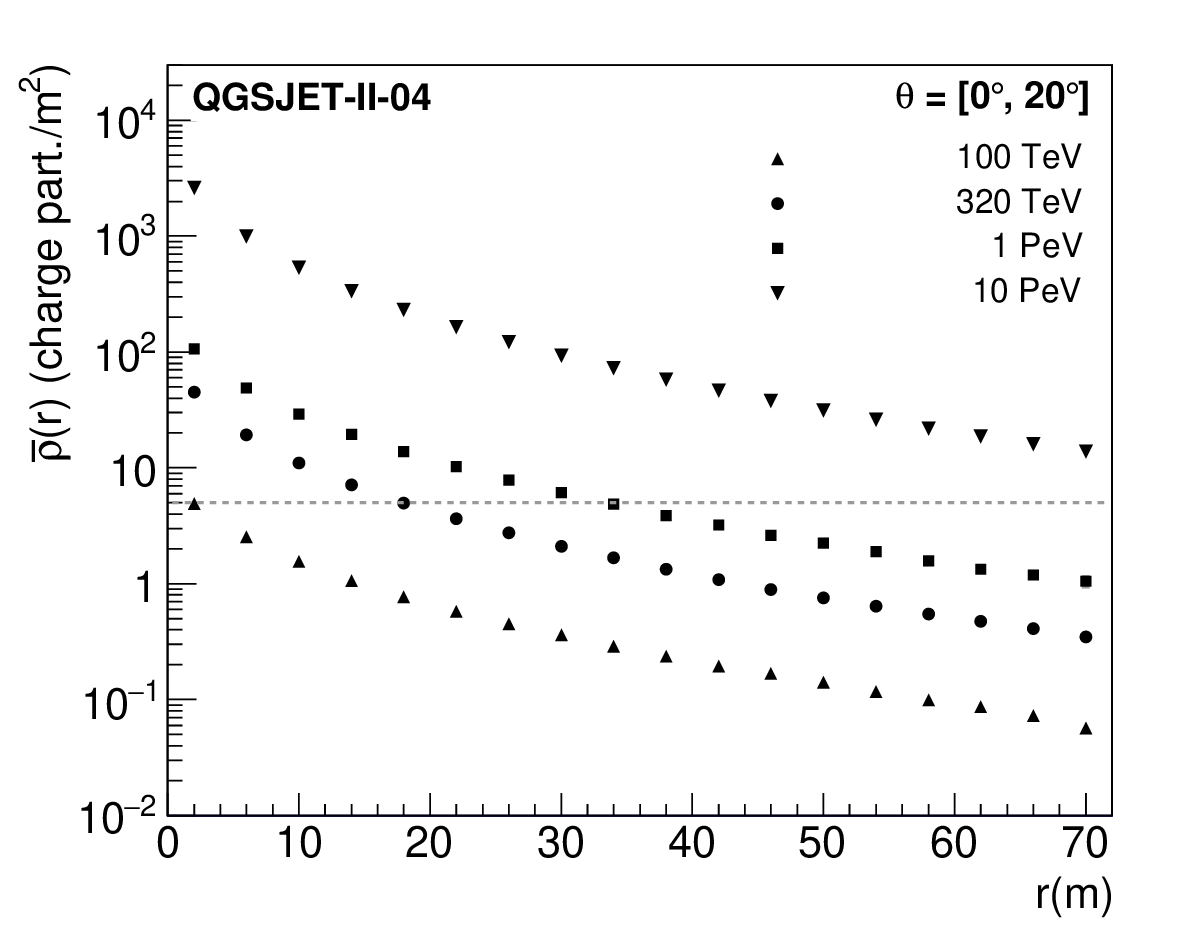}
  \includegraphics[width=2.5 in]{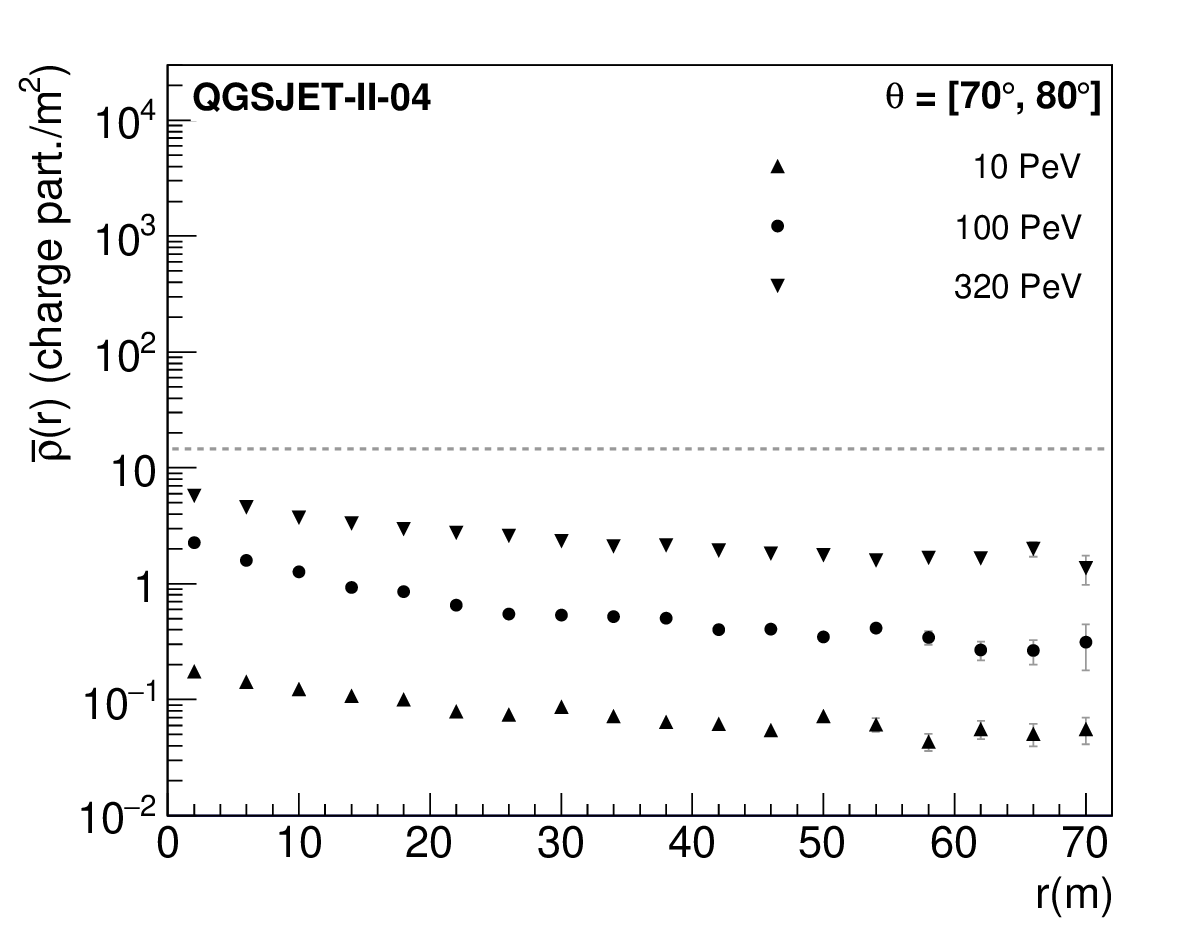}
  \caption{Mean radial densities of charged particles in the EAS front of events produced by protons of different energies. Left: Vertical events with energies of  $100 \, \tev$ (upward triangles), $320 \, \tev$ (circles), $1 \, \pev$ (squares) and  $10 \, \pev$ (downward triangles). Right: Inclined EAS with energies  $E = 10 \, \pev$ (upward triangles),  $100 \, \pev$ (circles) and  $320 \, \pev$ (downward triangles).  Distributions were calculated in shower disk coordinates. The densities were estimated for the MATHUSLA site using QGSJET-II-04. The dashed line shows the value of the average particle density for which the scintillating bars of MATHUSLA are saturated ($\sim 5\, \mbox{charged part.}/\mbox{m}^2$ for vertical events and $\sim 5\, \mbox{charged part.}/[\cos(70^\circ) \times \mbox{m}^2]$, for inclined showers). Vertical uncertainty bars represent statistical uncertainties  on the mean. Bins of $40 \, \mbox{m}$ in $r$ were employed to build the plots.}
 \label{density_sat_sci}
\end{figure}

  To investigate the EAS energy range that will be affected in MATHUSLA by the saturation of the scintillating bars, we have compared the average density threshold of this effect (after applying a geometrical correction factor $1/cos(\theta)$ due to the inclination of the shower axis) with  MC estimations for the lateral distributions of charged particles in air showers  (at shower disk coordinates) from primary protons of distinct energies. For vertical events, we used $E = 100 \, \tev$, $320 \, \tev$, $1 \, \pev$  and $10 \, \pev$, while for inclined EAS, we employed $E = 10 \, \pev$,   $100 \, \pev$ and $320 \, \pev$.  The results, presented in Fig.~\ref{density_sat_sci}, left, indicate that the saturation of the scintillation bars begins to affect the measurements of vertical EAS induced by protons at energies of $\sim 100 \, \tev$ near the shower core and that, at higher energies, the area affected by the saturation grows to cover the whole surface of the detector layers at about $10 \, \pev$. In Fig.~\ref{density_sat_sci}, right, the saturation of the scintillation bars is not dominant in MATHUSLA for inclined EAS with $\theta = [70^\circ, 80^\circ]$ up to primary energies of about $3.2 \times 10^{17} \, \ev$. 

\subsection{Muon bundles in MATHUSLA}
\subsubsection{Events from highly inclined EAS}
\label{muonbundlesInclined}

\begin{figure}[b!]
 \centering
  \includegraphics[width=3.0 in]{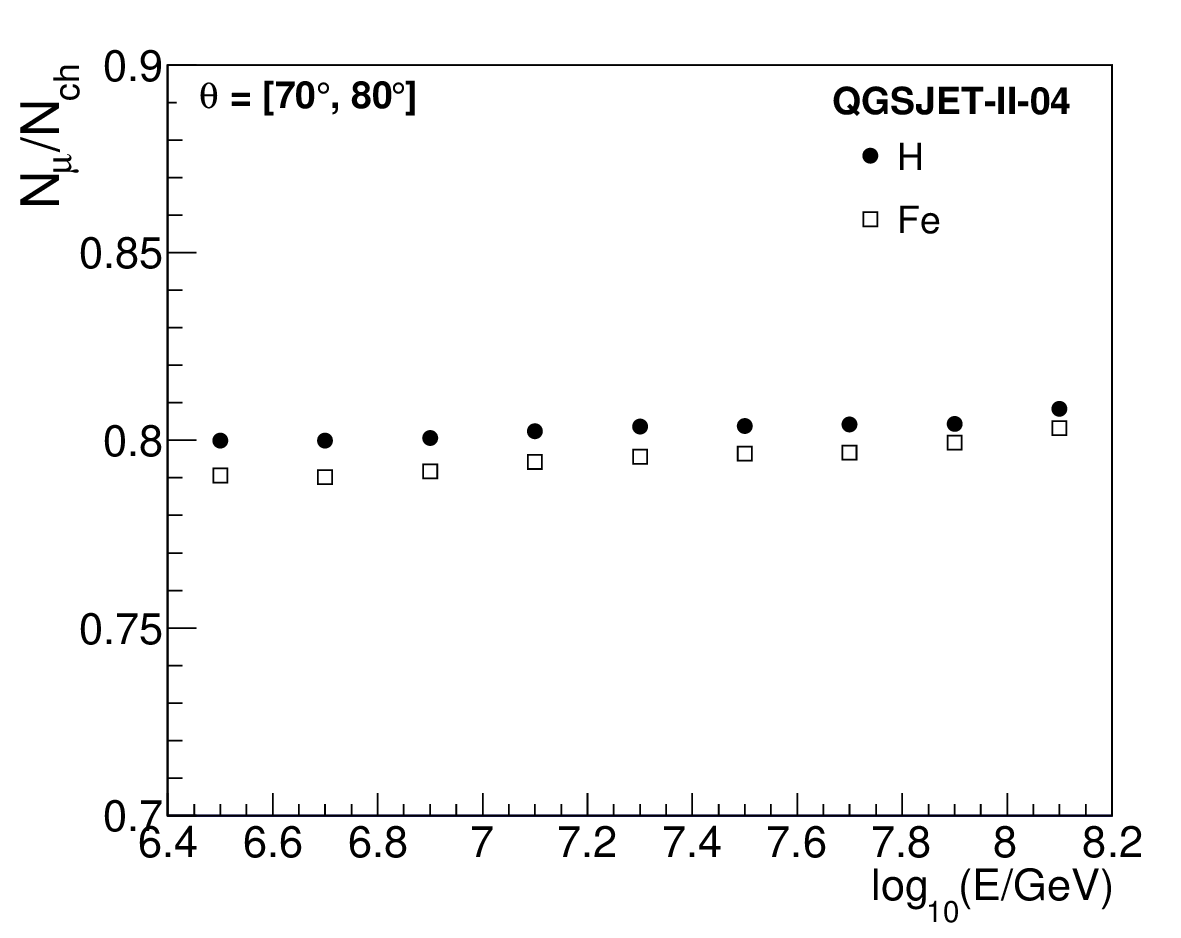}
 \caption{Predictions for the average ratio of muons to the total number of charged particles in hadronic EAS from inclined directions ($\theta = [70^\circ, 80^\circ]$) against the primary energy. The MC simulations were produced for H (circles) and Fe (open squares) primaries using QGSJET-II-04. Standard uncertainties on the mean were estimated and were displayed with vertical error bars, however, they can not be seen because they are smaller than the size of the symbols. }
 \label{muonfractionEAS}
\end{figure}

 At large zenith angles, the relative content of muons in EAS becomes increasingly larger due to the stronger absorption of electromagnetic and hadronic particles in the atmosphere. When the arrival direction of the events has zenith angles greater than $65^\circ$, muons become the most abundant component of EAS \cite{EASmodelsReview}. This phenomenon produces muon bundles \cite{Bogdanov20}, which are shower events dominated by multiple muon hits at the detectors. These types of showers are quite interesting to study because muons are sensitive to the physics of hadronic interactions in EAS, the primary energy, and the composition of primary cosmic rays \cite{EASmodelsReview}.  At the location of MATHUSLA, inclined EAS from primary nuclei with  $\theta$ between $70^\circ$ and $80^\circ$ have an average content of muons of about $(80 \pm 1)\%$ in the region of maximum reconstruction efficiency.  This can be seen in  Fig.~\ref{muonfractionEAS}, where we present QGSJET-II-04 predictions of the relative ratio of muons to charged particles in hadronic EAS for highly inclined events as a function of the primary energy. Therefore, local measurements with MATHUSLA of the total shower particle content of such events would be quite sensitive to the penetrating component of air showers.

\begin{figure}[t!]
 \centering
  \includegraphics[width=2.5 in]{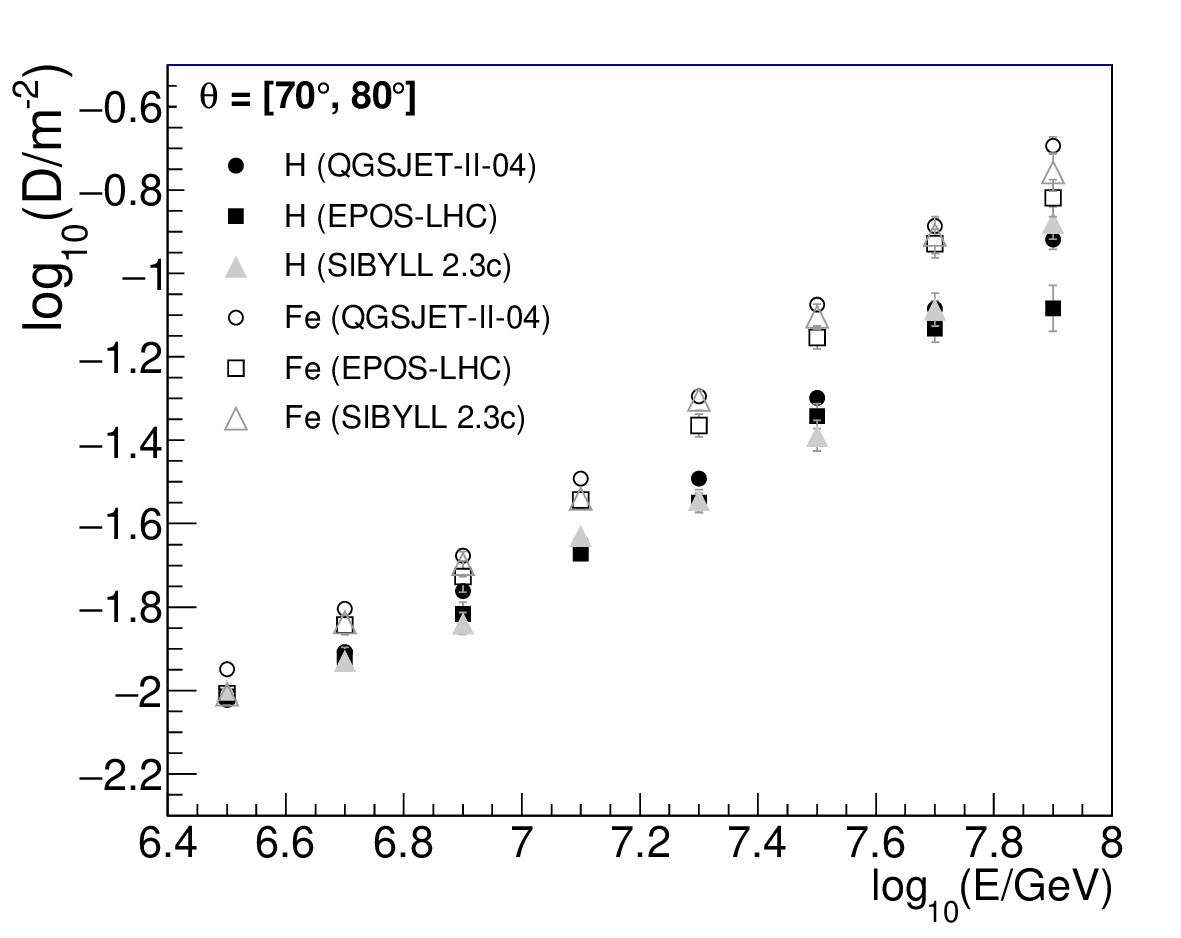}
  \includegraphics[width=2.5 in]{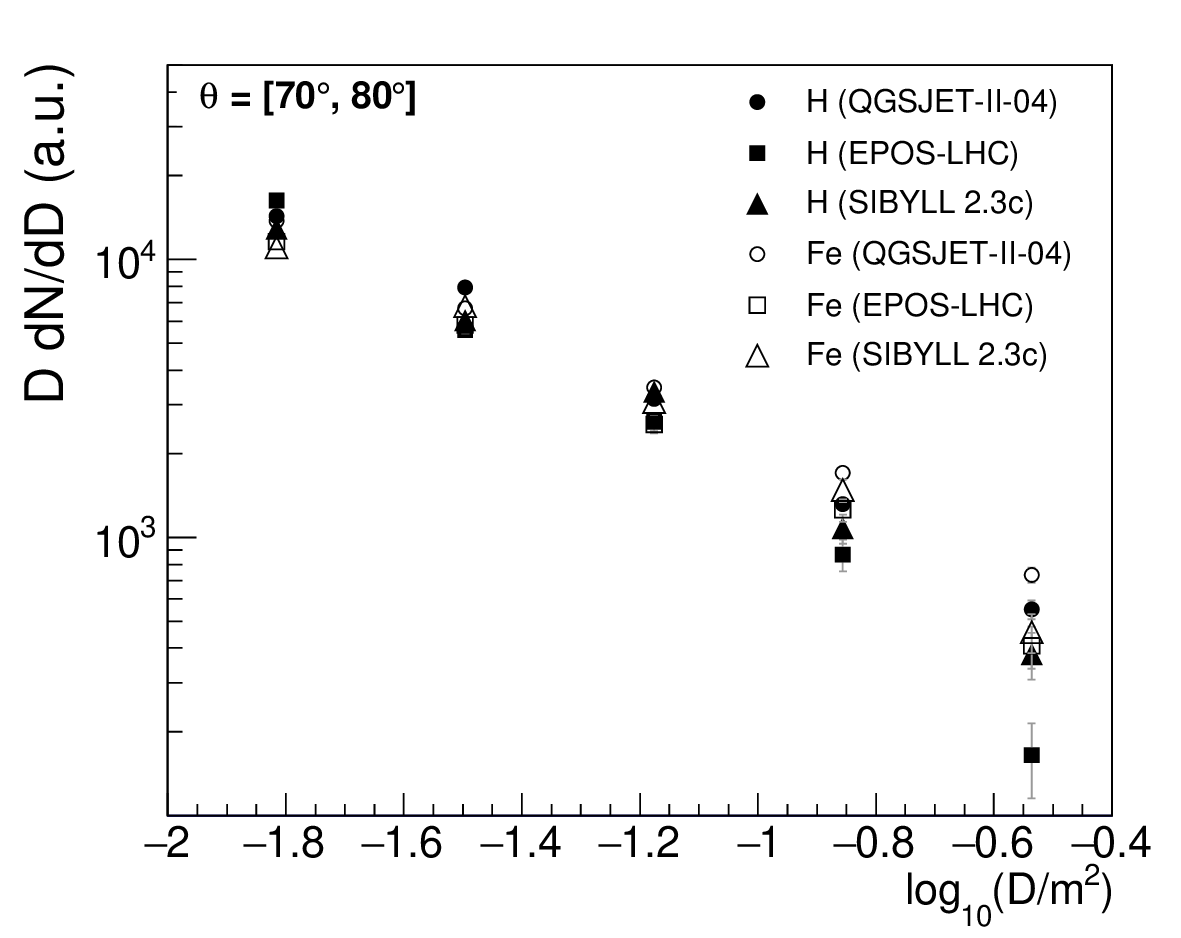}
  \caption{Left: The mean of the local muon density $D$ expected at the RPC of MATHUSLA for inclined events against the primary energy of the EAS according to QGSJET-II-04 (circles), EPOS-LHC (squares) and SIBYLL 2.3c (triangles) predictions for protons (filled symbols) and iron nuclei (hollowed symbols).  Vertical bars represent standard uncertainty of the mean. Right: Spectra of the local muon density $D$ of EAS predicted at the RPC of MATHUSLA and multiplied by a factor $D$. The curves were estimated with the same post-LHC high-energy hadronic interaction models and primary nuclei as in the plot on the left. The primary energy spectra of the models were weighted to have the same number of events and energy spectrum ($\propto E^{-2}$). Selection cuts were applied to the left and right plots.}
 \label{Muonbundle_1}
\end{figure}

To study the performance of MATHUSLA to these inclined events, we have estimated the expected spectrum and average value of the local muon density $D$ for inclined EAS at the RPC layer using the post-LHC models QGSJET-II-04, EPOS-LHC and SIBYLL 2.3c. The estimations were calculated as a function of the primary energy for H and Fe nuclei.   We define $D$ following  \cite{Bogdanov20}, as the ratio of the total number of particle hits per event to the area of the detector layer. For this study, we carried out a small production of MC events with EPOS-LHC and SIBYLL 2.3c consisting of $3.3 \times 10^4$  and $4.1 \times 10^4$ inclined  events, respectively, for each primary nucleus following the procedure described in section \ref{s.simulations}. We also weighted the primary energy spectra of these models to have the same number of events and overall spectral shape ($E^{-2}$) as QGSJET-II-04. 
 The model predictions for the average values and the spectra of $D$ are shown on the left and right plots of Fig.~\ref{Muonbundle_1}, respectively. We used bin widths of $\Delta \log_{10} (E/\gev) = 0.2$ and  $\Delta \log_{10} (D/\mbox{m}^2) = 0.32$ for the corresponding plots of the average value of $D$ and its spectrum. From Fig.~\ref{Muonbundle_1}, left, we observe that, at energies above a few $10^{16} \, \ev$, the local magnitude of $D$ increases  linearly with the primary energy in logarithmic scale, and it is larger for heavy primaries than for light cosmic-ray nuclei. The differences of $D$ for distinct primaries decrease at low energies due to the loss of  reconstruction efficiency. We note that the sensitivity of $D$ to the hadronic interaction model is not clear due to the small statistics of the MC data sets. However, it suggests that QGSJET-II-04 could produce inclined events with larger values of $D$ at the MATHUSLA RPCs. Whereas the plot on the right-hand-side of Fig.~\ref{Muonbundle_1} seems to indicate that the spectra of $D$ for EAS with a high content of muons are higher and harder than for events with a low multiplicity of muons according to the results with QGSJET-II-04, EPOS-LHC \cite{Pierog_2015} and SIBYLL 2.3c \cite{SIBYLL2.3c_2017}. These types of curves  could be used to test the predictions of high-energy hadronic interaction models with MATHUSLA RPCs  \cite{Bogdanov20}. In particular, if the hadronic  models predict fewer muons in EAS above $10^{16} \, \ev$ than the actual measurements as indicated by some experiments \cite{Whisp}, then we may expect that the predicted $D$ spectra, as estimated with a more realistic cosmic-ray composition model, e.g. \cite{Dembinski},  are smaller than the measured data. 

 Muon bundles induced by cosmic rays  at large zenith angles are expected to have low-particle densities at the MATHUSLA site, with average values smaller than the threshold for density saturation at the scintillating bars for primary energies below a few $10^{17} \, \ev$ (see Fig.~\ref{density_sat_sci}). Under these circumstances, MATHUSLA could be used not only to perform independent measurements of local muon densities $D$  but also 
 as a tracking detector to improve the pointing direction of the instrument. According to Fig. ~\ref{Maxhit}, fluctuations at the shower front are expected to saturate occasionally some scintillating bars. Despite that, such modules could be removed in a joint EAS analysis involving the scintillating tracking detector and the RPC detector layer.

 \subsubsection{Events in tandem with the CMS detector}
\label{muonbundlesCMS}

 As a stand-alone cosmic-ray detector, MATHUSLA will sample the front of hadronic EAS with its tracking system and the RPC detector layer. However, during the High-Luminosity LHC runs, measurements in combination with the underground detector CMS may also be carried out. This possibility is still under exploration with the CMS group. However, if  approved, the CMS detector would make possible EAS measurements of high-energy muons ($E_{\mu} \gtrsim 61 \, \gev$ for vertical incidence\footnote{Using the analytical formula for the range of muons in matter from \cite{LipariPRD44}, cosmic-ray muons of $61 \, \gev$ that propagate vertically in $90 \, \mbox{m}$ of standard rock will arrive to the top of the CMS cavern with $\sim 10 \, \gev$ of energy, which is enough to traverse the CMS detector \cite{CMS_cosmicmuons_2010}. The value for the muon energy threshold at ground level could be smaller since the material above the CMS cavern is not completely composed of rock \cite{CMS_cosmicmuons_2010b}.}) in  MATHUSLA at different lateral distances on the shower plane, ranging from $r = 0 \, \mbox{m}$ to $168 \, \mbox{m}$, depending of the zenith and azimuth of the shower axis and the impact point of the EAS core on the surface of MATHUSLA.  The data may be taken with independent triggers, then time correlations could be searched to find EAS events simultaneously recorded in both instruments.  With the CMS data on shower muons, MATHUSLA would have local, high-precision information about the multiplicity $N_{\mu}$, radial density $\rho_{\mu}$, charged ratio $\mu^{+}/\mu^{-}$, momentum $p_{\mu}$ and energy spectrum of muons for large threshold energies in the EAS front. 

For events observed in MATHUSLA and CMS, the most interesting ones would be those in which the shower cores traverse through the instrumented areas of both experiments. The muon data from these golden events would be collected directly from the forward region of the EAS, which is difficult to measure in dedicated particle physics experiments and is not well constrained by hadronic interaction models at high energies. In this case. The statistics are expected to be reduced due to the limited physical area of the CMS detector ($\sim 315 \, \mbox{m}^2$) and the reduced solid angle of observation for golden events. If we only consider the field of view subtended by the RPC detector layer of MATHUSLA from the CMS location, the zenith angle for such events is restricted from approximately $40^\circ$ to $64^\circ$ and the width of the azimuth interval to $\approx 71^\circ$ in the CMS reference system (using the geometry presented in figs.~\ref{fig:layout_v4} and \ref{fig:layout_P5}). Hence, we could expect an acceptance of around $77 \, \mbox{m}^2 \, \mbox{sr}$ in CMS for these events, which implies a reduction in statistics by a factor of approximately $2.5 \times 10^{-3}$ with respect to MATHUSLA. 

Simultaneous EAS measurements with MATHUSLA and CMS also  could help to investigate the connection between hadronic cosmic rays and the properties of muon bundles in underground experiments. These are air-shower events, which produce groups of penetrating muons with variable multiplicity in detectors buried underground under several meters of rock and soil. First measurements of muon bundles were reported by early deep underground experiments like LVD \cite{LVD_cosmicmuon_90}, Frejus \cite{Frejus_cosmicmuon_89} and MACRO \cite{MacroBundle}. Detailed studies of muon bundles with underground detectors began with the LEP experiments ALEPH \cite{aleph}, L3+C \cite{L3C_cosmicmuons_02} and DELPHI \cite{delphi}. The results from the LEP experiments showed that the measured rate of muon bundles with high multiplicity ($N_{\mu} \gtrsim 100$) from cosmic-ray showers was larger than the predictions of pre-LHC hadronic interaction models \cite{cosmoLep} like QGSJET01 \cite{qgsjet01}. In recent measurements at the LHC \cite{aliceref2}, the ALICE collaboration has shown that the frequency of muon bundle events with  $\theta < 50^\circ$ and multiplicities above $100$ are in better agreement with QGSJET-II-04 expectations if it is assumed that at primary energies  around $10^{17} \, \ev$  the composition of cosmic rays is dominated by heavy nuclei \cite{aliceref1}. Currently, there is no other experiment that confirms the ALICE result. Data from  MATHUSLA and CMS in tandem, if they become available, would provide an independent check of the ALICE observations.

\subsubsection{Searching for exotic shower events with MATHUSLA}
\label{exoticevents}

 ALICE results on muon bundles allow for the possibility that these underground events are due to exotic mechanisms in EAS  \cite{crCERN}. Some theoretical models propose that the high muon multiplicity events reported by ALICE are produced by strangelets, hypothetical nuclear matter composed of doublets or triplets of strange quarks \cite{Farhi84}, with possible astrophysical origin \cite{Kan2017}. In this context, the muon bundles detected by ALICE are thought to be created by a mixture of strange quark matter that loses mass in many successive interactions with air nuclei when penetrating the atmosphere. According to \cite{Maciej19}, strangelets could induce muon bundles in the atmosphere that could behave as EAS induced by atomic nuclei with an atomic mass larger than that of iron at a given energy. Hence, air shower events produced by strangelets would be characterised by smaller $X_{max}$ values and larger muon contents than hadronic air showers of the same primary energy. By exploiting the differences between exotic and hadronic EAS, it could be possible to perform an event-by-event search for strangelet-induced events in the MATHUSLA cosmic-ray data. 

\begin{figure}[t!]
 \centering
  \includegraphics[width=5.0 in]{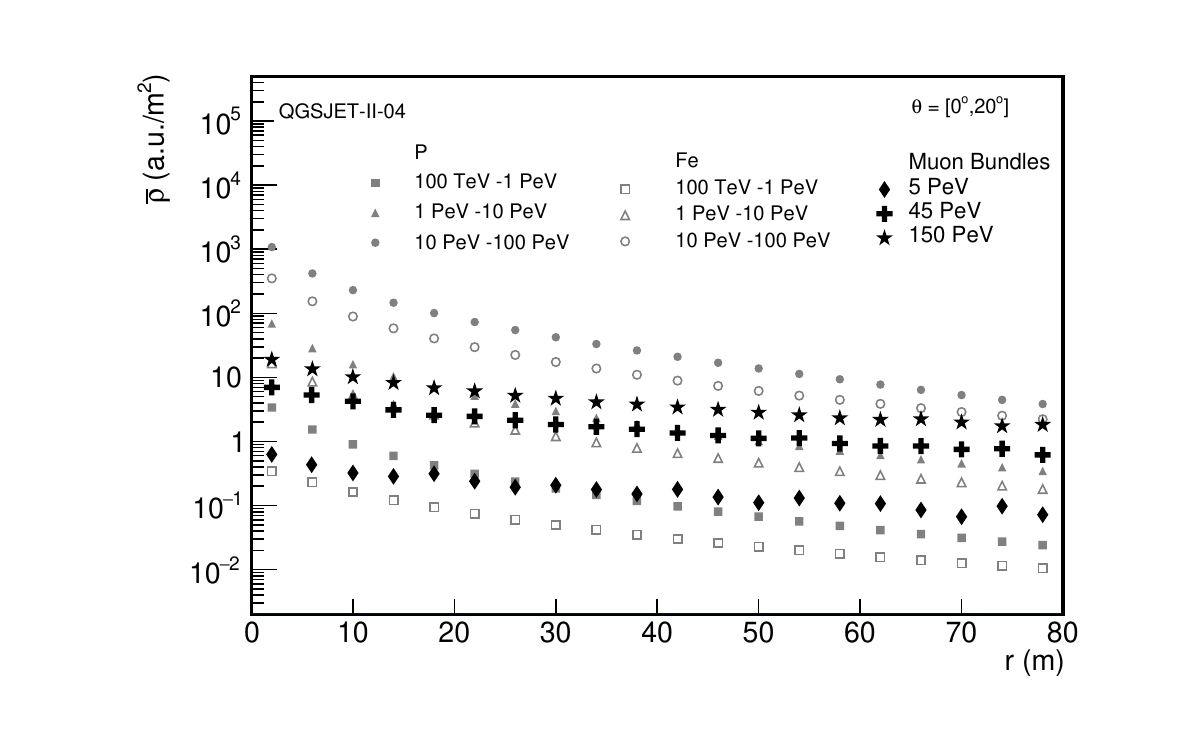}
  \caption{Predictions for the average of the lateral density distributions measured in the RPCs for muon bundles induced by strangelet events in the scenario of \cite{Maciej19} for primary energies of  $5 \, \pev$ (black diamonds), $45 \, \pev$ (black crosses) and $150 \, \pev$ (black stars) and zenith angles $\theta < 20^\circ$. For comparison, the averages of the lateral density profiles for EAS expected for primary protons and iron nuclei at different energy intervals according to QGSJET-II-04 are presented. Results for hadronic EAS for the energy intervals $100 \, \tev - 1 \,  \pev$, $1 \, \pev - 10 \,  \pev$ and $10 \, \pev - 100 \,  \pev$ are displayed with gray squares, triangles and circles. For proton and iron nuclei closed and hollow symbols are used, respectively. For hadronic events, we also used EAS with zenith angles below $20^\circ$. }
 \label{Muonbundle_1df}
\end{figure}

To investigate the potential of MATHUSLA to detect possible muon bundles from an exotic origin in the atmosphere, we simulated the signals of muon bundles at the MATHUSLA RPCs that are created by strangelets and we estimated the lateral density distributions of these events. These profiles were compared with the  corresponding predictions for protons and iron-induced EAS in order to look for possible differences in MATHUSLA between exotic and hadronic events. To simulate the production of muon bundles due to strangelets, we took the shower muon content of some QGSJET-II-04 EAS produced by iron nuclei with vertical incidence ($\theta < 20^\circ$) and with primary energies $1 \, \pev$, $10 \, \pev$ and $30 \, \pev$. These shower muon particles were passed to our MATHUSLA simulation and reconstruction programs, which are described in sections \ref{s.simulations} and \ref{EASreco}. Then we applied the selection cuts described in section \ref{selection} and, next, with the chosen data we built the average of the muon lateral density distributions of the events at different energies. The magnitudes of the lateral distributions were multiplied by an appropriate factor to reproduce the muon numbers that we should expect in strangelet events that have a similar $X_{max}$. To find this factor we used the predictions for the mean $X_{max}$ and muon content of EAS induced by protons, iron nuclei and strangelets given in \cite{Maciej19}. These plots indicate that EAS produced by iron nuclei with energies $1 \, \pev$, $10 \, \pev$ and $30 \, \pev$ have on average the same $X_{max}$ as muon bundles with energies $5 \, \pev$, $45 \, \pev$ and $150 \, \pev$, respectively, and that the muon content of these exotic events has approximately $1.3$ more particles than Fe induced events with the same primary energy. For QGSJET-II-04, we found $N_\mu \approx 7.3 \times 10^{-2} (E/\gev)^{0.92}$ for Fe nuclei with $\theta < 20^\circ$ and $E = [1 \, \pev, 1 \, \mbox{EeV}]$ at the MATHUSLA detector. Hence, we need to apply a scale factor\footnote{The procedure has several drawbacks but it gives us a naive estimation of the effect that we are looking for in MATHUSLA. In \cite{Maciej19}, for example, the EAS were simulated for the location of the Pierre Auger Observatory, which is situated at an altitude of $1400 \, \mbox{m}$ a.s.l. \cite{PAOInclined2014}. In addition, for EAS simulations in \cite{Maciej19}, a modified version of SHOWERSIM \cite{showersim84} was employed. Also $X_{max}$ for  hadronic EAS from vertical directions is dominated by photons and electrons and not by muons \cite{EASmodelsReview}.} of $1.3 \times (E_{str}/E_{Fe})^{0.92}$  to the lateral density of muons simulated for Fe nuclei of energy $E_{Fe}$ in order to obtain the respective profile for strangelets of primary energy $E_{str}$.

The average of the lateral density profiles for the simulated muon bundles are presented in Fig.~\ref{Muonbundle_1df} in comparison with QGSJET-II-04 predictions for vertical EAS resulting from protons and iron primaries. We note that the lateral distributions of muon bundles from strangelets decrease less rapidly as a function of $r$ than the respective distributions for hadronic EAS in the scenario described in \cite{Maciej19}. A more dedicated analysis is underway. However, this preliminary analysis suggests that MATHUSLA measurements could test the hypothesis on the strangelet origin of the high multiplicity muon bundles observed by ALICE by looking for anomalous events with flatter lateral density distributions than the cosmic-ray-induced EAS. The tests can be more restrictive if these analyses also include the golden events measured in correlation with the CMS detector in the hypothetical case that simultaneous measurements with the underground detector become available. Under these circumstances, by analysing MATHUSLA data from EAS correlated with muon bundles detected at the CMS detector, we could directly verify or discard the exotic nature of these underground events in the context of models like \cite{Maciej19}.

\section{Discussion}
\label{discussion}

  The purpose behind the proposal to install an RPC layer in MATHUSLA is to exploit the advantages for EAS studies of combining a large area and full coverage detector that offers a wide dynamical range for charged particle densities, with the benefits of the cosmic-ray tracking capability of MATHUSLA. The RPC layer would provide detailed information of cosmic-ray induced EAS about the  spatial and time structure of the shower front in EAS events with different inclinations, while the  tracking detector system of MATHUSLA would mainly help to measure the arrival times and radial density profiles of shower particles from inclined EAS. The combination of MATHUSLA with the RPC detector layer would result in an improved tracking detector for studying muon bundles from very inclined directions  and for measuring vertical EAS events of TeV energies. 

\begin{figure}[!b]
   \begin{center}
   \includegraphics[width=10cm]{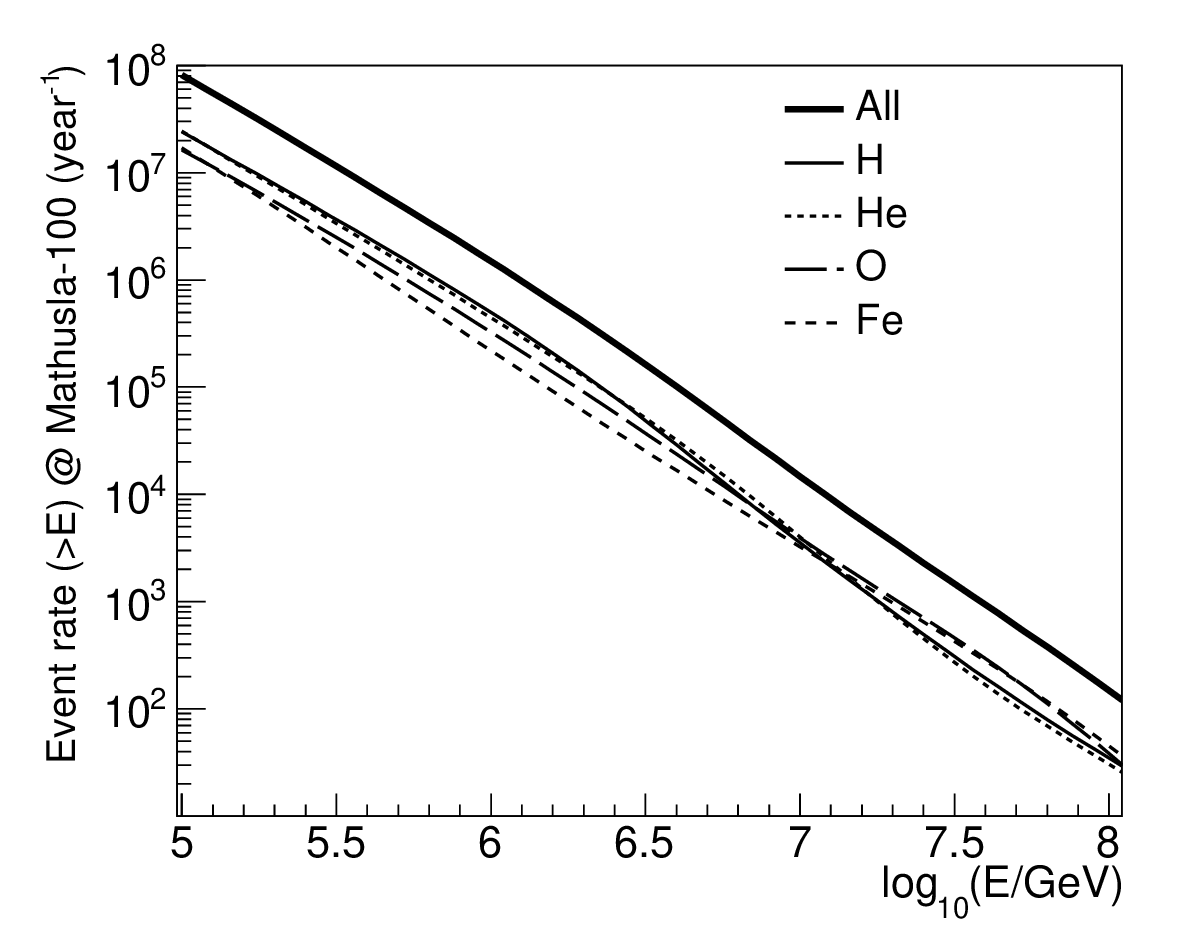}
   \caption{Number of cosmic-ray events per year above an energy $E$ expected at MATHUSLA from the zenith angle interval $\theta = [0^\circ, 90^\circ]$ and for an instantaneous field of view of $\pi$ sr considering  an horizontal detector geometry. These results assume that cores of the cosmic-ray induced EAS are located inside the $100 \, \mbox{m}^2 \times 100 \, \mbox{m}^2$ instrumented area and that the detector has full efficiency. In the calculations, the intensities of cosmic rays for different nuclei were taken from the Global Spline Fit model that was tuned to the latest measurements of cosmic rays in the energy range from  3 GeV to $10^{11}$ GeV \cite{Dembinski}. The black line represents the predictions for the sum of all-particle cosmic-ray nuclei and the colour  lines, the expectations for four representative mass groups.}
   \label{EASeventrate}
   \end{center}
   \end{figure}

According to the analyses of selection cut efficiency of the  RPC layer presented in subsection \ref{effrpc} the instrument would be sensitive to EAS from primary cosmic rays with energies as low as $10^{14} \, \ev$. The particular energy threshold will depend on the selection cuts and arrival direction of the shower. For the selection cuts here employed, the proposed design would have an energy threshold of $10^{14} \, \ev$ to vertically incident cosmic rays, while for inclined events with zenith angles between $70^\circ$ and $80^\circ$ the threshold would be $10^{16} \, \ev$. For higher energies, EAS detection at the RPC would be restricted  by the reduction in the intensity of cosmic rays and  the limited sampling size of the detector. This point is illustrated in  Fig.~\ref{EASeventrate} that shows the expected event rate of cosmic-ray showers at MATHUSLA as a function of the primary energy considering only events with shower cores impinging on the instrumented area and assuming full detector efficiency. From the rate shown on this plot and  considering a collection time of around $3 \, \mbox{yr}$, which is  similar to the projected time of the HL-LHC run, we estimate that MATHUSLA would collect $\sim 530$ air showers with $\theta \leq 70^\circ$ at energies $E > 10^{17} \, \ev$ and almost $52$ events with $\theta = [70^\circ, 80^\circ]$ in the same energy range.  Therefore, from the previous results, we could expect that the RPC detector layer would be  mainly constrained to the primary energy range $E = [10^{14} \, \ev, 10^{17} \, \ev]$ of the cosmic-ray spectrum, i.e. the region around the knee. Based on similar analyses for the MATHUSLA scintillating layers, see section \ref{sciperformance}, we found analogous energy thresholds to those for the RPC detector. 

   From Fig.~\ref{biasCoreRPC}, \ref{biasAngleRPC} and \ref{Mor3d}, we observe that the proposed RPC layer of MATHUSLA will make possible precise measurements of the core location and arrival direction of vertical events. The performance of the instrument, however, would be reduced when increasing the inclination of the shower axes of the events. In this case, our estimations for the angular resolution of the RPC layer might be improved by using the tracking capabilities of the MATHUSLA scintillating detectors. 
   This possibility needs further study, but it should apply also to  vertical events in the $\tev$ energy regime that have particle densities below the saturation level of the scintillating bars (see section \ref{saturation_sci}). A good angular resolution is important when building precise maps of the arrival direction of cosmic rays. These measurements are the basis of  studies of the anisotropy \cite{CosmicRays2,Deligny2016,Ahlers2017} and  clustering of these particles \cite{Antoni2004,Kang2015} that are done to get a better understanding of the propagation of this radiation in the interstellar medium and to constrain models on the spatial distribution of galactic cosmic-ray sources \cite{Pohl_2013}. Shower core measurements could also be possible with the MATHUSLA scintillating detectors for TeV showers and inclined events (see figs.~\ref{biasEAS} and \ref{biasCoreSci}). However, we did not observe an improvement here in comparison with the results for the RPC detector.

From figures \ref{3dMCevents}, \ref{3dMCeventsinclined} and  \ref{Mor3d}, we observe that the proposed RPC detector layer of MATHUSLA would provide  detailed information on the spatial and temporal structure of the EAS, which could be used to  estimate the primary energy of the event, to measure the distribution of charged particles at the shower front, and to investigate the primary mass of cosmic rays. This unique characteristic of the planned RPC detector could help to uncover hints of the internal hadronic processes that occur inside the EAS. Searches for clusters or multiple cores in the shower front of an event \cite{multicoreArgo2009,multicoreArgo2011}, as well as measurements of the amount and distribution of charged particles at the forward direction in EAS could be easily done with the RPC layer and would contribute to testing the validity of modern hadronic interaction models at high energies  \cite{qgsjet,qgsjet2,Pierog_2015,SIBYLL2.3c_2017}.

    We found that the amplitude of the lateral distribution of the deposited charge (see Fig.~\ref{logEvslogA}) and the value of the charge density at a fixed distance from the core (for example, Fig.~\ref{logEvsrho}) are the measured RPC variables that could be used to estimate the primary energy of the EAS.  The disadvantage here is that the energy calibration will be model dependent and some influence from the shower fluctuations may be expected in the result because the measurements will be made close to ground level. Notwithstanding, cross-calibration with other cosmic-ray observatories by using the knee feature of the energy spectrum or the intensity of cosmic rays could be exploited to accomplish this task, as it is done, for example, in \cite{Dembinski}. Alternative energy calibration methods, like neural networks \cite{ICECUBE19}, need to be also investigated. In this case, information on the time structure of the EAS from the RPC  and on the spatial structure of the shower front from the scintillating planes may increase the precision in the energy assignment of the shower events.

    Another observable that also depends on the primary energy of the EAS is  the fraction of Big Pads with signal, $f_{hit}$, as seen in Fig.~\ref{fhitrpc}, left and right. The sensitivity of this observable, however, seems to be constrained to energies smaller than  $8 \times 10^{15} \, \ev$  for vertical showers, since at higher energies, $f_{hit}$ at the RPC is saturated. In contrast, for inclined EAS, we do not see saturation effects on $f_{hit}$ at least up to $2.5 \times 10^{17} \, \ev$. In addition, this observable shows a smaller dependence on the nature of the primary particle for inclined events. Thus, $f_{hit}$ could also provide an energy estimate with reduced  sensitivity to the mass of the primaries for data with large zenith angles.  However, for vertical events, the mass sensitivity of the fraction hit increases.  Consequently, the fraction hit from the RPC layer may be also used to study  the mass composition of primary nuclei at least in the low-energy regime ($E < 10^{15} \, \ev$). From  Fig.~\ref{fhitsci}, we notice that the scintillating layers of MATHUSLA may also provide an estimation of the primary energy of the events by means of the fraction of scintillating bars with signals, which exhibits a larger energy threshold for vertical EAS (at around $\sim 4 \times 10^{16} \, \mbox{eV}$) than the corresponding one for the RPC detector.

    For the research of the composition of cosmic rays with the proposed RPC layer of MATHUSLA, a mass sensitive observable could be derived from the analysis of the slope of the lateral distribution of EAS. In this case, the slope of the lateral density of particles can be parameterised in terms of the lateral shower age using an NKG-like function (c.f. section \ref{LDFdescription}). Based on our studies in section \ref{ageandmass}, the lateral shower age seems to be sensitive to the type of primary nuclei in MATHUSLA (see Fig.~\ref{AgevslogA}) and therefore could be employed for analyses of the relative abundances of cosmic rays at least for vertical EAS. 

Further sensitivity to the mass composition could be gained by looking at the 2D morphology of the air shower, as measured with the RPC detector, by exploiting the fine-grained spatial resolution of the instrument along with its large coverage. According to our MC simulations, there appear to be mass-dependent differences in the 2D shape of the shower front of EAS as measured with the RPC (see section \ref{spatialstructure}). Fig.~\ref{Mor3d} illustrates this point. For example, the clumpiness in iron-induced EAS seems to be more pronounced than for protons of the same energy, and proton-induced events seem to produce more particles in the forward direction than iron nuclei of identical energy. A neural network or deep learning algorithm could be trained with this detailed information on the 2D spatial structure of the EAS to improve the mass classification of cosmic-ray events. In this regard, additional information from the scintillating layers on the  morphology of the EAS front could be useful. At TeV energies, the MATHUSLA tracking detector could give additional data on the particle densities at the EAS front in regions outside the shower core, where the scintillating bars are not saturated (see Fig.~\ref{density_sat_sci}). At higher energies ($> 1 \, \pev$), however, the measurements of the lateral particle densities with the tracking detector will be compromised because of the saturation of both the scintillating bars and the fraction of detector elements activated in the scintillating layers. 
  
     The role of the bottom tracking layers in MATHUSLA has not been investigated yet, but it may also provide valuable data for the determination of the primary composition of cosmic rays. For example, for inclined EAS the scintillating bars located close to the walls of the pit  that are shielded by the surrounding soil and rock could allow the measurement of local muon densities at the EAS front albeit for a narrow range of radial distances. 
    
     The cosmic-ray composition in the $\pev$ energy region is an interesting topic. There are still several open questions, which include, for example, the determination of the position of the  cut in the spectrum of $\mbox{H}+\mbox{He}$ primaries that is responsible for the structure known as the knee in the all-particle energy spectrum of cosmic rays. The KASCADE detector has located the  knee of the $\mbox{H}+\mbox{He}$ mass group at around $4 \, \pev$ \cite{Antoni2005,Apel2013}, but ARGO-YBJ recently measured this structure at energies below $1 \, \pev$ $(\sim 700 \, \tev$) \cite{Bartoli2015a}. It is not clear yet what is the origin of the discrepancy. It is an unsolved issue that is important to resolve, since the position of the cut depends on the reach of the acceleration mechanisms of cosmic rays \cite{Peters61}, the type of source \cite{Gaisser2013}, and the leakage of cosmic-ray nuclei from our galaxy \cite{Giacinti15,Hoerandel04}.

     At ground level, on average, almost $80\%$ of the particle content  of inclined EAS for primary energies above a few $\pev$ comprises muons according to our MC simulations with QGSJET-II-04 (see. Fig.~ \ref{muonfractionEAS}). This suggests that EAS at large zenith angles are basically muon bundles, which could be detected and studied with the scintillating detectors of MATHUSLA.
      As observed in Fig.~\ref{Muonbundle_1}, left, the RPC layer of MATHUSLA will make possible local density measurements of muon bundles from inclined EAS that together with similar measurements from the scintillating MATHUSLA detectors and the tracking capabilities of MATHUSLA could lead to the design of different tests of high-energy hadronic interaction models (see, for example, Fig.~\ref{Muonbundle_1}, right). This is also an interesting subject, whose study has revealed important discrepancies between the model predictions and data on the shower muon content that indicate problems in the understanding of the physics of hadronic interactions at high energies \cite{Whisp}.

       Detail RPC measurements on the lateral density distributions of EAS could be also used to look for exotic events (muon bundles) in the background of hadronic showers that could be created by new types of particles such as strangelets in scenarios such as those described in \cite{Maciej19} and discussed in section \ref{exoticevents}. These primaries could come from stars made of strange quark matter \cite{Alcock1986,Cheng2006,Kan2017}.

        If we could add to the fine spatial and temporal measurements of the shower front, the muon data from the CMS detector, then correlations between the content of charged and penetrating particles in the shower could be investigated and more interesting analyses of the EAS could be devised based in the local measurements of the density, charge ratio, momentum and energy spectrum of high-energy muons ($E_\mu \gtrsim 61 \, \gev$)  at different radial distances from the shower core, various zenith angle intervals, and distinct primary energies that could serve to give a more comprehensive picture of the hadronic physics in the EAS and also to investigate the primary composition of cosmic rays around the knee. 

The idea of probing underground shower muons in connection with EAS at CERN is not new. The L3+C experiment was conceived to investigate this issue \cite{L3C_cosmicmuons_02}, and there was also an early proposal at the LHC called ACME, above the ATLAS interaction point that was designed to continue this research \cite{ACME_ATLAS}. It is worth noting that if we could use the proposed RPC detector layer and the CMS  detector at CERN, it could also be possible to study the relation between underground muon bundles and EAS induced by cosmic-ray primaries of different energies and nuclear masses and to put limits to strangelets scenarios. It has been speculated that such particles may also produce large multiplicity muon bundles in underground detectors \cite{Farhi84}.  Correlated measurements between the MATHUSLA RPC layer and the CMS detector in this context would enable exploring these exotic physical possibilities and put limits on them. The investigation of these subjects is attractive, as their existence may reveal new kinds of astrophysical objects in the universe, new states of nuclear matter, and a new type of astrophysical radiation. 
  
        From this work, the cosmic-ray physics targets for the scintillating layers of MATHUSLA are mainly the study of the  energy spectrum and composition of cosmic rays between $10^{14} \, \ev$ and $10^{17} \, \ev$, the analysis of muon bundles, tests of hadronic interaction models of high energy with the muon content of inclined EAS and studies of the skymap distribution of the arrival directions of cosmic rays. The energy range from $10^{14} \, \ev$ to $10^{15} \, \ev$ in the energy spectrum of cosmic rays is very interesting as it has not been completely explored because it is located at the limit where direct and indirect detection techniques of cosmic rays can be applied. If the CMS experiment also joins this quest, the data from MAHUSLA and CMS would allow looking for connections of underground muon bundles and hadronic shower events. Even more, the CMS-MATHUSLA data on cosmic rays would make it possible to explore the  spectrum, local density, and charge ratio of shower muons of high energy at different lateral distances from the EAS core, including the forward region.

\subsection{Comparison with other EAS detectors}
  
  MATHUSLA will be a new type of instrument to investigate in detail the temporal and spatial structure of EAS at ground level and to study the physics of cosmic rays for primary energies between $100 \, \tev$ and $100 \, \pev$. The detector will sample a large area ($\sim 10^4 \, \mbox{m}^2$) of the EAS front with an RPC plane characterised by a huge coverage ($\sim 81 \%$) and an excellent spatial resolution ($\sim 1 \, \mbox{m}^2$ using the Big Pads, and smaller employing the RPC strips). It will also use a tower of scintillating detector planes, which will provide complementary information about the physical size and shower core location of vertical EAS at TeV energies and, for inclined events,  their arrival direction using the tracking properties of the MATHUSLA detector. There is a possibility that the CMS detector could join the MATHUSLA cosmic-ray program. If so, for some shower events measured during the high-luminosity runs at the LHC, the CMS underground detector would provide additional data on the local multiplicity, momentum, and charged ratio of high-energy muons at the EAS front.  Such a combination of air shower detectors has not existed to date in a single EAS experiment, and given the characteristics of MATHUSLA, it could provide novel information about the characteristics of the air showers induced by cosmic rays and put limits on the hadronic physics involved.

  In table \ref{T:ExperimentsCoverage}, we show a comparative table of the physical extension (A), sensitive area ($A_{s}$), and size of the detector modules ($A_{unit}$) of the MATHUSLA RPCs and the detection systems of other air showers installations that are optimised for cosmic-ray studies within the  $10^{14} \, \ev$ to $10^{17} \, \ev$ energy range: LHAASO \cite{LHAASO2019}, HAWC \cite{Alfaro2017}, ICETOP \cite{ICECUBE13,ICECUBE19,ICECUBE19b} and TALE \cite{TALE2016,TALETA,TALE2018}.  We also include details on ARGO-YBJ \cite{Bartoli2015a,Bartoli2015b}, which is the predecessor of the proposed RPC layer of MATHUSLA, and  KASCADE \cite{Antoni2003}, which was an important air shower detector array dedicated to studying the electromagnetic, muon and hadron components of EAS at ground level with the objective of investigating the energy spectrum and composition of cosmic rays from $1 \, \pev$ up to $100 \, \pev$. KASCADE incorporated several particle detector systems: an array of scintillating detectors, a muon tracking detector, and a central calorimeter.

\begin{table}[!t]
\caption{Comparison with other air shower detectors}
\label{T:ExperimentsCoverage}
\footnotesize
\centering
\begin{tabular}{l c  c  c  c}
\hline
\textbf{Experiment} & \textbf{A } & \textbf{$A_{s}$ } &  \textbf{$A_{unit}$ } & \textbf{Coverage} \\
 & \textbf{ ($10^4 \,  \mbox{m}^2$)} & \textbf{ ($10^4 \,  \mbox{m}^2$)} &  \textbf{ ($\mbox{m}^2$)}& \textbf{(\%)} \\
\hline 
MATHUSLA   &$1$&$0.81$& $1$& $81$\\
HAWC  \cite{Alfaro2017}  &&&& \\ 
\textit{central water cherenkov detector array}  &$2.2$&$1.26$&$41.9$& $57.07$\\ 
LHAASO \cite{LHAASO2019} &&&& \\ 
\textit{water cherenkov detector array}  &$7.8$&$7.8$& $25$& $100$\\
\textit{e.m. detector array}  &$10^2$&$0.52$& $1$& $0.52$\\ 
\textit{muon detector array}  &$10^2$&$4.22$& $36$& $4.22$\\
ICETOP/ICECUBE \cite{ICECUBE13,ICECUBE19,ICECUBE19b} &&&& \\
\textit{ice cherenkov detector array}  &$10^2$&$0.042$& $2.6$& $0.42$\\ 
TALE/Telescope array \cite{TALE2016,TALETA,TALE2018}  &&&& \\
\textit{TALE e.m. detector array}  &$7 \times 10^3$&$0.031$& $3$& $4.4 \times  10^{-4}$\\ 
\textit{TA e.m. detector array}  &$7 \times 10^4$&$0.152$& $3$& $2.2 \times  10^{-4}$\\
ARGO-YBJ \cite{Bartoli2015a,Bartoli2015b} &&&& \\ 
\textit{central carpet}  &$0.58$&$0.54$& $1.7$& $93.00$\\ 
\textit{guard ring}  &$1.1$&$0.0043$& $1.7$& $0.39$\\ 
KASCADE \cite{Antoni2003,Doll2002} &&&& \\ 
\textit{e.m. detector array}  &$4$&$0.049$& $1.94$& $1.23$\\  
\textit{muon detector array}  &$4$&$0.062$& $3.24$& $1.56$\\ 
\textit{muon tracking detector}  &$0.024$&$0.013$& $8$& $53.87$\\ 
\textit{central calorimeter}  &$0.032$&$0.031$& $0.06$& $97.66$\\
\hline
\end{tabular}
\end{table}   

From the list in table \ref{T:ExperimentsCoverage}, we can observe that only the water Cerenkov detector of LHAASO, the central detector carpet of ARGO-YBJ and the central calorimeter of KASCADE have a large coverage. The water Cerenkov array of LHAASO has a $100 \%$ coverage and a large detection surface. However, its spatial resolution is not comparable to the one expected for MATHUSLA. The  central carpet of ARGO-YBJ had a coverage of $93 \%$ and a spatial resolution of  $1.7 \, \mbox{m}^2$ (size of the Big Pads), but a physical area that was a factor of $0.58$ that of MATHUSLA. With regard to KASCADE, the central calorimeter had a coverage close to $98 \%$ and  better spatial resolution  ($0.06  \, \mbox{m}^2$), but a size $31$ times smaller than the one foreseen for the MATHUSLA detector. From these comparisons, we can conclude that MATHUSLA would provide improved time-spatial measurements of the EAS front with large coverage and physical area; conditions that have not  simultaneously existed in previous or existing air shower facilities yet.

 In comparison with the EAS facilities described above, except ARGO-YBJ and HAWC, MATHUSLA will measure only one shower component, i.e., charged particles. This can be seen in table \ref{T:Experiments}, where the EAS observables accessible to LHAASO, HAWC, ICETOP, TALE, ARGO-YBJ and the KASCADE experimental facilities are compared with MATHUSLA. We also included TAIGA-HiScore \cite{Taiga19,Taiga19a} in the list.  From table \ref{T:Experiments}, we observe that MATHUSLA has the disadvantage that it could not measure the muon component or the longitudinal profile of the EAS event-by-event. This restriction implies a smaller sensitivity to the mass of the primary particle and a larger dependence of the energy scale on the primary composition. However, by using good resolution measurements of other shower properties with the MATHUSLA RPCs, such as the arrival times of the shower front, the shape and the amplitude of the lateral density profile, the 2D spatial structure of the EAS and the fraction of hits, along with complementary data from the scintillating detector planes, the composition dependence of the energy assignment could be reduced and the mass sensitivity increased in MATHUSLA. The experiment may also use local muon measurements from the CMS detector. if this is the case, we could look for correlations between underground muon data from CMS and EAS events observed with the detector planes of MATHUSLA. These golden events would allow MATHUSLA to study the mass composition of cosmic rays below and above the energy region of the knee. The disadvantage of MATHUSLA in this regard will be the statistics since the measurements will be restricted to the data-taking periods of CMS for the HL-LHC.

 There are several open questions at energies just below the knee of the cosmic-ray spectrum, such as the shape of the energy spectrum of iron nuclei \cite{Antoni2005,Apel2013,Garyaka2007} and the discrepancy between ARGO-YBJ \cite{Bartoli2015a} and KASCADE \cite{Antoni2005,Apel2013} results on the location of the knee feature in the spectrum of light elemental nuclei. Measurements in this energy range are difficult for experiments like KASCADE, ICETOP, TALE, and TAIGA that have an energy threshold near the knee due to limitations of the indirect cosmic-ray detection techniques, and for instruments like HAWC and ARGO-YBJ, where the sensitivity ends at energies near the location of this feature because of restrictions of the size of the detectors. Generally, in order to achieve lower primary energy thresholds the large detector arrays must be located at high altitudes. LHAASO is one of a new generation of EAS detectors located at high altitudes that will allow for cosmic-ray composition studies from $10^{14} \, \ev$ up to $10^{18} \, \ev$ covering the knee region. Cosmic-ray measurements with EAS detectors starting at the $\tev$ region are scarce, hence MATHUSLA could become very useful to confirm or complement the measurements on the mass composition of cosmic rays performed with high-altitude EAS observatories such as LHAASO. 

 \begin{table}[!t]
\caption{Experimental comparison}
\label{T:Experiments}
\footnotesize
\centering
\begin{tabular}{l c  c  c  c c }
\hline
\textbf{Experiment} & \textbf{Energy range} & \textbf{Altitude} &  \textbf{Size} & \textbf{EAS} & \textbf{Energy}\\
& $\boldsymbol{(\pev)}$ & \textbf{(m a.s.l.)} &  $\boldsymbol{(10^{4}}$ \textbf{m}$\boldsymbol{^2)}$ &  \textbf{component} &  \textbf{calibration} \\
\hline
MATHUSLA  & $10^{-1} - 100$  & $374$ & $1$ & $N_{ch}$   & Hadronic model \\
HAWC  \cite{Alfaro2017} & $10^{-3} - 1$  & $4100$ & $2.2$ & $N_{ch}$  & Hadronic model \\ 
LHAASO \cite{LHAASO2019} & $10^{-1} - 10^{3}$  & $4410$ & $100$ & $N_{ch}$, $N_e$, $N_\mu$, $N_h$ & Cherenkov light  \\ 
ICETOP/ICECUBE \cite{ICECUBE13,ICECUBE19,ICECUBE19b} & $0.25 - 10^{3}$  & $2835$ & $100$ & $N_{ch}$, $N_\mu$  & Hadronic model \\
TALE/Telescope array \cite{TALE2016,TALETA,TALE2018} & $2 - 2 \times 10^{3}$  & $1400$ & $7 \times 10^{4}$ & $N_{ch}$, $N_e$  & Cherenkov/Fluorescence light \\ 
TAIGA-HiSCORE \cite{Taiga19,Taiga19a} & $0.2 - 3$  & $675$ & $50$ & $N_e$ & Cherenkov light  \\
ARGO-YBJ \cite{Bartoli2015a,Bartoli2015b} & $0.0003 - 3$  & $4300$ & $1.1$ & $N_{ch}$ & Hadronic model\\ 
KASCADE \cite{Antoni2003} & $1 - 100$  & $110$ & $4$ & $N_{ch}$, $N_\mu$, $N_e$, $N_h$  & Hadronic model \\ 
\hline
\end{tabular}
\end{table}  

MATHUSLA is not located at a high altitude as HAWC or LHAASO, as seen in table \ref{T:Experiments}. This means that shower fluctuations will have a stronger influence on the energy determination of the primary nuclei in MATHUSLA. However, this issue is not expected to spoil studies of the energy spectrum of cosmic rays. Analyses with KASCADE have shown that the spectrum can be investigated even at an altitude close to sea level \cite{Antoni2005,Apel2013}.

LHAASO \cite{LHAASO2019} and ICETOP/ICECUBE  \cite{ICECUBE13,ICECUBE19,ICECUBE19b} are some of the present EAS experiments that have the capabilities to measure the muon component of hadronic EAS in the $\pev$ region (see table \ref{T:Experiments}). LHAASO contains an underground muon detector array, which when completed will cover an enormous surface $100$ times larger than the physical area of MATHUSLA. This detector system will be able to perform measurements of muon densities ($E_\mu > 1 \, \gev$) at different lateral distances at the EAS front and from distinct zenith angles \cite{LHAASOMUON}.  ICETOP can make density measurements of $\gev$ muons in EAS for primary energies between $\sim 1 \, \pev$ and $\sim 100 \, \pev$ and radial distances between $600 \, \mbox{m}$ and  $800 \, \mbox{m}$ for vertical showers using  a $1 \, \mbox{km}^2$ detector array \cite{ICETOPMUON}. KASCADE also provided local density measurements of muons at different energy thresholds, radial distances, and zenith angles for cosmic-ray-induced EAS of energies from $1 \, \pev$ to $100 \, \pev$. One of the KASCADE muon detector systems was a shielded scintillating detector array, which had an area of $4$ times larger than MATHUSLA and allowed to get muon data at energies $E_\mu > 230 \, \mev$ \cite{Apel2006}. MATHUSLA will not have a muon detector system. However, if the CMS experiment could be added to the MATHUSLA cosmic-ray studies, the correlation with underground measurements of shower muons would provide relevant information about the high-energy muon component  ($E_\mu > 61 \, \gev$) in EAS in the $\pev$ energy region. The disadvantage of this combined detection system would be that the muon data would be restricted to limited radial distances, which would depend on the impact point of the shower core at the ground and  the arrival direction of the EAS. In addition, the CMS data would be available only for events registered during the running time of the underground detector. The correlation of MATHUSLA with CMS would have the advantage that could make possible not only measurements of the local densities and multiplicities but also of the energy spectrum and charge ratio of high-energy muons event-by-event, even from the forward region of the hadronic collision. No EAS facility has such possibilities at the moment. If the MATHUSLA detector becomes a reality and the CMS muon data is added, the correlations of underground shower muons and surface EAS measurements could provide new insights into air shower physics at very high energies.

Underground measurements of high-energy shower muons are not possible for the experimental facilities of table \ref{T:Experiments}, but it could be possible in MATHUSLA with CMS.  At the moment, there is only one underground EAS experiment, which has a similar energy threshold  for shower muons as CMS, it is called EMMA and it is located in the Pyhasalmi mine, Finland under $75 \, \mbox{m}$ of rock \cite{EMMA18,EMMA18a,EMMA20}. EMMA is aimed to study  underground muon bundles and the composition of cosmic rays in the $1 \, \pev - 10 \, \pev$ interval. It consists of an array of eleven tracking muon detectors, each with an area of  $15 \, \mbox{m}^2$. The experiment has an effective area of $100 \, \mbox{m}^2$, which is similar to the  physical area of CMS, but EMMA offers a larger monitoring time and, therefore, a larger exposure\footnote{The exposure is the effective area of the detector multiplied by the  effective time of observation and the solid angle covered by the experiment.} than  CMS \cite{EMMA20}. CMS, however, has the advantage that it is a high-precision muon detector, has $100\%$ coverage, and can provide additional event-by-event information about shower muons than EMMA, such as information about the energy spectrum and charge ratio of these penetrating particles. The MATHUSLA and the CMS measurements could be correlated with EAS data measured at ground level to increase the reach of the air shower investigations, an option that is not possible at the moment in EMMA. 

MATHUSLA will also have tracking capabilities that are not available in other EAS experiments. The tower of scintillating layers and the RPC detector plane of MATHUSLA could work as a giant EAS tracking detector for charged particles in the EAS, albeit with some restrictions due to saturation effects in the scintillating bars for vertical events above $10 \, \pev$. An instrument such as MATHUSLA is unprecedented in the study of cosmic rays. It could help, for instance, to improve the estimation of the arrival direction of EAS and to look for point sources in the small-scale anisotropy maps of cosmic rays. The main motivation for these studies comes from the possibility that there could exist nearby sources of cosmic-ray nuclei that suffer small deviations in the interstellar magnetic field before arriving to the Earth and/or could be emitting detectable neutral particles, such as high-energy neutrons (at $E \sim 100 \, \mbox{PeV}$ one $n$ has a decay length of $1 \, \mbox{kpc}$) or gamma-rays \cite{Antoni2004,Sven07,Kang2015}.  The presence  of such sources might explain the existence of small-scale anisotropy or clustering of events in the maps of the arrival direction of cosmic rays.  This is an interesting possibility that has been barely explored at energies above the knee because it requires not only a good angular resolution but also enough statistics to compensate the low-intensity flux of cosmic rays at this energy regime. The presence of nearby accelerators of very high-energy cosmic rays is still an open question in astrophysics and one that it is important to verify or constrain with the MATHUSLA detector's capabilities. A nearby source could be an interesting laboratory to study the particle physics acceleration at high energies and could help to study the propagation of cosmic rays in the galaxy, to identify the types of cosmic accelerators, etc.

By looking at other particle tracking detector systems in different EAS experiments, we can also envisage additional studies for MATHUSLA. KASCADE, for example, was equipped with two tracking detector systems for shower muons and hadrons but smaller in size than MATHUSLA. In KASCADE, the muon tracking detector was employed for measurements of the lateral distributions of muons ($E_\mu > 800 \, \mev$) \cite{APEL201555}, muon production heights and muon multiplicities \cite{APEL2011476}. The hadron tracking detector was dedicated to measuring the deposited energy, multiplicity and arrival directions, local densities and geometric distributions of hadrons ($E_h > 10 - 50 \, \gev$) in the EAS front \cite{ApelMu2007,AntoniHadron2005,AntoniHadron2001,AntoniHadron1999}. These tasks were specialised for the muon and hadron components of the EAS  and some of them could also be employed for MATHUSLA, for example, for shower muons of inclined EAS. 

For muon bundle studies above $1 \, \pev$ and zenith angles $> 30^\circ$ \cite{Saavedra_2013}, there is a large facility called NEVOD-DECOR located in Moscow, Rusia \cite{Bogdanov20}. The experiment consists of a  Cherenkov water calorimeter (NEVOD) with a volume of $9 \, \mbox{m (hight)} \, \times 9  \mbox{m (width)}\times 26 \, \mbox{m (length)}$ and a $70 \, \mbox{m}^2$ tracking detector (DECOR)  composed of eight modules of streamer tubes, which covers three of the lateral surfaces of the NEVOD detector. It has a spatial resolution of $1 \, \mbox{cm}$ and an angular precision of $1^\circ$ for individual tracks. In comparison, the tower of scintillating layers of MATHUSLA of the intermediate and upper detector layers would have a vertical dimension similar to the NEVOD-DECOR experiment, but a larger sensitive area, which could reduce the statistical uncertainties on the measurements of the local densities of muon bundles. This advantage of MATHUSLA could help to improve the investigations of muon bundles between $1 \, \pev$ and $1 \, \eev$. Particularly important are the measurements close to the ultra high-energy regime, where a muon excess has been reported in EAS from inclined directions by  cosmic-ray detectors like the Pierre Auger observatory \cite{PAOInclined2014}. The origin of this anomaly has not yet been well understood and is still under investigation by different experimental collaborations \cite{Whisp}. In addition,  MATHUSLA could also provide additional spatial-time information on a larger scale than the NEVOD-DECOR experiment that could help to study the structure of the EAS front. However, the NEVOD detector can make calorimetric measurements of the shower for muon bundles from inclined directions \cite{NEVOD2021}, which MATHUSLA cannot. In this comparison, we have not considered the bottom scintillating layers of MATHUSLA, which could increase the effective area of the detector and provide additional data of inclined events, for example, information on high-energy muons by using the surrounding rock as an absorbing layer.
   
In table \ref{T:Experiments2}, we summarise the shower core and angular resolutions for vertical EAS in ARGO-YBJ, HAWC,  ICETOP, TALE, the main array of KASCADE and MATHUSLA at about $1 \, \pev$ for vertical events. From these values we note that the core resolution expected for MATHUSLA is comparable to LHAASO, ARGO-YBJ and KASCADE, and better than HAWC and ICETOP. While the angular resolution is similar to the resolution of ICETOP. It also appears to be better than the TALE angular resolution, but worse than that for the experiments ARGO-YBJ, HAWC, KASCADE and LHAASO. We expect to be able to improve the shower core and angular direction precision of MATHUSLA. For this,  we employed simple EAS reconstruction techniques, which could be further refined. By exploiting the fine granularity of the RPC detector improvements are also possible. In addition, we are investigating alternative EAS reconstruction methods.

\begin{table}[!t]
\caption{Core and angular resolution of EAS in different cosmic-ray experiments at $1 \, \pev$ in comparison with MATHUSLA.}
\label{T:Experiments2}
\footnotesize
\centering
\begin{tabular}{l  c  c  c  }
\hline
\textbf{Experiment} & 
\textbf{Core}     & \textbf{Pointing}  \\
 & 
\textbf{position}  & \textbf{direction} \\
\hline
ARGO-YBJ \cite{Bartoli2017b} &  
$\lesssim 2 \, \mbox{m}$     & $\lesssim 0.3^\circ$ \\
HAWC  \cite{Alfaro2017} &  
$\lesssim 8 \, \mbox{m}$   & $\lesssim 0.3^\circ$\\
ICETOP/ICECUBE \cite{ICECUBE19} &  
 $\lesssim 16 \, \mbox{m}$          & $\lesssim 0.7^\circ$ \\
TALE (Telescope array) \cite{TALETA}  &
 $-$           & $\lesssim 5^\circ$  \\
KASCADE main array \cite{Antoni2003}  &  
$\lesssim 1.5 \, \mbox{m}$          & $\lesssim 0.25^\circ$ \\ 
LHAASO \cite{Aharonian2021}  &  
$\lesssim 2.5 \, \mbox{m}$          & $\lesssim 0.16^\circ$ \\ 
MATHUSLA-100  & 
 $\lesssim 2 \, \mbox{m}$       & $\lesssim 1^\circ$ \\ 
\hline
\end{tabular}
\end{table} 

\section{Conclusions}
\label{s.conclusions}

  MATHUSLA comprises a large tower of tracking detector planes\footnote{“During the final stages of this work, the MATHUSLA collaboration added an additional layer of scintillating bars at the top of the detector. After evaluation, we conclude that the additional layer has no effect on the conclusions of this study} that will monitor a volume of $25 \, \mbox{m} \times 100 \, \mbox{m}^{2}$ to look for signals of exotic long-lived-particles produced at the interaction point of the CMS detector during the High-Luminosity LHC runs. The instrument can also serve as an extensive air shower detector to primary cosmic rays in the energy range from  $10^{14} \, \ev$ to $10^{17}  \, \ev$. The lower limit might be relaxed by reducing the number of hits required by the selection cuts. This way
  MATHUSLA will enable measurements of the  energy spectrum of TeV cosmic rays as well as of  muon bundles associated with inclined events.  Adding an RPC layer significantly enhances the  capabilities of MATHUSLA for cosmic-ray physics. The RPC detector system would provide fine space-time information about the EAS front at ground level that can be used to calibrate the primary energy of the event and study the mass composition of cosmic rays, as well as the structure of the EAS front and the arrival directions of the primary particles. The operation of the proposed RPC detector together with the scintillating tracking layers of MATHUSLA would improve the reconstruction of EAS induced by TeV cosmic rays, making possible the exploration of the energy spectrum of cosmic rays at the frontier between direct and indirect cosmic-ray experiments. At the moment, TeV measurements of cosmic rays with EAS detectors are done at high altitudes, but with MATHUSLA they could also be carried out close to sea level. There exists a possibility of including in MATHUSLA simultaneous observations of EAS with the  CMS detector. In this case, the CMS-MATHUSLA instrument would provide novel data on the muon content of EAS with energies above $61 \, \gev$ that would help to perform tests of high-energy hadronic interaction models and to explore the relation between underground muon bundles and cosmic rays. In conclusion, an experimental detector like MATHUSLA has the potential to address questions in particle physics and high-energy astrophysics.

\section*{Acknowledgments}
M.R.C. thankfully acknowledges the permission to use computer resources, the technical advice, and the support provided by the Laboratorio Nacional de Superc\'omputo del Sureste de M\'exico (LNS), a member of the Consejo Nacional de Humanidades, Ciencias y Tecnologías (CONACyT) national network of laboratories, with resources from grant number 201701035C. Financial support for this work has been received from CONAHCyT, grant number A1-S-13525. Also many thanks to the High Performance Computing Laboratory LARCAD-UNACH Mexico, and the Digital Research Alliance of Canada, for the use of the Cedar computing cluster. CA and HJL thank the U. S. National Science Foundation for their support. O.G. Morales-Olivares was supported by a CONACyT postdoctoral fellowship. J.C. Arteaga-Velázquez and D. Rivera thank the support from CONAHCyT, grant number A1-S-46288.

\bibliography{references}
\bibliographystyle{JHEP}


\end{document}